\documentclass[twocolumn]{aastex631}
\usepackage{CJK} 

\usepackage{amsmath}
\usepackage{amssymb}
\usepackage{epsfig}
\usepackage{apjfonts}
\usepackage{natbib}
\usepackage{hyperref}
\usepackage{epstopdf}
\usepackage{verbatim}
\usepackage{enumitem}
\usepackage{color}
\setlist[enumerate]{itemsep=-1mm}
\usepackage{soul, xcolor}
\setstcolor{blue}

\usepackage{mathtools}

\usepackage{bigints}

\newcommand{\Ks}{\mbox{$K_{S}$}}
\newcommand{\Jtms}{\mbox{$J_{\rm 2MASS}$}}
\newcommand{\Htms}{\mbox{$H_{\rm 2MASS}$}}
\newcommand{\Ktms}{\mbox{$K_{S, \rm 2MASS}$}}
\newcommand{\Teffstar}{\mbox{$T_{\rm eff,\star}$}}
\newcommand{\Rstar}{\mbox{$R_{\star}$}}
\newcommand{\Mstar}{\mbox{$M_{\star}$}}

\newcommand{\logLbolstar}{\mbox{$\log(L_{\rm bol,\star}/\Lsun)$}}
\newcommand{\loggstar}{\mbox{$\log(g_{\star})$}}
\newcommand{\Teff}{\mbox{$T_{\rm eff}$}}
\newcommand{\R}{\mbox{$R$}}
\newcommand{\M}{\mbox{$M$}}
\newcommand{\Fbol}{\mbox{$F_{\rm bol}$}}
\newcommand{\logFbol}{\mbox{$\log{(F_{\rm bol})}$}}
\newcommand{\Lbol}{\mbox{$L_{\rm bol}$}}
\newcommand{\logLbol}{\mbox{$\log(L_{\rm bol}/\Lsun)$}}

\newcommand{\logLxbol}{\mbox{$\log{(L_{X}/L_{\rm bol})}$}}
\newcommand{\logRx}{\mbox{$\log{(R_{X})}$}}
\newcommand{\Fx}{\mbox{$F_{X}$}}

\newcommand{\logLx}{\mbox{$\log{(L_{X})}$}}
\newcommand{\logFxj}{\mbox{$\log{(F_{X}/F_{J})}$}}
\newcommand{\logFxk}{\mbox{$\log{(F_{X}/F_{K_{S}})}$}}
\newcommand{\logFxv}{\mbox{$\log{(F_{X}/F_{V})}$}}
\newcommand{\logFnuvj}{\mbox{$\log{(F_{NUV}/F_{J})}$}}
\newcommand{\ergscm}{\hbox{erg~s$^{-1}$~cm$^{-2}$}}
\newcommand{\YMKO}{\mbox{$Y_{\rm MKO}$}}
\newcommand{\JMKO}{\mbox{$J_{\rm MKO}$}}
\newcommand{\HMKO}{\mbox{$H_{\rm MKO}$}}
\newcommand{\KMKO}{\mbox{$K_{\rm MKO}$}}
\newcommand{\fsed}{\mbox{$f_{\rm sed}$}}

\newcommand{\kms}{\hbox{km~s$^{-1}$}}
\newcommand{\masyr}{\hbox{mas~yr$^{-1}$}}

\newcommand{\Lsun}{\mbox{$L_{\odot}$}}
\newcommand{\Rsun}{\mbox{$R_{\odot}$}}
\newcommand{\Msun}{\hbox{M$_{\odot}$}}
\newcommand{\Mjup}{\mbox{$M_{\rm Jup}$}}

\newcommand{\logg}{\mbox{$\log(g)$}}
\def\sun{\hbox{$\odot$}}
\def\earth{\hbox{$\oplus$}}

\usepackage{tikz}

\makeatletter
\global\let\tikz@ensure@dollar@catcode=\relax
\makeatother

\received{Nov 2nd, 2021; Revised Jun 25, 2022; Accepted Jun 28, 2022}
\submitjournal{ApJ}

\begin{document}


\begin{CJK*}{UTF8}{gbsn}

\title{COol Companions ON Ultrawide orbiTS (COCONUTS).\\ III. A Very Red L6 Benchmark Brown Dwarf around a Young M5 Dwarf}

\author[0000-0002-3726-4881]{Zhoujian Zhang (张周健)}
\affiliation{The University of Texas at Austin, Department of Astronomy, 2515 Speedway, C1400, Austin, TX 78712, USA}
\affiliation{Institute for Astronomy, University of Hawaii at Manoa, Honolulu, HI 96822, USA}
\affiliation{Visiting Astronomer at the Infrared Telescope Facility, which is operated by the University of Hawaii under contract 80HQTR19D0030 with the National Aeronautics and Space Administration.}

\author[0000-0003-2232-7664]{Michael C. Liu}
\affiliation{Institute for Astronomy, University of Hawaii at Manoa, Honolulu, HI 96822, USA}

\author[0000-0002-4404-0456]{Caroline V. Morley}
\affiliation{The University of Texas at Austin, Department of Astronomy, 2515 Speedway, C1400, Austin, TX 78712, USA}

\author[0000-0002-7965-2815]{Eugene A. Magnier}
\affiliation{Institute for Astronomy, University of Hawaii at Manoa, Honolulu, HI 96822, USA}

\author[0000-0002-2471-8442]{Michael A. Tucker}
\affiliation{Institute for Astronomy, University of Hawaii at Manoa, Honolulu, HI 96822, USA}

\author[0000-0002-0853-3464]{Zachary P. Vanderbosch}
\affiliation{The Division of Physics, Mathematics, and Astronomy, California Institute of Technology, Pasadena, CA 91125, USA}

\author[0000-0003-3429-7845]{Aaron Do}
\affiliation{Institute for Astronomy, University of Hawaii at Manoa, Honolulu, HI 96822, USA}

\author[0000-0003-4631-1149]{Benjamin J. Shappee}
\affiliation{Institute for Astronomy, University of Hawaii at Manoa, Honolulu, HI 96822, USA}

\begin{abstract}
We present the third discovery from the COol Companions ON Ultrawide orbiTS (COCONUTS) program, the COCONUTS-3 system, composed of a young M5 primary star UCAC4~374$-$046899 and a very red L6 dwarf WISEA~J081322.19$-$152203.2. These two objects have a projected separation of 61$''$ (1891~au) and are physically associated given their common proper motions and estimated distances. The primary star, COCONUTS-3A, has a mass of $0.123\pm0.006$~M$_{\odot}$ and we estimate its age as 100~Myr to 1~Gyr based on its stellar activity (via H$\alpha$ and X-ray emission), kinematics, and spectrophotometric properties. We derive its bulk metallicity as $0.21 \pm 0.07$~dex using empirical calibrations established by older and higher-gravity M dwarfs, and find this [Fe/H] could be slightly underestimated according to PHOENIX models given COCONUTS-3A's younger age. The companion, COCONUTS-3B, has a near-infrared spectral type of L6$\pm1$ \textsc{int-g}, and we infer physical properties of \Teff$= 1362^{+48}_{-73}$~K, \logg$= 4.96^{+0.15}_{-0.34}$~dex, $R = 1.03^{+0.12}_{-0.06}$~R$_{\rm Jup}$, and $M = 39^{+11}_{-18}$~M$_{\rm Jup}$, using its bolometric luminosity, its host star's age, and hot-start evolution models. We construct cloudy atmospheric model spectra at the evolution-based physical parameters and compare them to COCONUTS-3B's spectrophotometry. We find this companion possesses ample condensate clouds in its photosphere ($f_{\rm sed}=1$) with the data-model discrepancies likely due to the models using an older version of the opacity database. Compared to field-age L6 dwarfs, COCONUTS-3B has fainter absolute magnitudes and a 120~K cooler \Teff. Also, the $J-K$ color of this companion is among the reddest for ultracool benchmarks with ages older than a few 100~Myr. COCONUTS-3 likely formed in the same fashion as stellar binaries given the companion-to-host mass ratio of $0.3$ and represents a valuable benchmark to quantify the systematics of substellar model atmospheres.
\end{abstract}

\section{Introduction}
\label{sec:introduction}
Wide-field sky surveys have been a powerful means to construct a large census of planetary-mass and substellar objects in the solar neighborhood, allowing us to investigate star formation at the very low-mass end. These surveys also have helped to establish a photometric sequence of ultracool dwarfs spanning M, L, T, and Y dwarfs \citep[e.g.,][]{2004AJ....127.2948V, 2021AJ....161...42B, 2021ApJS..253....7K}, as well as to find peculiar objects, including L dwarfs with much redder infrared colors and fainter absolute magnitudes than field objects with similar spectral types \citep[e.g.,][]{2012AJ....144...94G, 2015AJ....150..182K, 2014AJ....147...34S, 2017AJ....153..196S, 2016ApJS..225...10F, 2013ApJ...777L..20L, 2016ApJ...833...96L}. 

The anomalous photometry of these unusually red L dwarfs suggest they have more dusty atmospheres \citep[e.g.,][]{2008ApJ...689.1327S} or a stronger thermo-chemical instability in their atmospheres \citep[e.g.,][]{2016ApJ...817L..19T} than normal field dwarfs. Many unusually red L dwarfs have low surface gravities and their anomalous spectrophotometry thus suggests a gravity dependence of ultracool dwarfs' atmospheric properties during the L/T tradition \citep[e.g.,][]{2006ApJ...651.1166M, 2016ApJS..225...10F, 2016ApJ...833...96L, 2017ApJ...850...46T}. However, a number of red L dwarfs likely have old ages ($\gtrsim 1$~Gyr) as they are not associated with any young stars or moving groups and have no low-gravity spectral features \citep[e.g.,][]{2008ApJ...686..528L, 2010ApJS..190..100K, 2013ApJ...772...79A}.

\begin{figure*}[t]
\begin{center}
\includegraphics[height=5in]{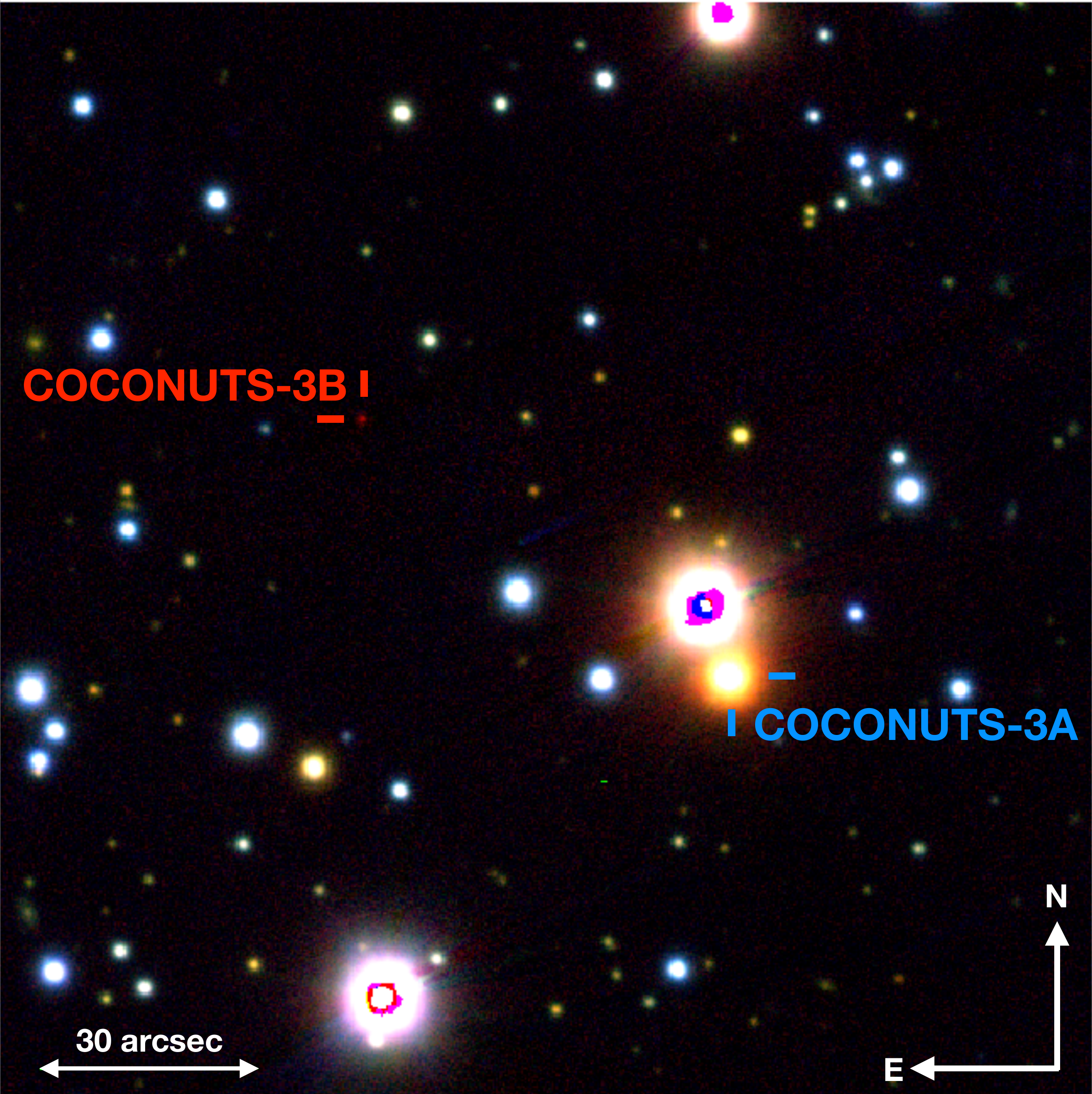}
\caption{M dwarf primary (COCONUTS-3A) and the L dwarf companion (COCONUTS-3B) in the tri-color Pan-STARRS1 image ($g_{\rm P1}$: blue, $i_{\rm P1}$: green, $y_{\rm P1}$: red). The A and B components are separated by $61.22\arcsec \pm 0.03\arcsec$, corresponding to a projected physical separation of $1898 \pm 2$~au at the primary star's distance. The bright star next to COCONUTS-3A is a background giant star (see Appendix~\ref{app:xray_bg}).}
\label{fig:finderchart}
\end{center}
\end{figure*}

Spectroscopic characterization of unusually red L dwarfs rely on cloudy and dis-equilibrium model atmospheres. However, these models are challenged by the uncertainties in opacity line lists, the assumption of radiative-convective equilibrium, and the presence of complex, patchy, time-evolving clouds \citep[e.g.,][]{2015ARA&A..53..279M, 2020SSRv..216..139S, 2020RAA....20...99Z}. Also, modeling efforts are still needed to reconcile the discrepancies between model predictions and the observed photometric sequence of ultracool dwarfs, particularly near the L/T transition \citep[e.g., Figures~22-24 of][]{2016ApJ...833...96L}. To improve the performance of these models, ultracool dwarfs that are either wide-orbit companions to stars or members of nearby associations are essential benchmarks. The ages and metallicities of these objects can be independently inferred from their associated stars and can therefore identify specific shortcomings of model assumptions and directly quantify the systematic errors of model predictions \citep[e.g.,][]{2020ApJ...891..171Z, 2021ApJ...916...53Z}.

To establish a large census of wide-orbit ($\gtrsim 500$~au) benchmark companions, we are conducting the COol Companions ON Ultrawide orbiTS (COCONUTS) program, by targeting 300,000 stars within 100~pc selected from Gaia. Mining astrometry and photometry from wide-field sky surveys (e.g., Pan-STARRS1 [PS1; \citealt{2016arXiv161205560C}], AllWISE \citep[][]{2010AJ....140.1868W, 2014yCat.2328....0C}, and CatWISE2020 [CatWISE hereafter; \citealt{2021ApJS..253....8M}]), we have searched for candidate companions which have consistent proper motions as their host stars (with projected separations $<10^{4}$~au) and have optical and infrared colors and magnitudes expected for planetary-mass objects and brown dwarfs \citep[e.g.,][]{2018ApJS..234....1B}. We have then conducted ground-based follow-up observations to confirm the companionship and substellar nature of new wide-orbit companions. Our first discovery COCONUTS-1AB \citep[][]{2020ApJ...891..171Z} is composed of a very old ($7.3^{+2.8}_{-1.6}$~Gyr) white dwarf primary and a T4 brown dwarf companion ($69^{+2}_{-3}$~\Mjup), which we use to test the high-gravity regime of cloudless model atmospheres \citep[][]{2021arXiv210707434M}. Our second discovery COCONUTS-2Ab \citep[][]{2021ApJ...916L..11Z} is composed of a young ($100-800$~Myr) M3 primary star with a T9 super-Jupiter companion ($6 \pm 2$~\Mjup), which is the nearest imaged planetary-mass object to our Solar System and also the second imaged exoplanet whose physical properties overlap both hot-start and cold-start formation models. 

Here, we report the third COCONUTS discovery, a system with a young M-dwarf primary star (UCAC4~374$-$046899; hereafter COCONUTS-3A) with a very red L-dwarf companion (WISEA~J081322.19$-$152203.2; hereafter COCONUTS-3B), which was previously identified as a free-floating object by \cite{2017AJ....153..196S}. We describe the system in Section~\ref{sec:system} and our spectroscopic follow-up observations in \ref{sec:obs}. We then study the physical properties of the primary star and the companion in Sections~\ref{sec:primary} and \ref{sec:companion}, respectively, followed by a brief summary in Section~\ref{sec:conclusion}.

\section{The COCONUTS-3 System}
\label{sec:system}
Figure~\ref{fig:finderchart} shows the COCONUTS-3 system and its neighborhood. Based on Gaia~EDR3 \citep{2016AandA...595A...1G, 2020arXiv201201533G}, COCONUTS-3A has a parallax of $32.33 \pm 0.02$~mas \citep[distance $= 30.88 \pm 0.02$~pc;][]{2021AJ....161..147B} and a proper motion of $(\mu_{\alpha}\cos{\delta}, \mu_{\delta}) = (-131.02 \pm 0.02, +93.39 \pm 0.02)$~\masyr, consistent with the value of $(-120.9 \pm 4.4, +92.1 \pm 5.3)$~\masyr\ from CatWISE \citep{2021ApJS..253....8M}\footnote{The offset correction in coordinates and proper motions suggested by \cite{2021ApJS..253....8M} have been incorporated in all CatWISE proper motions used in this work.} and $(-126.4 \pm 4.4, +84.3 \pm 3.3)$~\masyr\ from PS1 \citep{2016arXiv161205560C, 2020ApJS..251....6M}. The Gaia astrometry of this star is reliable and consistent with the five-parameter single-star model, since it is determined from 20 visibility periods with a mild Renormalised Unit Weight Error (RUWE) of 1.14 \citep[][]{Lindegren2018}, suggesting COCONUTS-3A is single. Based on BANYAN~$\Sigma$ \citep[][]{2018ApJ...856...23G} and Gaia astrometry, this star is not associated with any known young moving groups. Also, COCONUTS-3A is a mid-M dwarf, given its optical and near-infrared spectroscopy (Section~\ref{sec:primary_spt}).

COCONUTS-3B is a previously known, very red L dwarf, WISEA~J081322.19$-$152203.2 (WISE~J0813$-$1522), identified by \cite{2017AJ....153..196S} using 2MASS and AllWISE photometry. \cite{2017AJ....153..196S} assigned an L7 spectral type using $R \approx 3500$ CTIO/ARCoIRIS near-infrared ($0.8-2.4$~$\mu$m) spectra with a signal-to-noise ratio (S/N) of only $8$~per pixel near the $J$-band peak. Based on proper motions computed from 2MASS and AllWISE astrometry, they suggested this L dwarf is a possible member of the Argus \citep[with a $87\%$ membership probability from BANYAN~II;][]{2013ApJ...762...88M, 2014ApJ...783..121G} or Carina-Near (with a $97\%$ membership probability from the \citealt{2013ApJ...774..101R} convergent point tool) moving group. However, \cite{2017AJ....153..196S} cautioned the distances predicted by BANYAN~II ($15 \pm 2$~pc) and the \cite{2013ApJ...774..101R} convergent point tool ($17$~pc), assuming young moving group membership, are both significantly closer than the object's photometric distance of $31 \pm 6$~pc (derived using \Ktms, spectral type, and empirical relations between absolute magnitudes and spectral types by \citealt{2012ApJS..201...19D}). We update this object's photometric distance to be $32 \pm 7$~pc based on our analysis in Section~\ref{subsec:companion_lbol}, which is independent from the physical association between this object and COCONUTS-3A (see below). Using BANYAN~$\Sigma$ \citep{2018ApJ...856...23G} and this L dwarf's photometric distance, we find its CatWISE proper motion, $(\mu_{\alpha}\cos{\delta}, \mu_{\delta}) = (-120.8 \pm 6.9, +92.5 \pm 7.6)$~\masyr, corresponds a $30\%$ Carina-Near and a $28\%$ Argus membership. In addition, this object's PS1 proper motion, $(-112.6 \pm 21.0, +101.1 \pm 21.0)$~\masyr, corresponds a $19\%$ Carina-Near and $57\%$ Argus membership. Thus, young moving group membership for WISE~J0813$-$1522 seems unlikely, but this object's photometric distance and proper motions agree very well with those of COCONUTS-3A.

\begin{figure*}[t]
\begin{center}
\includegraphics[height=6.5in]{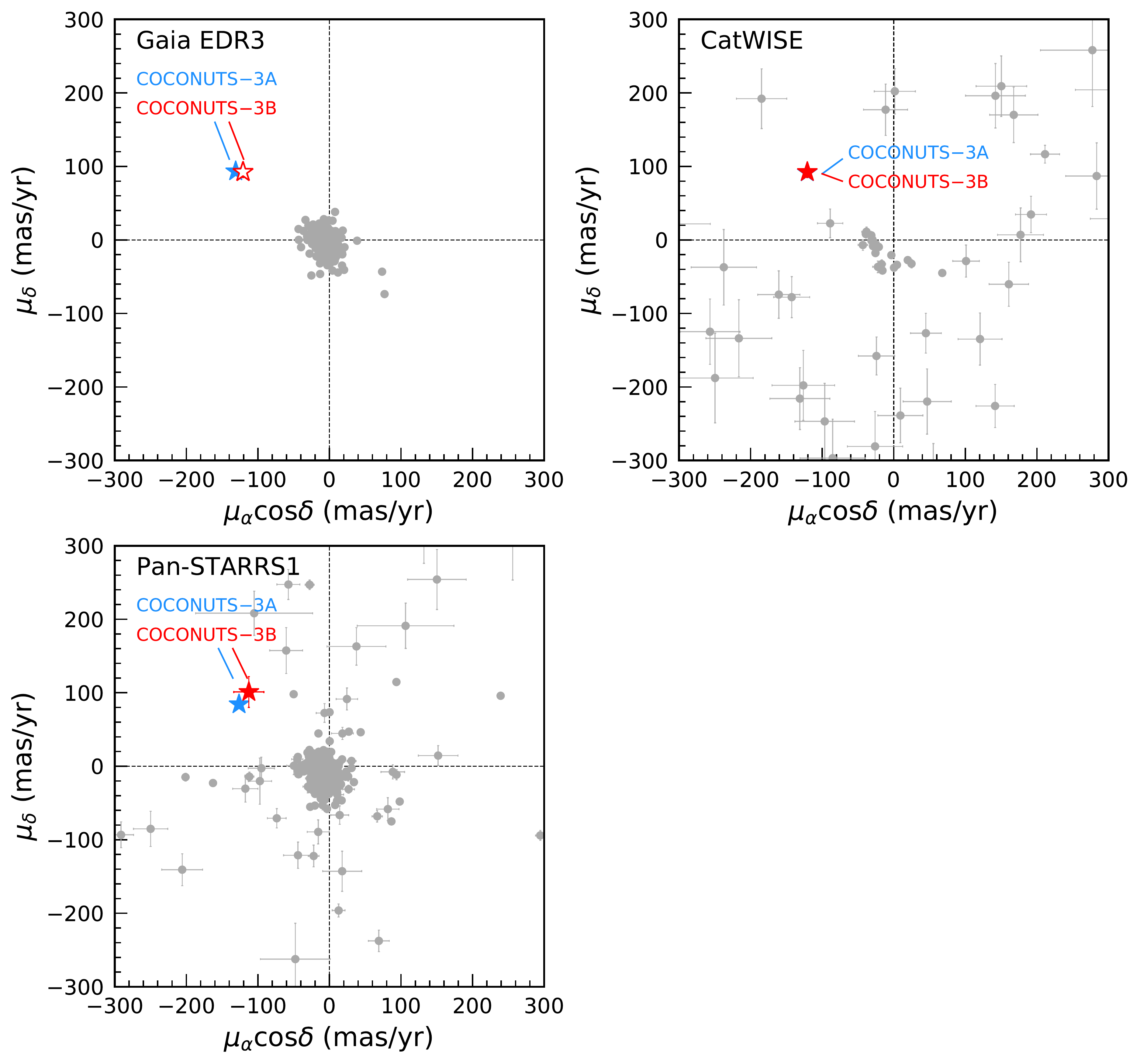}
\caption{Proper motions of COCONUTS-3A (blue) and 3B (red) from Gaia EDR3 (top left), CatWISE (top right), and PS1 (bottom left). Since the companion has no Gaia detection, we plot its CatWISE proper motion (red open symbol) in the top left panel. The two objects have very consistent CatWISE proper motions so their symbols overlap in the top right panel. Proper motions (if S/N$>5$) of nearby stars (grey) within 15\arcmin\ are overlaid in each panel. The common proper motions between A and B components is demonstrated by these diagrams and validates their physical association. }
\label{fig:common_pm}
\end{center}
\end{figure*}

We conclude WISE~J0813$-$1522 is a co-moving companion to the field dwarf COCONUTS-3A (Figure~\ref{fig:common_pm}). Based on their CatWISE coordinates, the A and B components are separated by $61.46'' \pm 0.04''$, meaning a projected physical separation of $1891 \pm 2$~au at the primary star's distance. Properties of COCONUTS-3AB are summarized in Table~\ref{tab:info}.

\section{Observations}
\label{sec:obs}

\subsection{Spectroscopy of the M-Dwarf Primary: COCONUTS-3A}
\label{subsec:spec_primary}

\subsubsection{UH~2.2m/SNIFS}
\label{subsubsec:primary_uh88}
We obtained optical ($3500-9100$~\r{A}) spectra of COCONUTS-3A on UT 2019 October~7 using the SuperNova Integral Field Spectrograph \citep[SNIFS;][]{2002SPIE.4836...61A, 2004SPIE.5249..146L} mounted on the University of Hawai`i's 2.2m telescope. SNIFS is a $6'' \times 6''$ integral field spectrograph and provides a spectral resolution of $R \approx 1200$. We took one 1800-second exposure for our target and then followed the pipeline described in \cite{2001MNRAS.326...23B} to extract the one-dimensional spectrum, comprising dark, bias, and flat-field corrections, wavelength calibration, and sky subtraction. The dispersion in wavelength calibration is $0.55$~\r{A} for the blue channel ($3500-4700$~\r{A}) and $0.08$~\r{A} for the red channel ($5300-9100$~\r{A}). We flux-calibrated the 1D spectrum using spectrophotometric standard stars observed during the same night, LTT~2415, Feige~110, HR~3454, and GD~71. Although the night was non-photometric, only relative fluxes of SNIFS spectra (and GMOS spectra in Section~\ref{subsubsec:primary_gmos}) are used in our subsequent analysis and are thus reliable. Our resulting SNIFS spectrum has air wavelengths, with S/N $=28$ at $6700$~\r{A} and 74 at $8200$~\r{A}, and also exhibits H$\alpha$ emission. 

\subsubsection{Gemini/GMOS}
\label{subsubsec:primary_gmos}
The optical ($6400-8900$~\r{A}) spectra of COCONUTS-3A was also acquired with the GMOS spectrograph \citep[][]{2004PASP..116..425H} at the Gemini-North Telescope (queue program GN-2019B-Q-139; PI: Z. Zhang) on UT 2019 December~12, in order to search for lithium $\lambda\lambda$6708 absorption. The R831 grating was used in conjunction with the GG-455 filter and the $0.5''$ long slit ($R \approx 4400$). One 330-second exposure was taken with a central wavelength of 7600~\r{A}, 7650~\r{A}, and 7700~\r{A} (i.e., 3 exposures in total). Such wavelength dithering compensates for the detectors' inter-chip gaps ($\approx 38$~\r{A} in our data), enabling continuous wavelength coverage. Using the Gemini IRAF data reduction package, we performed dark, bias, and flat-fielding corrections, cosmic-ray rejection, wavelength calibration, and sky subtraction. The dispersion in the wavelength calibration is $0.01$~\r{A}. We then flux-calibrated our extracted spectra by using a spectrophotometric standard, Feige~34, observed on UT 2019 November~28. This calibration produces reliable relative fluxes for the GMOS spectra for our subsequent analysis, which does not require absolute fluxes. We computed the weighted average of all three spectra with flux uncertainties propagated to provide the final GMOS spectrum of COCONUTS-3A, with air wavelengths and S/N $=125$ at $6700$~\r{A} and $380$ at $8200$~\r{A}.

\subsubsection{IRTF/SpeX}
\label{subsubsec:primary_irtf}
Near-infrared ($0.7-2.5$~$\mu$m) spectra of COCONUTS-3A were observed using the NASA Infrared Telescope Facility (IRTF) on UT 2019 April~26. We used the SpeX spectrograph \citep{2003PASP..115..362R} in the short wavelength cross-dispersed (SXD) mode with the $0.3''$ slit ($R \approx 2000$) and took six 120-second exposures in a standard ABBA nodding pattern. We contemporaneously observed an A0V standard star HD~67725 within $0.01$ airmass of COCONUTS-3A for telluric correction. We reduced the data using version~4.1 of the Spextool software package \citep{2004PASP..116..362C} and obtained a $0.122$~\r{A} dispersion in the wavelength calibration. Our resulting SpeX SXD spectrum has vacuum wavelengths, with a median S/N of $135$ and $146$ per pixel in $J$ and $K$ band, respectively.

\subsection{Spectroscopy of the L-Dwarf Companion: COCONUTS-3B}
\label{subsec:spec_companion}

\subsubsection{IRTF/SpeX}
\label{subsubsec:spex_companion}
We acquired near-infrared ($0.7-2.5$~$\mu$m) spectra of COCONUTS-3B using IRTF/SpeX in prism mode with the $0.8''$ slit ($R \approx 75$) on UT 2019 April~07 and UT 2019 November~02. We took a total of 74~exposures with 120 seconds each in a ABBA pattern, and contemporaneously observed multiple A0V standard stars (HD~78955, HD~82724, HD~89911, HD~72366, and HD~71580) to perform the telluric correction, given that the airmass of the companion spanned a wide range over long exposures. We divided raw spectroscopic data of COCONUTS-3B into five subsets, with each set calibrated by a telluric standard with a $<0.1$~airmass difference. The dispersion in wavelength calibration of the prism data is 5.9~\r{A}. Combining all the five telluric-corrected spectra using a robust weighted mean, we obtained the SpeX prism spectrum with vacuum wavelengths and a median S/N$= 53$ per pixel in $J$ band.

\subsubsection{Gemini/GNIRS}
\label{subsubsec:gnirs_companion}
The near-infrared ($0.9-2.5$~$\mu$m) spectra of COCONUTS-3B was observed with the Gemini-North GNIRS spectrograph \citep{2006SPIE.6269E..4CE} on UT 2019 November~12 (queue program GN-2019B-Q-139; PI: Z. Zhang) with a higher spectral resolution ($R \approx 750$) than the SpeX prism data. The cross-dispersed (XD) mode was used with the 32~I/mm grating, short camera ($0.15''$ per pixel), $0.675''$ slit. Eight exposures were taken with 222~seconds in an ABBA pattern. Also, an A0V telluric standard star, HIP 37787, was contemporaneously observed within 0.05 airmass of COCONUTS-3B. We reduced the data using a modified version of Spextool (K. Allers, private communication; also see \citealt{2003PASP..115..389V, 2004PASP..116..362C}; Section~4.4 of \citealt{2011ApJS..197...19K}) and obtained a dispersion of $0.8$~\r{A} in wavelength calibration. Our resulting GNIRS XD spectrum has vacuum wavelengths and a median S/N$=45$ per pixel in $J$ band.

\begin{figure*}[t]
\begin{center}
\includegraphics[height=6.5in]{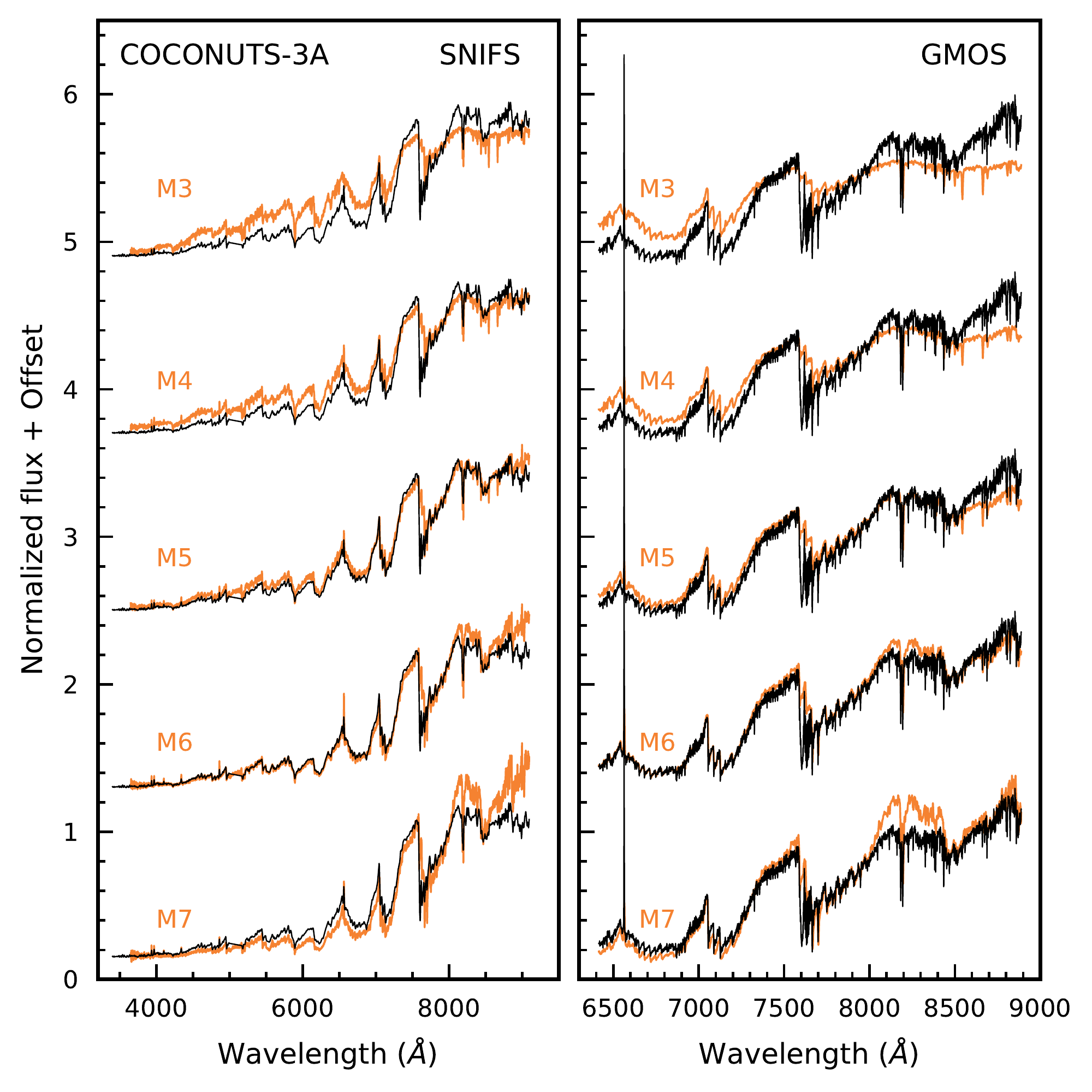}
\caption{The SNIFS and GMOS spectra of COCONUTS-3A (black) normalized by the flux at $8000$~\r{A}. PyHammer M3$-$M7 spectral templates \citep[][]{2017ApJS..230...16K} are overlaid and scaled by the averaged flux at $6500-9000$~\r{A}. We only show templates with the solar metallicity (orange) since the overall morphology of these templates at a given spectral type do not vary significantly with metallicities from $-0.5$ to $+0.5$~dex. The spike shown at the short wavelength of our observed GMOS spectrum is the H${\alpha}$ emission line discussed in Section~\ref{subsubsec:halpha}. We derive a visual optical spectral type of M5.5$\pm$0.5 for COCONUTS-3A. }
\label{fig:primary_opt_spt}
\end{center}
\end{figure*}

\section{COCONUTS-3A: the M-Dwarf Primary Star}
\label{sec:primary}

\subsection{Radial Velocity}
\label{subsec:restframe}
We convert spectra of COCONUTS-3A to the stellar rest frame by comparing with spectral templates. For the optical spectrum, we use templates from PyHammer \citep{2017ApJS..230...16K}, a modified version of the Hammer spectral classification tool \citep{2007AJ....134.2398C}. These templates are constructed using the Baryon Oscillation Spectroscopic Survey of the Sloan Digital Sky Survey \citep{2013AJ....145...10D}, spanning $3650-10200$~\r{A} wavelengths in vacuum ($R \sim 2000$), O5$-$L3 spectral types, and [Fe/H]$=-2$ to $+1$~dex (with intervals of 0.5~dex). We first put our SNIFS and GMOS spectra in vacuum following \cite{1996ApOpt..35.1566C}. Given that COCONUTS-3A has an optical spectral type of M$5.5\pm0.5$ (Section~\ref{sec:primary_spt}), we cross-correlate with all M5 and M6 PyHammer templates (with [Fe/H]$=-0.5, 0, +0.5$~dex at each spectral type) over a wavelength range of $6500-8800$~\r{A} for both SNIFS and GMOS spectra. We then shift these spectra using the averaged radial velocity of $120.4$~\kms\ (SNIFS) and $15.0$~\kms\ (GMOS), after confirming no $>5\sigma$ outliers exist. 

For the near-infrared spectrum, we use Gl~213 (M4), Gl~268AB (M4.5), and Gl~51 (M5) from the IRTF Spectral Library \citep{2005ApJ...623.1115C, 2009ApJS..185..289R} as templates, given that COCONUTS-3A has a near-infrared spectral type of M$4.5\pm0.5$ (Section~\ref{sec:primary_spt}). We cross-correlate the SpeX SXD spectrum of COCONUTS-3A with each template over individual orders~$3-7$ and then shift the spectrum by using the averaged value of $58.8$~\kms\ among all computed radial velocities (no $>5\sigma$ outliers are detected).

Based on the size of one resolution pixel, we assign an uncertainty of $250$~\kms, $68$~\kms, and $150$~\kms\ to the estimated radial velocity by SNIFS, GMOS, and SpeX SXD spectra, respectively. We further apply the barycentric correction to each measured radial velocity using Astropy \citep{2013A&A...558A..33A, 2018AJ....156..123A} and compute their weighted mean and weighted error, leading to a radial velocity of $41 \pm 60$~\kms\ for COCONUTS-3A.

\subsection{Spectral Type}
\label{sec:primary_spt}
We determine the optical spectral type of COCONUTS-3A using PyHammer \citep{2017ApJS..230...16K}, which measures a suite of indices and then compares to those of spectral templates via a weighted least-squares approach. PyHammer contains templates with a range of metallicities, thereby enabling a metallicity estimate along with the spectral type. \cite{2017ApJS..230...16K} suggested the spectral types and metallicities estimated by PyHammer are accurate to $\pm1.5$~subtypes and $\pm0.4$~dex, respectively. We find COCONUTS-3A has an index-based spectral type of M6 using both SNIFS and GMOS spectra, with a metallicity of $-0.5$~dex and $0$~dex, respectively. Visually comparing our observed spectra with PyHammer spectral templates, we assign a visual type of M5.5$\pm$0.5 (Figure~\ref{fig:primary_opt_spt}) and adopt this as the final optical spectral type of the primary star. Given the large metallicity uncertainties from PyHammer, we subsequently determine the bulk metallicity of COCONUTS-3A using narrow-band spectroscopic features (Section~\ref{sec:primary_metallicity}).

\begin{figure*}[t]
\begin{center}
\includegraphics[height=5.in]{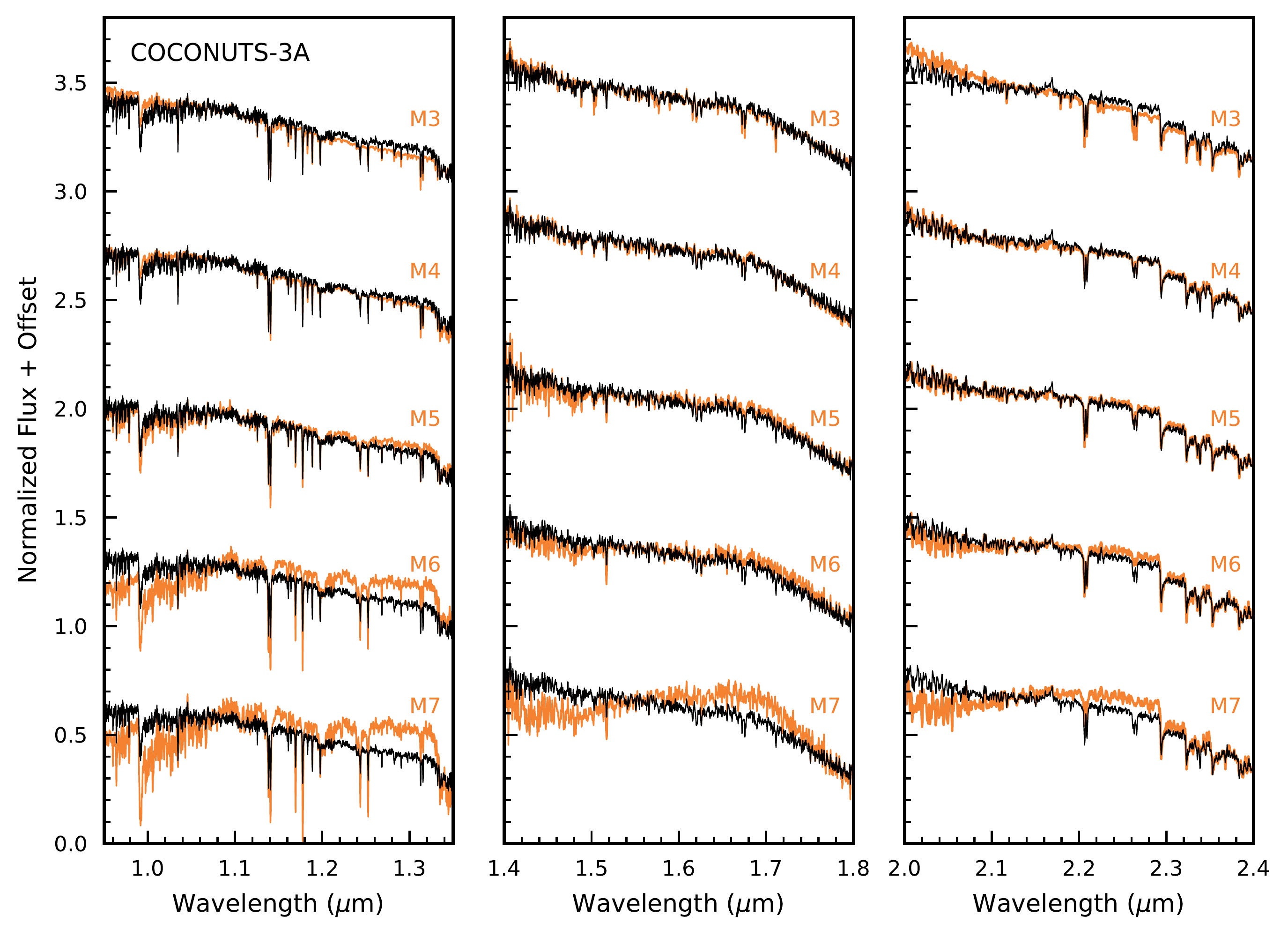}
\caption{The SpeX SXD spectrum of COCONUTS-3A (black), as compared to the M3$-$M7 spectral standards (orange) from the IRTF Spectral Library \citep{2005ApJ...623.1115C, 2009ApJS..185..289R} in $J$ (left), $H$ (middle), and $K$ (right) bands: Gl~388 (M3), Gl~213 (M4), Gl~51 (M5), Gl~406 (M6), and Gl~644C (M7). All these spectra are normalized by their average fluxes in each band. We derive a visual near-infrared spectral type of M4.5$\pm$0.5 for COCONUTS-3A. }
\label{fig:primary_nir_spt}
\end{center}
\end{figure*}

We determine the near-infrared spectral type of COCONUTS-3A following \cite{2013ApJ...772...79A}, who derived empirical polynomials between optical spectral types and four H$_{2}$O-band indices using IRTF/SpeX spectra of young and field-age M4$-$L8 ultracool dwarfs. \cite{2013ApJ...772...79A} also includes a gravity classification system, based on strengths of gravity-sensitive spectral features. We compute an index-based spectral type of M$4.2 \pm 0.5$ and an intermediate gravity class \textsc{int-g} (see Section~\ref{subsubsec:nir_grav} for further discussion) and caution such a spectral type is near the margin of the applicable range (M4$-$L8) for the \cite{2013ApJ...772...79A} classification. We also measure the H$_{2}$O-K$2$ index, a spectral type indicator for M0$-$M9 dwarfs calibrated by \cite{2012ApJ...748...93R} and derive M$4.5 \pm 1.3$. We further compare our SpeX SXD spectrum to M3$-$M7 spectral standards from the IRTF Spectral Library \citep{2005ApJ...623.1115C, 2009ApJS..185..289R} in $J$, $H$, and $K$ bands and assign a visual type of M$4.5\pm0.5$ (Figure~\ref{fig:primary_nir_spt}), which we adopt as the final near-infrared spectral type of COCONUTS-3A.

\subsection{Metallicity}
\label{sec:primary_metallicity}
We compute the bulk metallicity of COCONUTS-3A by using empirical calibrations established using binaries composed of FGK primary stars (with independently measured [Fe/H]) and M dwarf secondaries. Such M-dwarf metallicity calibrations were originally developed for photometry \citep[e.g.,][]{2005A&A...442..635B, 2010A&A...519A.105S, 2012AJ....143..111J, 2012A&A...538A..25N} and then extended to moderate-resolution spectroscopy \citep[$R \sim 2000$; e.g.,][]{2010ApJ...720L.113R, 2012ApJ...748...93R, 2012ApJ...747L..38T, 2013AJ....145...52M, 2014AJ....147..160M, 2014AJ....147...20N}. Using a sample of 112 binaries, \cite{2013AJ....145...52M} derived metallicity calibrations for objects with optical spectral types of K5$-$M5 using either optical, $J$-, $H$-, or $K$-band spectra as observed by UH~2.2m/SNIFS or IRTF/SpeX SXD. They also used their sample to update calibrations in previous work \citep{2007ApJ...669.1235L, 2012AJ....143...67D, 2012AJ....143..111J, 2012ApJ...747L..38T}. \cite{2014AJ....147..160M} further extended the calibration to near-infrared spectral types of M4.5$-$M9.5 using sodium and calcium features in $K$-band SXD spectra. A metallicity calibration for near-infrared spectral types of M1$-$M5 was also established by \cite{2014AJ....147...20N} using the sodium doublet in $K$-band SXD spectra.

Given that COCONUTS-3A has an optical spectral type of M5.5 and a near-infrared spectral type of M4.5 (Section~\ref{sec:primary_spt}), we apply both the \cite{2014AJ....147..160M} and \cite{2014AJ....147...20N} calibrations and obtain metallicities of $0.23 \pm 0.08$~dex and $0.16\pm0.12$~dex, respectively, with spectral flux uncertainties and calibration errors propagated in a Monte Carlo fashion. We adopt the weighted mean and weighted standard deviation of [Fe/H]$_{\star}=0.21 \pm 0.07$~dex as the final metallicity of the COCONUTS-3 system (also see Table~\ref{tab:metal_age_primary}). Our adopted empirical calibrations were established using field M dwarfs with slightly higher surface gravities and older ages than COCONUTS-3A (see Sections~\ref{sec:primary_age}), so we explore the potential surface-gravity dependence of our derived [Fe/H]$_{\star}$ in Appendix~\ref{app:logg_effect_metal}. Based on the PHOENIX stellar models, our bulk metallicity could be under-estimated by $0.2-0.3$~dex. A more accurate estimate of [Fe/H]$_{\star}$ would benefit from an empirical calibration customized for young low-mass stars, which is unfortunately not available to date, given that the existing sample of nearby young stars lacks sufficiently large variations in [Fe/H] and mostly has solar metallicity. 

While the spectral type of our primary star is just outside the applicable range (K5$-$M5) of the \cite{2013AJ....145...52M} calibrations, we applied them to our SNIFS and SpeX SXD spectra and obtained [Fe/H]$_{\star} = -0.22 \pm 0.16$~dex in the optical, $0.09 \pm 0.13$~dex in $J$ band, $-0.17 \pm 0.11$~dex in $H$ band, and $0.12 \pm 0.09$~dex in $K$ band. Using the \cite{2012ApJ...747L..38T} $H$- and $K$-band calibrations as updated by \cite{2013AJ....145...52M}, we obtain [Fe/H]$_{\star}= 0.11 \pm 0.15$~dex and $0.15 \pm 0.14$~dex, respectively. Most of these values are $<1\sigma$ consistent with our adopted metallicity. We find the spectral features traced by the \cite{2013AJ....145...52M} optical- and $H$-band calibrations are not prominent in COCONUTS-3A's spectra, likely leading to less accurate metallicity estimates.

\subsection{Physical Properties}
\label{subsec:phys}
We derive physical properties of COCONUTS-3A using its broadband photometry and empirical relations. We first synthesize its 2MASS and MKO $JHK$ photometry using our SpeX SXD spectra and measured $J$ and \Ks\ from the VISTA Hemisphere Survey \citep[VHS;][]{2013Msngr.154...35M}, given that this star lacks MKO photometry and its 2MASS photometry is contaminated by a diffraction spike from a nearby source. We use the Vega spectrum and obtain 2MASS, MKO, and VHS filters from \cite{2003AJ....126.1090C}, \cite{2006MNRAS.367..454H}, and the ESO VISTA instrument webpage\footnote{\url{http://www.eso.org/sci/facilities/paranal/instruments/vircam/inst/}.}, respectively. We propagate the uncertainties in VHS magnitudes and spectral fluxes in a Monte Carlo fashion, and when the VHS photometry and the synthesized ones are in different bandpasses, we add a $0.05$~mag error in quadrature \citep[e.g.,][]{2012ApJS..201...19D}. We find using VHS $J$ and \Ks\ lead to very similar magnitudes in each band and we adopt their average and standard deviation as the final synthetic photometry (see Table~\ref{tab:info}).

Following empirical calibrations of \cite{2015ApJ...804...64M} and \cite{2019ApJ...871...63M}, we convert the COCONUTS-3A's \Ktms\ absolute magnitude and metallicity into a radii of \Rstar$= 0.151 \pm 0.004$~\Rsun\ and a mass of \Mstar$= 0.123 \pm 0.006$~\Msun, leading to a surface gravity of \loggstar$= 5.17 \pm 0.03$~dex. We also compute a bolometric luminosity of \logLbolstar$= -2.80 \pm 0.04$~dex by using the \cite{2020A&A...642A.115C} empirical relation between \Lbol\ and \Jtms-band absolute magnitudes, which is robust over a range of stellar metallicities spanning from $-1.0$~dex to $+0.6$~dex (see their Section~4). We then use the Stefan-Boltzmann law to compute an effective temperature of \Teffstar$=2966 \pm 85$~K. All measurement uncertainties and calibration errors are propagated in a Monte Carlo fashion.

\subsection{Age}
\label{sec:primary_age}

\subsubsection{H$\alpha$ Emission}
\label{subsubsec:halpha}
The chromospheric H$\alpha$ emission probes stellar activity which is correlated with stellar ages \citep[e.g.,][]{2010ARA&A..48..581S}. Based on H$\alpha$ measurements of a large M-dwarf sample from SDSS, \cite{2008AJ....135..785W} noted the activity fraction of these objects decreases with vertical distance from the Galactic midplane, and the slope of such a decrease depends on spectral type. Using one-dimensional dynamical simulations encoded with a correlation between stellar ages and Galactic positions, \cite{2008AJ....135..785W} derived a model-based activity lifetime for a given M subtype. More recently, an empirical calibration between stellar age and H$\alpha$-based activity has been established by \cite{2021AJ....161..277K} using M dwarfs that are either members of young moving groups or companions to white dwarfs. \cite{2021AJ....161..277K} fitted a broken power-law to the fractional H$\alpha$ luminosity ($L_{\rm H\alpha}/L_{\rm bol}$) as a function of age and found that $L_{\rm H\alpha}/L_{\rm bol}$ stays in a saturation regime with a little evolution at $<1$~Gyr but decreases more rapidly at older ages.  

To assess the H$\alpha$ emission of COCONUTS-3A, we measure the equivalent width (EW) using SNIFS and GMOS spectra (with rest-frame vacuum wavelengths), following definitions of the H$\alpha$ line and continuum by \cite{2015AJ....149..158S}. We integrate the line flux over $6557.61-6571.61$~\r{A} with the pseudo-continuum approximated by the mean flux across two surrounding wavelength regions of $6530-6555$~\r{A} and $6575-6600$~\r{A}. We measure EW(H$\alpha$) $= -2.9 \pm 0.2$~\r{A} (SNIFS) and $-7.18 \pm 0.02$~\r{A} (GMOS) with flux uncertainties propagated in a Monte Carlo fashion. Using slightly different line and continuum definitions by \cite{2011AJ....141...97W} and \cite{2017ApJ...834...85N}, as well as fitting a Gaussian function for the H$\alpha$ line, we obtain similar EW$_{\rm H\alpha}$ values with a $<1$~\r{A} difference for each spectrum. Therefore, COCONUTS-3A is active, with its H$\alpha$ emission from the stellar chromosphere rather than disk accretion \citep[e.g.,][]{2003AJ....126.2997B}. We attribute the EW(H$\alpha$) difference of $4.3$~\r{A} between the SNIFS and GMOS spectra to variability and note such a strong variability tends to occur for relatively young stars \citep[$\lesssim 100$~Myr; see Figure~7 of][]{2021AJ....161..277K}.

We compute $L_{\rm H\alpha}/L_{\rm bol} = - {\rm EW(H\alpha)}  \times \chi$, with $\chi$ being a factor calibrated against broadband colors and spectral types \citep[e.g.,][]{2004PASP..116.1105W, 2008PASP..120.1161W}. We use the \cite{2014ApJ...795..161D} relation between $\chi$ and optical spectral types and obtain $\chi = (2.1065 \pm 0.4167) \times 10^{-5}$, with both the value and error computed as the average of those at M5 and M6 types. We derive $\log(L_{\rm H\alpha}/L_{\rm bol}) = -4.22 \pm 0.09$~dex (SNIFS) and $-3.79 \pm 0.09$~dex (GMOS), and then estimate the stellar age using the \cite{2021AJ....161..277K} H$\alpha$ activity-age relation in a Bayesian framework \citep[also see][]{2021ApJ...916L..11Z}. We evaluate the age ($t$) based on the following log-likelihood function:
\begin{equation}
\ln{\mathcal{L}(t)} = -\frac{\left[\log(L_{\rm H\alpha}/L_{\rm bol})_{\rm model} - \log(L_{\rm H\alpha}/L_{\rm bol})_{\rm obs}\right]^{2}}{2 \times (\sigma_{\rm obs}^2 + \sigma_{V}^{2})}
\end{equation}
where $\log(L_{\rm H\alpha}/L_{\rm bol})_{\rm obs} = -4.22$ (SNIFS) or $-3.79$ (GMOS), and $\sigma_{\rm obs} = 0.09$. We use the best-fit model parameters of \cite{2021AJ....161..277K} to compute $\log(L_{\rm H\alpha}/L_{\rm bol})_{\rm model}$ as a function of stellar age $(t)$, with $\sigma_{V} = 0.22$~dex. We use the Markov Chain Monte Carlo (MCMC) algorithm {\it emcee} \citep{2013PASP..125..306F} and assume a uniform prior of $[1.5\ {\rm Myr}, 10\ {\rm Gyr}]$ in age. Running MCMC with 20 walkers and $5000$ iterations (with the first $100$ iterations excluded as burn-in), we obtain an age estimate of $2.9^{+2.4}_{-1.4}$~Gyr and $870^{+940}_{-570}$~Myr using the SNIFS and GMOS spectra, respectively. Both these estimates are consistent with M5 and M6 dwarfs' model-based activity lifetime of $<7 \pm 0.5$~Gyr computed by \cite{2008AJ....135..785W}.

We caution that the older derived age from the SNIFS spectrum is primarily constrained by a power-law component of \cite{2021AJ....161..277K} models at ages $>$1~Gyr, which is not well constrained by their very small sample size of 6 old M dwarfs. Also, the measured $\log(L_{\rm H\alpha}/L_{\rm bol})_{\rm obs} = -4.22 \pm 0.09$~dex from the SNIFS spectrum is in fact consistent with young M dwarfs spanning ages of 10~Myr to 1~Gyr, considering the large scatter in the H$\alpha$ activity-age relation. Therefore, we adopt an H$\alpha$-based age of $870^{+940}_{-570}$~Myr as determined from the GMOS spectrum.

\subsubsection{X-ray Emission}
\label{subsubsec:xray}
As another indicator of stellar magnetic activity, coronal X-ray emission has a qualitatively similar behavior with stellar age as H$\alpha$ emission. At young ages ($\lesssim 100$~Myr$-$1~Gyr), the X-ray-to-bolometric luminosity ratio \logRx$=$\logLxbol\ exhibits a plateau and the specific saturation level tends to increase with lower stellar masses \citep[e.g.,][]{2012MNRAS.422.2024J}. This plateau is then followed by a power-law decline toward older ages, with a slope weakly dependent on spectral type \citep[e.g.,][]{2017MNRAS.471.1012B}. 

To estimate the X-ray emission of COCONUTS-3A, we query the Second ROSAT All-Sky Survey \citep{2016A&A...588A.103B} using a 30\arcsec\ matching radius and follow \cite{1995ApJS...99..701F} to convert the measured count rate and hardness ratio (HR1) into a flux \Fx$= (1.70 \pm 0.96) \times 10^{-13}$~\ergscm. Multiple background sources are near the ROSAT X-ray detection, which we find make a negligible contribution to our measured X-ray emission (see Appendix~\ref{app:xray_bg}). We derive an X-ray luminosity of \logLx$=28.3 \pm 0.2$~dex, similar to M dwarfs in the Pleiades \citep[$112\pm5$~Myr;][]{2015ApJ...813..108D} and the Hyades \citep[$750 \pm 100$~Myr;][]{2015ApJ...807...58B} according to \cite{2005ApJS..160..390P}. Based on \cite{2014ApJ...788...81M}, the \logLx\ of COCONUTS-3A is fainter than members of AB~Doradus moving group \citep[$149^{+51}_{-19}$~Myr;][]{2015MNRAS.454..593B} and younger associations but brighter than the inactive field population. We compute \logRx$=-2.7 \pm 0.3$~dex, suggesting COCONUTS-3A is in the saturation regime of the X-ray activity-age relation. We also derive flux ratios between X-ray and 2MASS $J$ and \Ks\ bands as \logFxj$= \log{F_{X}} + 0.4 J + 6.3 = -1.6 \pm 0.2$~dex and \logFxk$= \log{F_{X}} + 0.4 K_{S} + 7.0 = -1.2 \pm 0.2$~dex, which are comparable with members of the $\beta$~Pictoris moving group \citep[$24\pm3$~Myr;][]{2015MNRAS.454..593B} and the Pleiades \citep[e.g.,][]{2009ApJ...699..649S, 2012AJ....143...80S}. The coronal X-ray activity of COCONUTS-3A thus suggests an age of $\lesssim 850$~Myr.

\subsubsection{UV Emission}
\label{subsubsec:uv}
We also study the stellar activity of COCONUTS-3A through UV emission as measured by the Galaxy Evolution Explorer \citep[GALEX;][]{2005ApJ...619L...1M}. Using the Milkulski Archive for Space Telescopes (MAST), we find COCONUTS-3A was observed in the near-UV ($NUV$) band by the GALEX All-Sky Imaging Survey (AIS) but not detected. We also note the sources within 1\arcmin\ are all fainter than the GALEX $5\sigma$ detection limit of 20.5~mag in $NUV$. To estimate an upper limit of $NUV$ emission from COCONUTS-3A, we extract all GALEX sources within 10\arcmin\ and exclude problematic photometry with a bright star window reflection, dichroic reflection, detector run proximity, or bright star ghost. We then fit a linear function between these objects' $NUV$ magnitudes and the logarithmic photometric S/N. We adopt an upper limit of $NUV>22.3$~mag, which corresponds to the predicted value at S/N$=2$ \citep[e.g.,][]{2018AJ....155..122S}. 

We compute the flux ratio between $NUV$ and 2MASS $J$ band as \logFnuvj$= -0.4\times(NUV - J) + 0.358 < -3.7$~dex, as well as colors of $NUV-G>8.3$~mag and $NUV-J>10.1$~mag. These upper limits for the $NUV$ emission are broadly consistent with association members spanning ages of $10-750$~Myr and older field dwarfs \citep[e.g.,][]{2011AJ....142...23F, 2018AJ....155..122S, 2018ApJ...862..138G}.

\subsubsection{Kinematics}
\label{subsubsec:kinematics}
Combining Gaia EDR3 astrometry and our estimated radial velocity (Section~\ref{subsec:restframe}), we compute the space position and motion of COCONUTS-3A as $XYZ = (-16.899 \pm 0.011, -25.311 \pm 0.016, 5.542 \pm 0.004)$~pc and $UVW = (-41 \pm 33, -22 \pm 49, -2 \pm 11)$~\kms. Such a space motion is outside but consistent within uncertainties with the ``good box'' of young stars as assigned by \cite{2004ARA&A..42..685Z}, with $-15 \leqslant U \leqslant 0$~\kms, $-34 \leqslant V \leqslant -10$~\kms, and $-20 \leqslant W \leqslant +3$~\kms. To revisit the ages of objects within the good box, we study the mean space motions of 28 nearby young moving groups from \cite{2018ApJ...856...23G} and \cite{2020ApJ...903...96G}. We find all associations with ages $\gtrsim 150$~Myr have mean space motions outside the good box, while all younger associations fall inside the box, except for IC~2391 \citep[$50\pm5$~Myr;][]{2004ApJ...614..386B}, Octans \citep[$35\pm5$~Myr;][]{2015MNRAS.447.1267M}, and Taurus ($1-2$~Myr by \citealt{1995ApJS..101..117K}, or $1-30$~Myr as suggested by, e.g., \citealt{2017ApJ...838..150K, 2018ApJ...858...41Z}). We note the Taurus mean space motion is only 0.7~\kms\ outside the border. The fact that the mean $UVW$ of COCONUTS-3A is outside the good box suggests it has an age of $\gtrsim 150$~Myr, although a more precise radial velocity measurement is needed for verification. We also run BANYAN~$\Sigma$ \citep{2018ApJ...856...23G} and LACEwING \citep{2017AJ....153...95R}, finding COCONUTS-3A is not associated with any nearby young moving groups.

\begin{figure*}[t]
\begin{center}
\includegraphics[height=3.1in]{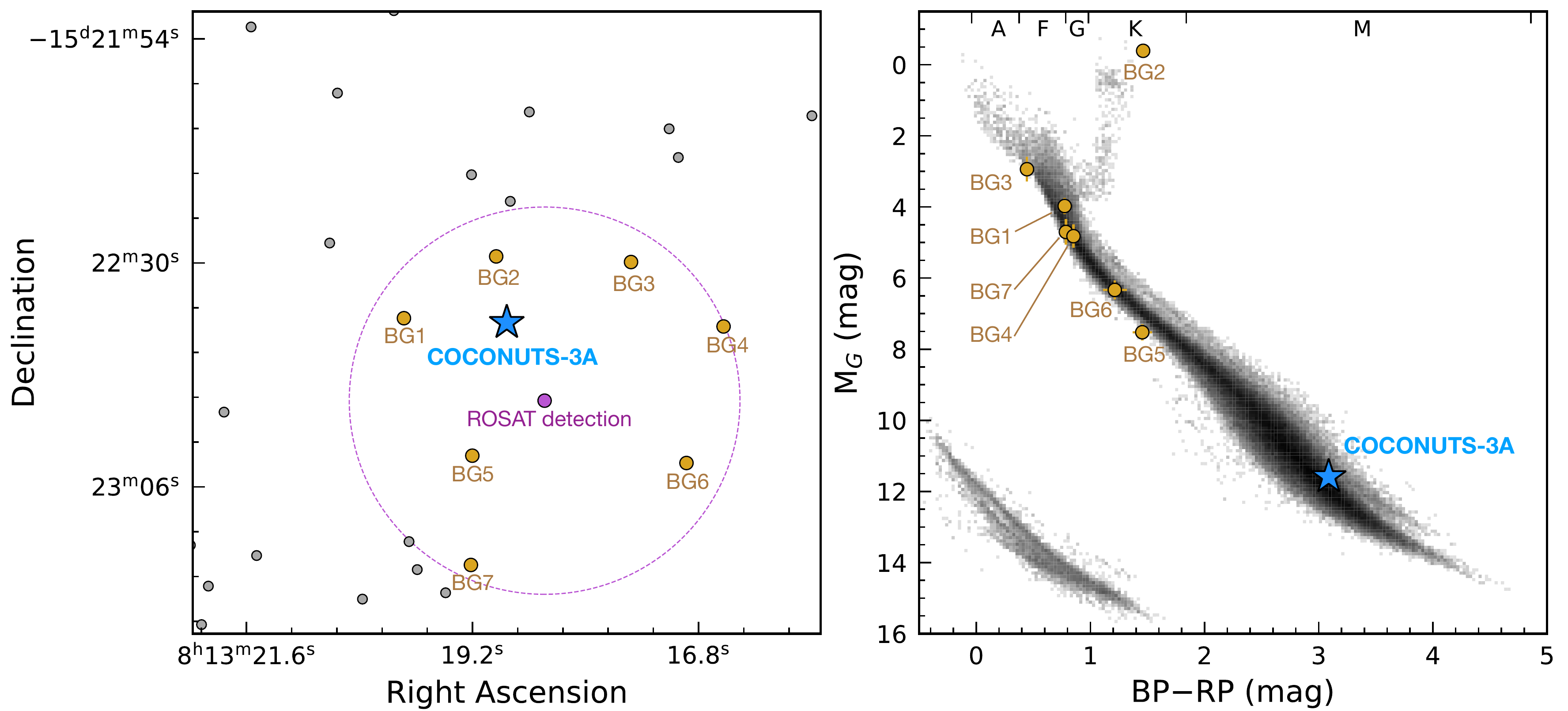}
\caption{{\it Left}: Gaia EDR3 coordinates (at epoch J2000) of COCONUTS-3A (blue) and nearby sources (grey and brown). The seven background stars within 30\arcsec\ (purple dashed circle) of the ROSAT detection 2RXS~J081318.4$-$152252 (purple) are highlighted in brown. {\it Right}: Gaia photometry of COCONUTS-3A (blue) and background stars (brown). We also present the number density of Gaia DR2 stars within 100~pc (grey), after applying the photometric and astrometric filters suggested by the Appendix~B of \cite{2018A&A...616A..10G}. }
\label{fig:xray_neighbors}
\end{center}
\end{figure*}

\subsubsection{HR Diagram Position}
\label{subsubsec:hrdiagram}
According to the MESA isochrones \citep[][]{2011ApJS..192....3P, 2016ApJS..222....8D, 2016ApJ...823..102C}, M dwarfs with comparable masses and metallicities as COCONUTS-3A remain in the pre-main sequence evolutionary stage for $\sim 1.5$~Gyr. As shown in Figure~\ref{fig:xray_neighbors}, COCONUTS-3A is located near the upper envelope of the main sequence, indicating it is younger than a few Gyrs. Comparing the Gaia $G$-band absolute magnitude ($11.5826 \pm 0.0015$~mag) and BP$-$RP color ($3.088 \pm 0.004$~mag) to empirical isochrones of nearby young moving group members compiled by \cite{2021AJ....161..277K}, we find COCONUTS-3A has similar photometry as members of the Pleiades ($112 \pm 5$~Myr), the AB~Doradus moving group ($149^{+51}_{-19}$~Myr), Coma Berenices \citep[$562^{+98}_{-84}$~Myr;][]{2014A&A...566A.132S}, Praesepe \citep[$650 \pm 50$~Myr;][]{2019ApJ...879..100D}, and the Hyades ($750\pm100$~Myr), and has a fainter absolute magnitude than younger association members. This suggests an age estimate of $100$~Myr$-$1.5~Gyr.

\subsubsection{Lithium Absorption}
\label{subsubsec:lithium}
Approaching the main sequence, young M dwarfs contract and increase their core temperatures, which then burns their initial lithium via convective mixing \citep[e.g.,][]{2010ARA&A..48..581S}. Based on the \cite{1997A&A...327.1039C} models, the lithium-burning would occur after $\sim 10-15$~Myr for M2$-$M3 dwarfs and $\sim 100$~Myr for M5$-$M6 dwarfs \citep[also see][]{2008ApJ...689.1127M, 2014MNRAS.438L..11B, 2016MNRAS.455.3345B}. We measure the equivalent width of lithium EW$_{\rm Li}$ for COCONUTS-3A using the GMOS spectrum. To approximate the pseudo continuum at the Lithium absorption feature, we fit a PyHammer template (M6 and [Fe/H]$=0$) to the GMOS spectrum, with the scaling factor determined by Equation~2 of \cite{2008ApJ...678.1372C}. Both GMOS and template spectra have vacuum wavelengths in the rest frame, and we compute EW(Li) over $6708.8-6710.8$~\r{A}, leading to an upper limit of $4$~m\r{A}. The absence of lithium absorption in COCONUTS-3A suggests an age of $\gtrsim 100$~Myr.

\begin{figure*}[t]
\begin{center}
\includegraphics[height=6in]{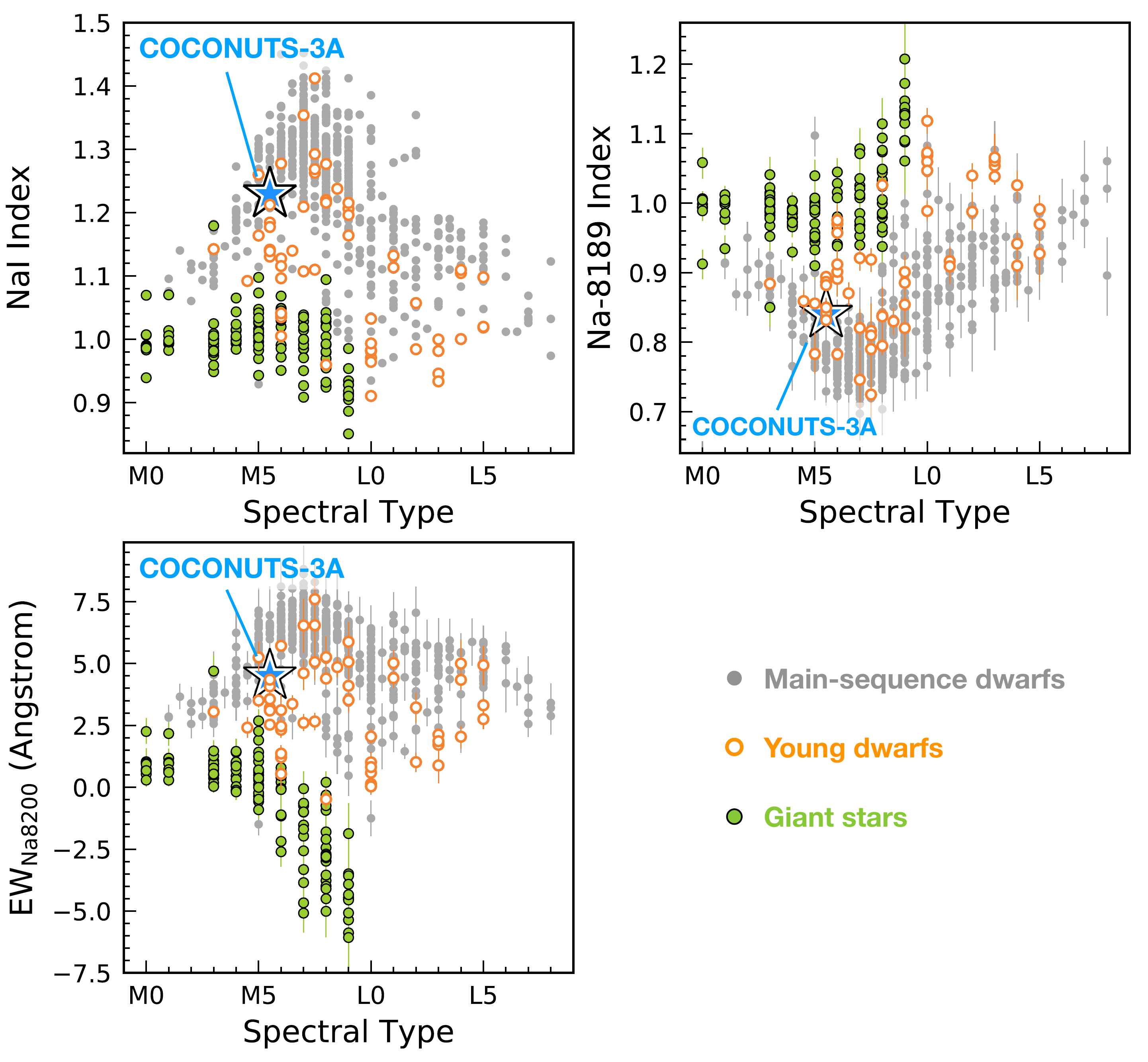} 
\caption{The computed \cite{2004MNRAS.355..363L} Na~I index, \cite{2006AJ....131.3016S} Na-8189 index, and \cite{2012AJ....143..114S} EW$_{\rm Na8200}$ of COCONUTS-3A (blue), as well as the main-sequence dwarfs (grey), young dwarfs (orange), and giant stars (green) from the Ultracool RIZzo Spectral Library \citep[][]{cruz_gagne_2014} and the Montreal Spectral Library (Section~\ref{subsubsec:na}). These young dwarfs either have $\beta$, $\gamma$, or $\delta$ gravity classes or are kinematic members of $\lesssim 200$~Myr young moving groups. Uncertainties are shown if they exceed the symbol size. We note the strength of the Na doublet of COCONUTS-3A falls in the low-gravity envelope of the main-sequence stars and is similar to several young M dwarfs. }
\label{fig:primary_na}
\end{center}
\end{figure*}

\subsubsection{Optical Sodium Doublet}
\label{subsubsec:na}
The sodium doublet ($\lambda\lambda 8183,8195$; all wavelengths discussed in this subsection are in the air) is gravity-sensitive and thereby an indicator of stellar age. To quantify the strength of this doublet, \cite{2004MNRAS.355..363L} proposed a Na~I index$= \langle F_{\lambda8148-8172} \rangle / \langle F_{\lambda8176-8200} \rangle$, computed as the ratio between the averaged fluxes over $8148-8172$~\r{A} and $8176-8200$~\r{A}. They found such indices for giant stars are significantly lower than those of higher-gravity main-sequence dwarfs for early-to-mid-M types, with young pre-main sequence stars falling between these two classes of objects. \cite{2014AJ....147...85R} revisited the Na~I index with the addition of nearly 50 main-sequence stars and 25 association members, and they cautioned the interpretation of this index at $V-K_{S} \lesssim 5$~mag ($\lesssim$M3.5). Also, \cite{2006AJ....131.3016S} proposed a Na-8189 index$=F_{\lambda8174-8204}/F_{\lambda8135-8165}$, defined as the ratio between integrated fluxes over $8174-8204$~\r{A} and $8135-8165$~\r{A}. This index is quantitatively equivalent to the inverse of the \cite{2004MNRAS.355..363L} Na~I index and is gravity-sensitive, particularly for M2$-$M7 spectral types \citep[e.g.,][]{2006AJ....132.2665S, 2017ApJ...838..150K}. In addition, \cite{2012AJ....143..114S} defined an equivalent width of the Na doublet EW$_{\rm Na8200}$, with the absorption line computed from the integrated flux over $8179-8201$~\r{A} and the pseudo-continuum approximated by the mean flux over $8149-8169$~\r{A} and $8236-8258$~\r{A}. Based on stellar model atmospheres, they suggested EW$_{\rm Na8200}$ can robustly distinguish $\lesssim100$~Myr stars from the field population at $V-K_{S}\geqslant5$~mag.

Here we perform the largest examination of all these Na spectral indices and equivalent widths by combining spectra of giant and dwarf stars from the Ultracool RIZzo Spectral Library \citep[][]{cruz_gagne_2014} and the Montreal Spectral Library.\footnote{\url{https://jgagneastro.com/the-montreal-spectral-library/}.} We make use of all optical and near-infrared spectra which cover the sodium feature and have S/N$>$10 at 8190~\r{A}, including 131 giants (K4$-$M9), 581 main-sequence field dwarfs (K0$-$T2), and 56 young dwarfs (M3$-$L5), which either have $\beta$, $\gamma$, or $\delta$ gravity classes or are kinematic members of $\lesssim 200$~Myr young moving groups (based on a cross-match with the young moving group census from \citealt{2018ApJ...856...23G} and \citealt{2018ApJ...862..138G}). We homogeneously determine the Na~I index, Na-8189 index, and EW$_{\rm Na8200}$ values of these objects, with spectral flux uncertainties propagated in a Monte Carlo fashion and with the vacuum wavelengths of near-infrared spectra put in the air following \cite{1996ApOpt..35.1566C}. These spectra were obtained by various instruments that span spectral resolutions of $R \sim 100-2000$, with the most common instruments being the Ritchey-Chr\'{e}tien (RC) spectrograph (504 objects), the GoldCam spectrograph (148 objects), and the SpeX spectrograph (47 objects). Within each dataset of giants, main-sequence dwarfs, and young dwarfs, we find our measured Na indices and equivalent widths exhibit no systematic offsets between the different spectrographs.

Figure~\ref{fig:primary_na} presents the \cite{2004MNRAS.355..363L} Na~I index, the \cite{2006AJ....131.3016S} Na-8189 index, and the \cite{2012AJ....143..114S} EW$_{\rm Na8200}$ as a function of spectral types. We note the Na doublet strengths of giants and main-sequence dwarfs are distinct, while those of pre-main sequence dwarfs bridge these two populations with a very large scatter. We also measure these Na features for COCONUTS-3A using its SNIFS, GMOS, and SpeX SXD spectra and find a $<3\%$ discrepancy in a given feature among different spectra. Compared to the library M dwarfs, the strength of COCONUTS-3B's Na doublet $\lambda\lambda8183,8195$ is located near the low-gravity envelope of main-sequence stars and similar to several young pre-main sequence M dwarfs with $\beta$ and $\gamma$ gravity classes. These suggest an age of a few $100$~Myr \citep[e.g.,][]{2009AJ....137.3345C} but cannot rule out the much older ages given the large scatter of Na indices and equivalent widths of main-sequence stars.

\subsubsection{Near-Infrared Gravity-Sensitive Features}
\label{subsubsec:nir_grav}
For mid-M to L dwarfs, there are several gravity-sensitive features in the near-infrared (see \citealt{2013ApJ...772...79A} and reference therein), including FeH (0.99~\micron, 1.20~\micron, 1.55~\micron), VO (1.06~\micron), Na~I (1.138~\micron), K~I (1.169~\micron, 1.177~\micron, 1.244~\micron, 1.253~\micron), and the shape of the $H$-band continuum ($1.46-1.68$~\micron). At a given near-infrared spectral type within M5$-$L7, \cite{2013ApJ...772...79A} provided critical values in the equivalent widths or spectral indices of these features, leading to three gravity classes: \textsc{vl-g} (very low gravity), \textsc{int-g} (intermediate gravity), and \textsc{fld-g} (field gravity). These gravity classes roughly correspond to ages of $\lesssim 30$~Myr, $\approx 30-200$~Myr, and $\gtrsim 200$~Myr, respectively, with considerable uncertainties \citep[e.g.,][]{2013ApJ...772...79A, 2016ApJ...833...96L}. 

COCONUTS-3A is classified as \textsc{int-g} based on the \cite{2013ApJ...772...79A} scheme (see Table~\ref{tab:al13_index}). Such gravity class is determined primarily by its low Na~I equivalent width at 1.138~\micron. All the other gravity-sensitive features cannot be applied to the M4.5 near-infrared spectral type (Section~\ref{sec:primary_spt}), but we find those measured values mostly fall in the low-gravity envelope of the field sequence \citep[also see][]{2015ApJ...806..254A} and line up with an \textsc{int-g} classification. These suggest COCONUTS-3A likely has an age of $\approx 30-200$~Myr but could be younger or older, given its too early spectral type and the uncertainties in the age-gravity class relation.

\begin{figure*}[t]
\begin{center}
\includegraphics[height=3in]{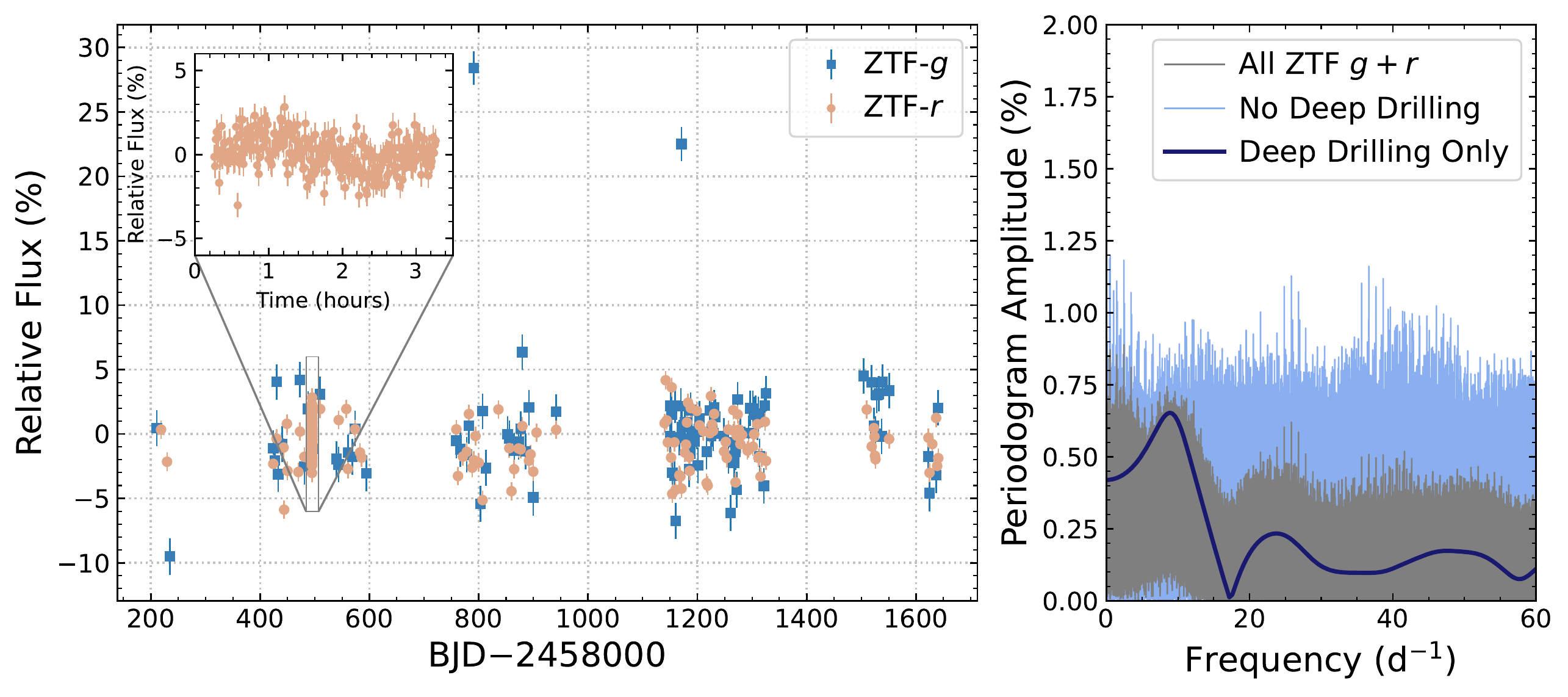} 
\caption{ {\it Left}: The ZTF light curve of COCONUTS-3A in $g$ (blue) and $r$ (light red) bands. Relative flux is the difference in magnitude with respect to the median magnitude in each band, converted to percentage flux units, while BJD is the barycentric-corrected julian date. An inset plot presents a zoom-in view of the high-cadence $r$-band data taken on 2019 Jan 11 with a timespan of 3~hours. It is unclear if the two bright $g$-band points are due to flare activity or contamination by the nearby bright source, but regardless are ignored during the calculation of periodograms. {\it Right}: Periodograms for the ZTF light curve with all combined $g$ and $r$ photometry (grey), the light curve with all but the high-cadence, deep-drilling data (light blue), and the light curve with only high-cadence photometry (dark blue). We detect a tentative period of $2.7 \pm 0.2$~hours from the $r$-band deep-drilling light curve and adopt this value as a lower limit for the stellar rotation period of COCONUTS-3A. }
\label{fig:ztf_lightcurve}
\end{center}
\end{figure*}

\subsubsection{Rotation}
\label{subsubsec:rotation}
This object (TIC~125245420) was observed by the Transiting Exoplanet Survey Satellite (TESS) in sectors 7 and 34 with a 30-minute cadence. However, according to \cite{2019AJ....158..138S}, $83\%$ of the measured flux from the TESS pixels covering COCONUTS-3A is in fact from nearby contaminating sources, especially the close ($10\arcsec$) background giant star, which is $\approx 3$~mag brighter in Gaia $G$ band (see Appendix~\ref{app:xray_bg}). The All-Sky Automated Survey for SuperNovae \citep[ASAS-SN;][]{2014ApJ...788...48S} provided visible light curves with a cadence of 2-3 days, but its $7.8\arcsec$ pixel scale and $\sim 15\arcsec$ FWHM point spread function are too large to avoid the contaminating flux from the background giant star.

We also query the Infrared Science Archive (IRSA) to obtain time-series photometry from DR11 of the Zwicky Transient Facility survey \citep[ZTF;][]{2019PASP..131a8002B, 2019PASP..131a8003M} in $g$ and $r$ bands (Figure~\ref{fig:ztf_lightcurve}). The ZTF images have 1.0$''$ pixels and a FWHM typically less than 3.0$''$, allowing for COCONUTS-3A and the nearby bright star to be resolved on most nights. Following recommendations in the ZTF Science Data System explanatory supplement\footnote{\url{http://web.ipac.caltech.edu/staff/fmasci/ztf/ztf_pipelines_deliverables.pdf}}, we removed poor-quality detections from the ZTF light curve by requiring {\em catflags}~$=1$. The resulting light curve has 478 data points ($g=98$, $r=380$) spanning 1430 days, although 274 of the $r$-band data come from the 3-hour continuous, deep-drilling observations on 2019 Jan 11 as part of the high-cadence galactic plane survey \citep{2021MNRAS.505.1254K}. All ZTF images have exposure times of 30~seconds, and we applied barycentric corrections to the UTC timestamps of each image using the Astropy python package \citep{2013A&A...558A..33A, 2018AJ....156..123A}.

We do not detect a significant stellar rotation period based on the Lomb-Scargle periodograms derived from the light curves of combined $g$ and $r$ photometry or from the ones of individual bands, likely because the observed time-series photometry is sparsely sampled. Excluding the high-cadence data, the ZTF light curve for COCONUTS-3A has an average observing cadence of 4.7 days. However, the $r$-band deep-drilling light curve itself exhibits variations that are consistent with a period of $2.7 \pm 0.2$~hours. This period is very tentative, given that the entire baseline of the high-cadence light curve is only 3~hours. We therefore place a lower limit of $P_{\rm rot, \star} \gtrsim 2.7$~hours for COCONUTS-3A, meaning its stellar rotation can be broadly consistent with M dwarfs spanning ages of a few Myr to Gyr \citep[e.g.,][]{2018AJ....155..196R}.

\subsubsection{Age Summary}
\label{subsubsec:age_summary}
Table~\ref{tab:metal_age_primary} summarizes the age estimates from stellar activity (H$\alpha$, X-ray, and UV emission), kinematics, photometry, and spectroscopic features. We adopt a final age of 100~Myr$-1$~Gyr for the COCONUTS-3 system.

\section{COCONUTS-3B: The Unusually Red L-Dwarf Companion}
\label{sec:companion}

\subsection{Spectral Type and Surface Gravity Classification}
\label{subsec:comp_spec}
We first determine the quantitative near-infrared spectral type of COCONUTS-3B using two index-based methods from \cite{2006ApJ...637.1067B} and \cite{2013ApJ...772...79A}. \cite{2006ApJ...637.1067B} established a near-infrared classification system for T dwarfs based on five H$_{2}$O-band and CH$_{4}$-band features, and \cite{2007ApJ...659..655B} later extended this system to spectral types of L0$-$T8 using polynomial fits to these indices. Following this system, we determine a spectral type of L6.1$\pm$1.8 and L6.0$\pm$1.3 for COCONUTS-3B based on its SpeX prism and GNIRS XD spectra, respectively, with the uncertainties in spectral fluxes and polynomial calibrations propagated in an analytic fashion (Table~\ref{tab:b06_index}).

The \cite{2013ApJ...772...79A} classification system is based on four H$_{2}$O-band indices and is applicable to spectral types of M4$-$L8. Only one index (H$_{2}$OD) provides a reliable classification for $>$L2 dwarfs as the other three indices saturate at L2$-$L8 types \citep[see Figure~6 of][]{2013ApJ...772...79A}. We determine an H$_{2}$OD-based spectral type of L5.0$\pm$0.8 and L5.2$\pm$0.7 for COCONUTS-3B using its SpeX and GNIRS spectra, respectively, which are consistent with those derived from the \cite{2006ApJ...637.1067B} system.

\begin{figure*}[t]
\begin{center}
\includegraphics[height=5.5in]{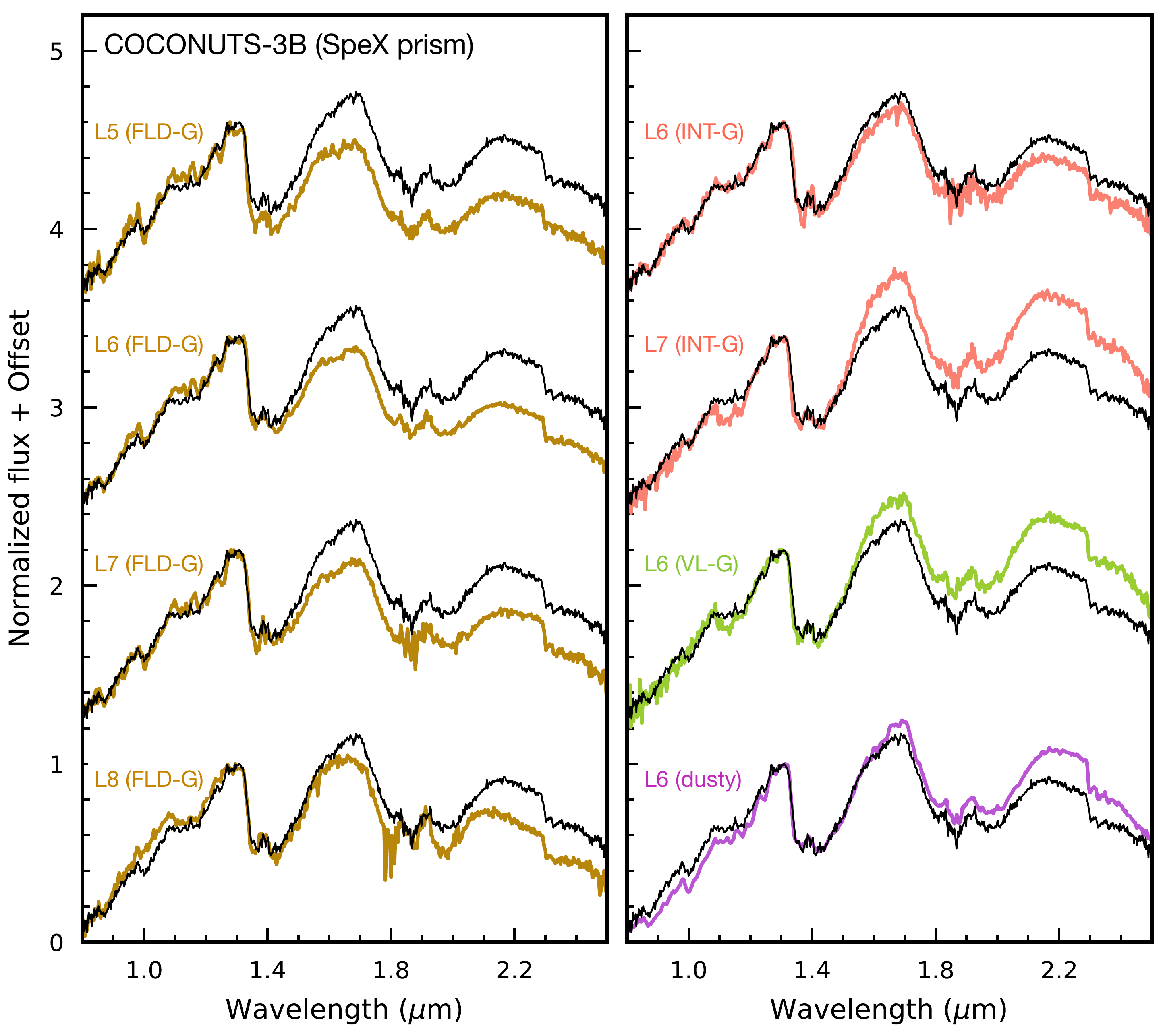} 
\caption{Left: The SpeX prism spectrum of COCONUTS-3B (black) as compared to L5--L8 spectral standards (brown) from the SpeX Prism Library \citep[][]{2014ASInC..11....7B}: 2MASS~J08350622$+$1953050 \citep[L5; brown;][]{2006AJ....131.2722C}, 2MASS~J10101480$-$0406499 \citep[L6; brown;][]{2003AJ....126.2421C}, 2MASS~J0028208$+$224905 \citep[L7; brown;][]{2010ApJ...710.1142B}, 2MASS~J16322911$+$1904407 \citep[L8; brown;][]{1999ApJ...519..802K}. All these standards have \textsc{fld-g} gravity class. Right: We also compare COCONUTS-3B's spectrum to the SpeX prism spectra of an L6 \textsc{int-g} dwarf, 2MASSI~J0103320$+$193536 \citep[2MASS~J0103+1935; red;][]{2004ApJ...604L..61C}, an L7 \textsc{int-g} dwarf, WISEP~J004701.06+680352.1 \citep[WISE~J0047+6803; red;][]{2012AJ....144...94G, 2015ApJ...799..203G}, an L6 \textsc{vl-g} dwarf, 2MASSW~J2244316$+$204343 \citep[2MASS~J2244+2043; green;][]{2008ApJ...686..528L}, and a dusty, unusually red L6 \textsc{fld-g} dwarf, 2MASS~J21481628+4003593 \citep[2MASS~J2148+4003; purple;][]{2008ApJ...686..528L}. All spectra shown in this figure are normalized by their $J$-band peak fluxes. The spectrum of COCONUTS-3B is best matched by that of 2MASS~J0103$+$1935, leading to a visual spectral type of L6 \textsc{int-g}.}
\label{fig:companion_spt_spex}
\end{center}
\end{figure*}

\begin{figure*}[t]
\begin{center}
\includegraphics[height=5.in]{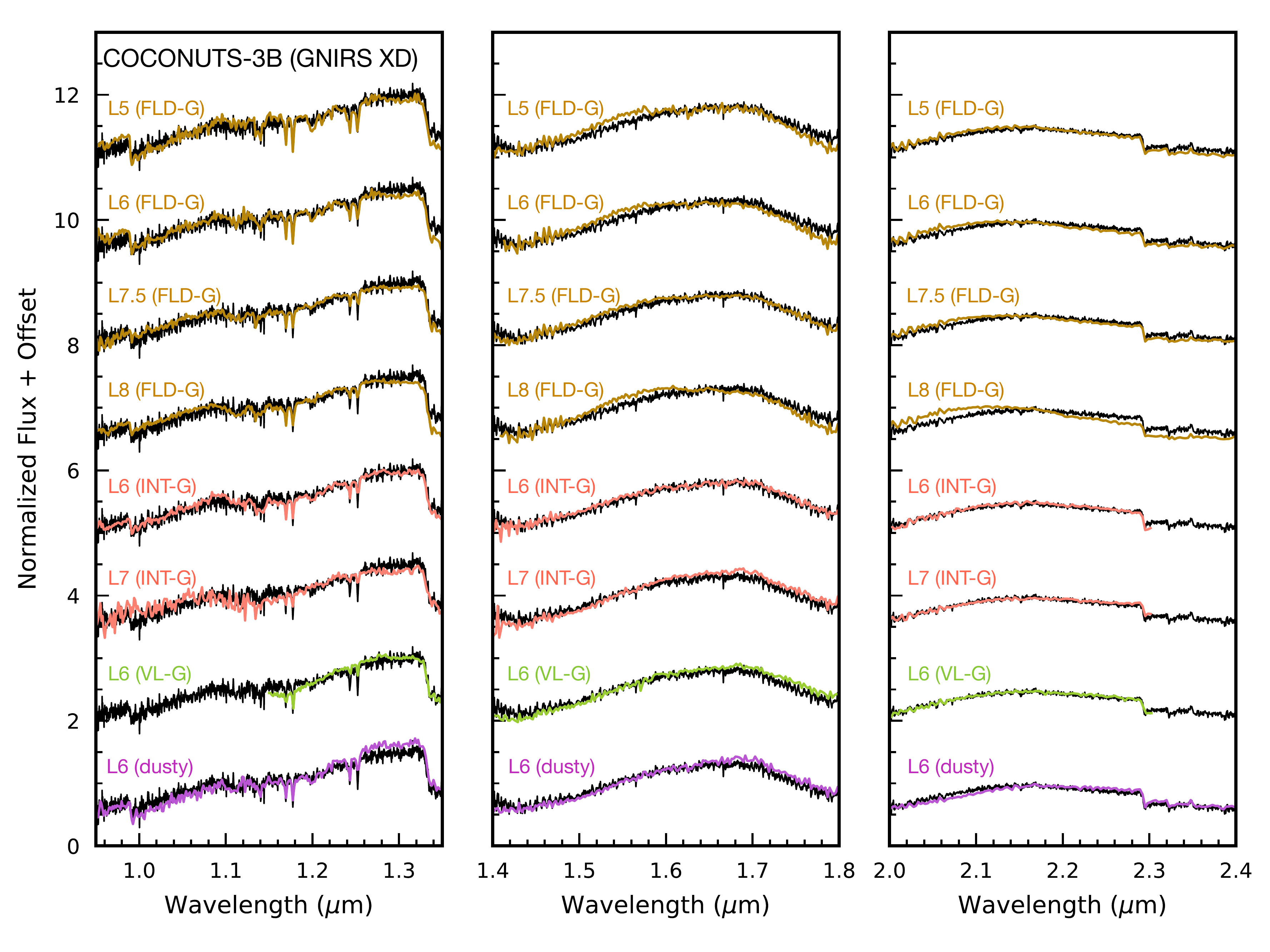} 
\caption{The GNIRS XD spectrum of COCONUTS-3B (black) as compared to the SpeX SXD spectra ($R \sim 2000$) of L5--L8 standards from the IRTF Spectral Library \citep{2005ApJ...623.1115C, 2009ApJS..185..289R} in $J$, $H$, and $K$ bands: 2MASS~J15074769$-$1627386 (L5 \textsc{fld-g}; brown), 2MASS~J15150083+4847416 (L6 \textsc{fld-g}; brown), 2MASS~J08251968+2115521 (L7.5 \textsc{fld-g}; brown), and DENIS-P~025503.3$-$470049.0 (L8 \textsc{fld-g}; brown). We also include the moderate-resolution spectra ($R \sim 2000$) of 2MASS~J0103+1935 \citep[L6~\textsc{int-g}; red; Keck/NIRSPEC data by][]{2003ApJ...596..561M}, WISE~J0047+6803 \citep[L7~\textsc{int-g}; red; SpeX SXD data by][]{2015ApJ...799..203G}, 2MASS~J2244+2043 \citep[L6~\textsc{vl-g}; green; NIRSPEC data by][]{2003ApJ...596..561M}, and 2MASS~J2148+4003 \citep[purple; SpeX SXD data by][]{2008ApJ...686..528L}, whose low-resolution spectra are in Figure~\ref{fig:companion_spt_spex}. We have downgraded the spectral resolution of all standards to match that of the GNIRS XD data ($R \sim 750$) and we have normalized all spectra by their average fluxes in each band. The L6 \textsc{int-g} standard 2MASS~J0103+1935 provides the best match to COCONUTS-3B.}
\label{fig:companion_spt_gnirs}
\end{center}
\end{figure*}

We then derive the \cite{2013ApJ...772...79A} gravity classification (see Section~\ref{subsubsec:nir_grav}) of \textsc{int-g} for both SpeX prism and GNIRS XD spectra. The derived gravity class lines up with the moderately young age of the primary star COCONUTS-3A (Section~\ref{sec:primary_age}). While field-age, dusty L dwarfs can have similar spectral slopes as their young, low-gravity counterparts (e.g., \citealt{2008ApJ...686..528L, 2010ApJS..190..100K}; also see Figure~\ref{fig:companion_spt_spex}), the \cite{2013ApJ...772...79A} gravity classification system can effectively distinguish these two classes of objects. Indeed, we find the FeH$_{z}$ index and the K~\textsc{i} equivalent widths of COCONUTS-3B are distinct from those of high-gravity, dusty L dwarfs with similar spectral types (see Figures~20 and 23 of \citealt{2013ApJ...772...79A}). 

In Figures~\ref{fig:companion_spt_spex} and \ref{fig:companion_spt_gnirs}, we determine the visual spectral type and gravity class of COCONUTS-3B by comparing its SpeX prism and GNIRS XD data to the following spectral standards:
\begin{enumerate}
\item[(1)] L5$-$L8 \textsc{fld-g} dwarfs \citep{1999ApJ...519..802K, 2003AJ....126.2421C, 2005ApJ...623.1115C, 2006AJ....131.2722C, 2009ApJS..185..289R, 2010ApJ...710.1142B}, 
\item[(2)] an L6 \textsc{int-g} dwarf, 2MASSI~J0103320$+$193536 \citep[2MASS~J0103+1935;][]{2003ApJ...596..561M, 2004ApJ...604L..61C},
\item[(3)] an L7 \textsc{int-g} dwarf, WISEP~J004701.06+680352.1 \citep[WISE~J0047+6803;][]{2012AJ....144...94G, 2015ApJ...799..203G},
\item[(4)] an L6 \textsc{vl-g} dwarf, 2MASSW~J2244316$+$204343 \citep[2MASS~J2244+2043;][]{2008ApJ...686..528L},
\item[(5)] and an L6 \textsc{fld-g} dwarf with dusty atmosphere and unusually red colors, 2MASS~J21481628+4003593 \citep[2MASS~J2148+4003;][]{2008ApJ...686..528L}.
\end{enumerate}
The spectra of COCONUTS-3B are best matched by those of 2MASS~$0103+1935$, suggesting a visual classification of L6 \textsc{int-g}. 

\cite{2017AJ....153..196S} determined a near-infrared spectral type of L7 for COCONUTS-3B by comparing this object's CTIO/ARCoIRIS spectra to the \cite{2010ApJS..190..100K} spectral standards based on $\chi^{2}$ values. Their derived spectral type was based on the best match, 2MASS~$0103+1935$, which had an L7 spectral type initially assigned by \cite{2010ApJS..190..100K} but then became the L6 \textsc{int-g} standard later suggested by \cite{2013ApJ...772...79A}. The \cite{2017AJ....153..196S} spectral type is therefore the same as our visual classification result.

\cite{2018AJ....155...34C} have suggested that 2MASS~$0103+1935$ has no low-gravity spectral features in the optical wavelengths and thus might not be an appropriate L6 \textsc{int-g} standard. This will only impact our visual gravity classification of COCONUTS-3B, but the low surface gravity of our companion is also supported by the quantitative spectral indices and equivalent widths of its gravity-sensitive features based on \cite{2013ApJ...772...79A}, its very red near-infrared colors, and the moderately young age of COCONUTS-3A.

Combining our index-based and visual spectral classification, we adopt an L6$\pm$1 \textsc{int-g} as the final near-infrared spectral type for COCONUTS-3B.

\subsection{Bolometric Luminosity and Synthesized Photometry}
\label{subsec:companion_lbol}

We determine the bolometric luminosity of COCONUTS-3B by integrating its spectral energy distribution (SED). We first flux-calibrate both SpeX prism and GNIRS XD spectra of the companion using its observed VHS \Ks\ magnitude. Then we construct an SED by combining the object's $0.9-2.4$~$\mu$m SpeX prism spectrum with broadband fluxes from $z_{\rm P1}$ and CatWISE $W1$ and $W2$ photometry. We linearly interpolate fluxes to fill in the wavelength gaps between spectra and broadband fluxes. At shorter wavelengths than $z_{\rm P1}$, we linearly extrapolate the SED to zero flux at zero wavelength, and at longer wavelengths beyond $W2$ (with a cut at 1000~$\mu$m), we append a Rayleigh-Jeans tail. We compute the bolometric luminosity of \logLbol$= -4.451 \pm 0.002$~dex, with the uncertainties of SED fluxes and the host star's parallax propagated in a Monte Carlo fashion. We also construct the SED using the companion's GNIRS XD data with wavelengths of $0.95-1.35$~$\mu$m, $1.45-1.8$~$\mu$m, and $1.95-2.4$~$\mu$m to avoid spectra with low S/N and significant telluric features. Integrating this GNIRS-based SED leads to a consistent \Lbol\ with the SpeX-based value within $1\sigma$.

To examine the systematic error of our computed \Lbol, we apply the same SED analysis to a set of \cite{2008ApJ...689.1327S} atmospheric model spectra whose fluxes are scaled at the substellar surface, spanning \Teff$ = 1000-1600$~K (100~K intervals), \logg$= 4$ and $5$~dex, and \fsed$= 1$ and $2$. These physical parameters are close to properties of COCONUTS-3B (see Sections~\ref{subsec:companion_phys} and \ref{subsec:discuss}). We synthesize $z_{\rm P1}$, $W1$, and $W2$ broadband fluxes for each model spectrum, and then we tailor the spectral resolution and wavelength range of the models to match those of our SpeX and GNIRS data used for the aforementioned SED construction. We also conduct the same linear interpolation and extrapolation to generate model-based SED fluxes spanning 0$-$1000~$\mu$m. By integrating our emulated SED at each model grid point, we compute the bolometric flux \Fbol\, and then compare with the original model value of $F_{\rm bol,true} = \sigma T_{\rm eff}^{4}$, where $\sigma$ is the Stefan-Boltzmann constant. Among all grid points, we find the mean and standard deviation of $\log{(F_{\rm bol,true})} -$\logFbol\ is $-0.003 \pm 0.033$~dex for SpeX-based SEDs, and is $-0.010 \pm 0.033$~dex for GNIRS-based SEDs. These differences in logarithmic bolometric fluxes between the computed values from emulated SEDs and the original values equal to the differences in logarithmic bolometric luminosities. We therefore adopt a systematic error of $0.03$~dex for our SED-based \Lbol\ and derive a bolometric luminosity of \logLbol$= -4.45 \pm 0.03$~dex for COCONUTS-3B.

We also synthesize 2MASS and MKO magnitudes for COCONUTS-3B using its spectra and VHS photometry following Section~\ref{subsec:phys}. For each broadband $X$ in \{\Jtms, \Htms, \Ktms, \YMKO, \JMKO, \HMKO, \KMKO\}, we use each of the companion's SpeX prism and GNIRS XD spectra to compute two colors, $X-J$ and $X-$\Ks, with $J$ and \Ks\ bands from VHS. Spectral flux uncertainties are propagated into these computed colors in a Monte Carlo fashion. For this calculation, we use the Vega spectrum and obtain 2MASS, MKO, and VHS filters from \cite{2003AJ....126.1090C}, \cite{2006MNRAS.367..454H}, and the ESO VISTA instrument webpage (see Section~\ref{subsec:phys}), respectively. We thus obtain four estimates of the $X$-band magnitude by computing $J + (X-J)$ and $K_{S} + (X-K_{S})$ using the SpeX and GNIRS spectra. We confirm all these four estimates in a given band $X$ are consistent within $1\sigma$ and compute their average and the standard deviation. For $X$ that is not $J$ or $K$ band, we follow \cite{2012ApJS..201...19D} and incorporate an additional 0.05~mag systematic error to the final synthetic photometry (see Table~\ref{tab:info}). 

We find our synthesized \Htms$=15.89 \pm 0.06$~mag is brighter than the observed \Htms$=16.25 \pm 0.18$~mag by 0.36~mag ($2.0\sigma$) and our synthesized \Ktms$=15.02 \pm 0.03$~mag is fainter than the observed \Ktms$=14.86 \pm 0.13$~mag by 0.16~mag ($1.2\sigma$). These indicate that the observed \Htms$-$\Ktms\ is $\sim 0.5$~mag redder than the color from our SpeX prism and GNIRS XD data, although these two spectra have consistent morphology and do not exhibit noticeable spectral variability. In addition, if we compute the VHS $J-$\Ks\ color using our SpeX and GNIRS spectra, then these values are both consistent with the observed $J-$\Ks\ color within 0.05~mag. The observed \Htms\ and \Ktms\ magnitudes of COCONUTS-3B are not blended or affected by artifact, but have modest S/N ($\approx 5-8$). We therefore recommend using our synthetic photometry in 2MASS bands. Photometric variability monitoring of this object will be valuable to investigate this discrepancy.

We compute the photometric distance of COCONUTS-3B as $32 \pm 7$~pc, by using its synthetic \JMKO\ magnitude and the typical \JMKO-band absolute magnitude of L6 \textsc{int-g} dwarfs provided by Table~10 of \cite{2016ApJ...833...96L}. As discussed in Section~\ref{sec:system}, this distance for COCONUTS-3B is consistent with that of COCONUTS-3A ($d = 30.88 \pm 0.02$~pc) and validates their physical association.

\begin{figure*}[t]
\begin{center}
\includegraphics[height=4.5in]{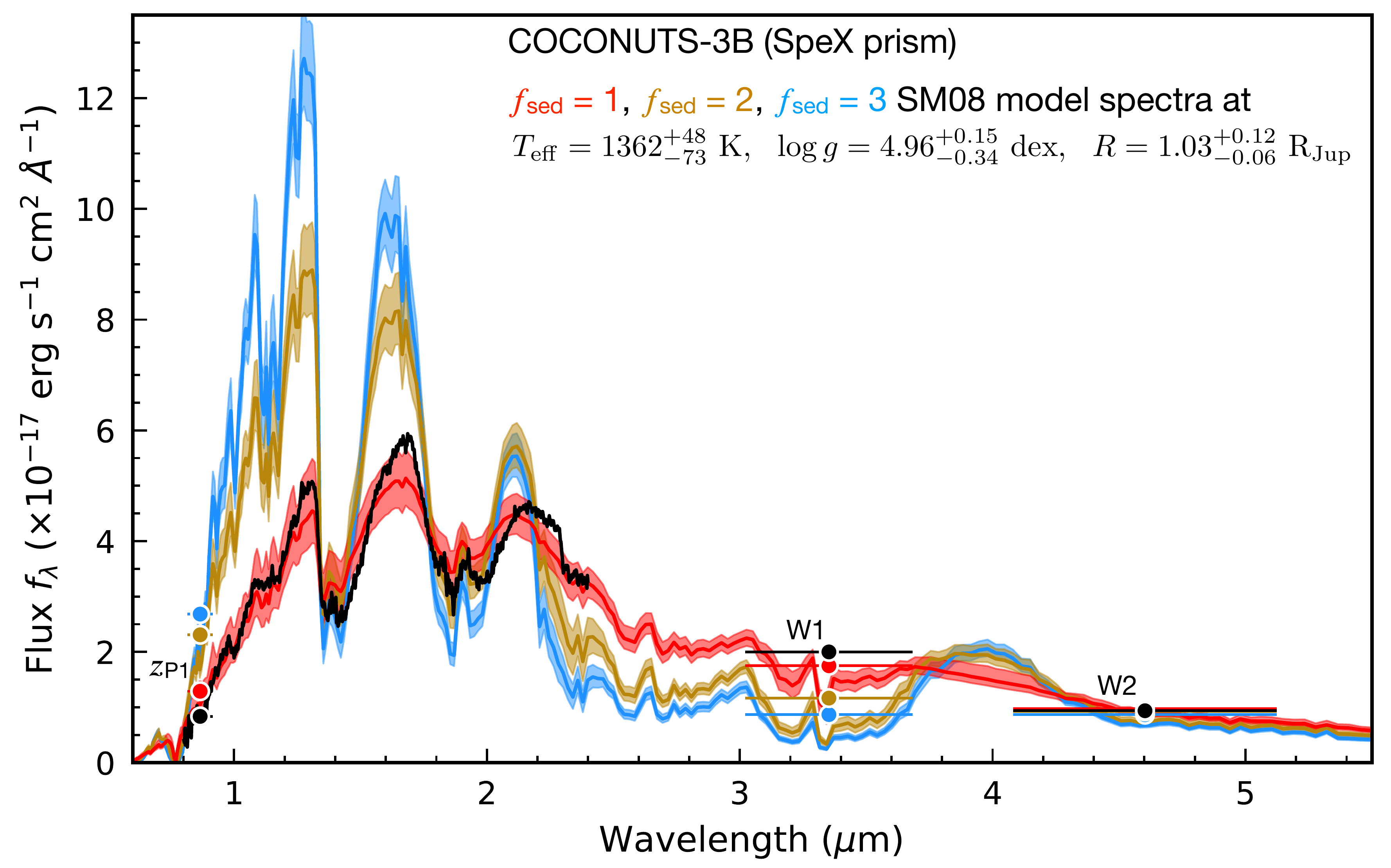} 
\caption{The SpeX prism spectrum of COCONUTS-3B (black) as compared to the evolution-based \cite{2008ApJ...689.1327S} model spectra described in Section~\ref{subsubsec:companion_sm08spec}, with \fsed$=1,2,3$ models displayed in red, brown, and blue, respectively. We use colored solid lines and shadows to present the median and $68\%$ confidence intervals of the model spectra constructed at the evolution-based \Teff, \logg, $R$ of COCONUTS-3B. We also synthesize $z_{\rm P1}$, $W1$, and $W2$ fluxes using these models (colored circles) and compare them with the observed photometry (black circles). The SED of COCONUTS-3B can be best matched by atmospheric models with \fsed$=1$, suggesting the companion retains ample condensate clouds in its photosphere.}
\label{fig:companion_sm08_spex}
\end{center}
\end{figure*}

\begin{figure*}[t]
\begin{center}
\includegraphics[height=4.5in]{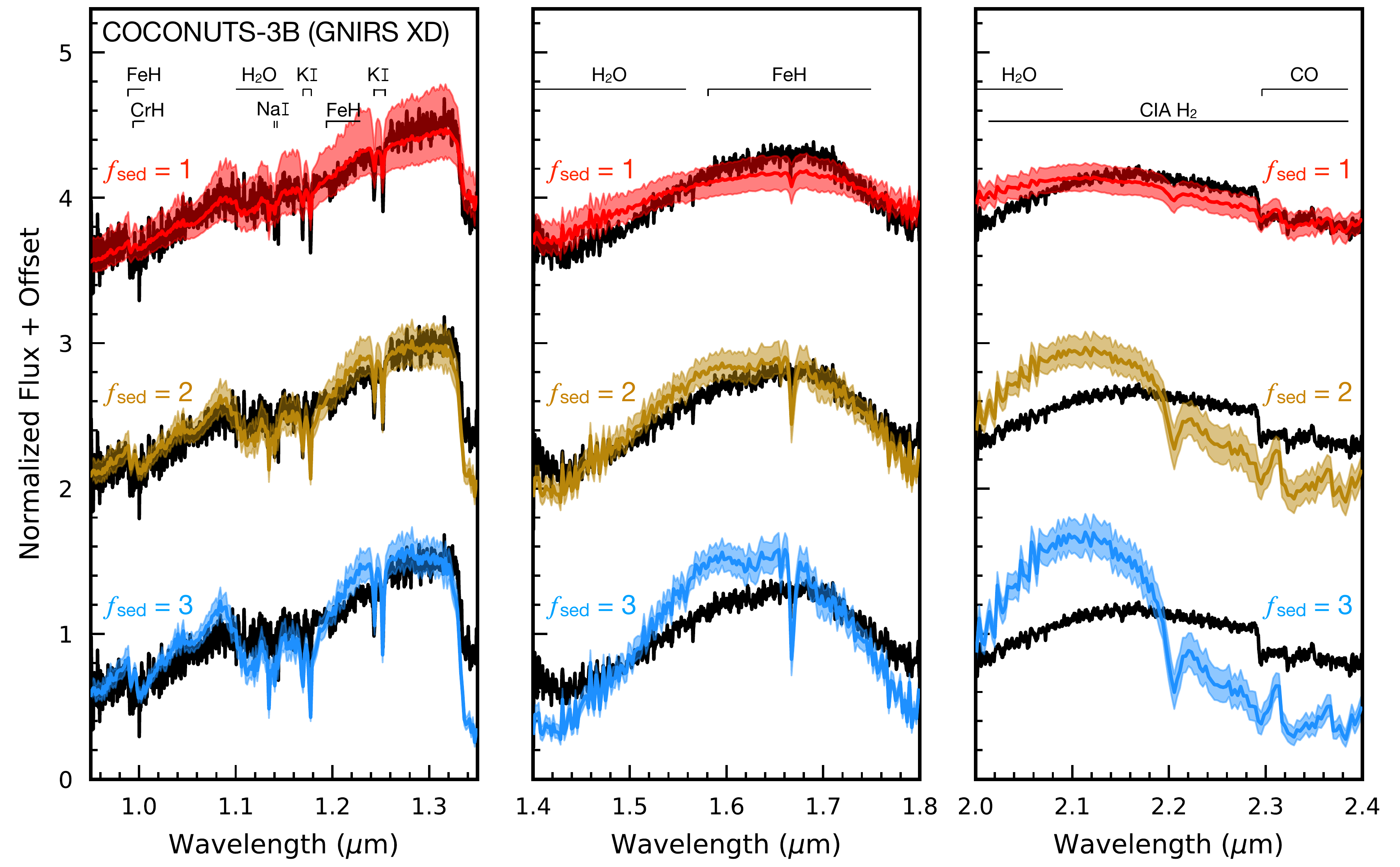} 
\caption{The GNIRS XD spectrum of COCONUTS-3B (black) as compared to the evolution-based \cite{2008ApJ...689.1327S} model spectra (colored; same as those shown in Figure~\ref{fig:companion_sm08_spex}) in individual $JHK$ bands. All spectra are normalized by their mean values within each band. The data are best matched by \fsed$=1$ models, although these models predict too shallow depths in Na~\textsc{i} and K~\textsc{i} resonance lines in $J$ band and slightly bluer spectral slope in $K$ band than the observation. The absorption feature at $\sim 1.66$~$\mu$m predicted by models is likely due to the incomplete CH$_{4}$ line list used by \cite{2008ApJ...689.1327S} models.}
\label{fig:companion_sm08_gnirs}
\end{center}
\end{figure*}

\subsection{Physical Properties based on Evolution Models}
\label{subsec:companion_phys}

We infer COCONUTS-3B's effective temperature (\Teff), surface gravity (\logg), radius (\R), and mass (\M) by using the hot-start \cite{2008ApJ...689.1327S} cloudy evolutionary models. These models parameterize the sedimentation efficiency of condensate particles via \fsed\ using the \cite{2001ApJ...556..872A} framework and include a grid of \fsed$=1,2,3,4$, with higher \fsed\ meaning a larger average cloud particle size, higher sedimentation efficiency, and thereby less cloud effect on the photosphere. To reproduce the photometric sequence of MLT-type ultracool dwarfs, \cite{2008ApJ...689.1327S} also produced a suite of ``hybrid'' models with \fsed$=1$ for $T_{\rm eff} \geqslant 1400$~K and \fsed$=4$ for $T_{\rm eff} \leqslant 1200$~K. For $1200 < T_{\rm eff} < 1400$~K, they interpolated the surface boundary condition in $T_{\rm eff}$ using the \fsed$=2$ models at $T_{\rm eff}=1400$~K and the cloudless models at $T_{\rm eff}=1200$~K. 

We assume COCONUTS-3B's age follows a uniform distribution spanning $[0.1,1.0]$~Gyr (Section~\ref{sec:primary_age}) and its \logLbol\ follows a normal distribution of $N(\mu=-4.45, \sigma^{2}=0.03^{2})$. We draw random age and \logLbol\ values from these distributions to linearly interpolate the grid of physical parameters predicted by the \cite{2008ApJ...689.1327S} hybrid evolution models at these ages and \logLbol. Such interpolation is conducted logarithmically for the $T_{\rm eff}$ values. Computing the median and 16-to-84 percentile intervals of the interpolated model parameters, we derive \Teff$= 1362^{+48}_{-73}$~K, \logg$= 4.96^{+0.15}_{-0.34}$~dex, $R = 1.03^{+0.12}_{-0.06}$~R$_{\rm Jup}$, and $M = 39^{+11}_{-18}$~M$_{\rm Jup}$ as the evolution-based physical properties of COCONUTS-3B.

Our derived physical properties will inevitably carry any systematic errors in the evolution model predictions. \cite{2009ApJ...692..729D, 2014ApJ...790..133D} measured dynamical masses of two binaries composed of mid-L brown dwarfs, HD~130948BC (L4$+$L4) and Gl~417BC (L4.5$+$L6), based on orbit monitoring. These two binaries are both companions to solar-mass host stars and thus have independently known ages. Similar to our analysis of COCONUTS-3B, \citeauthor{2009ApJ...692..729D} used these L dwarfs' ages and observed bolometric luminosities to infer their masses from several evolution models. However, they found the evolution-based masses are higher than their directly measured values by $\approx 25\%$ for the \cite{2008ApJ...689.1327S} hybrid evolution models. Recently, \cite{2019AJ....158..140B} and \cite{2021arXiv210907525B} further expanded the dynamical mass sample by combining radial velocities, relative astrometry, and the Hipparcos-Gaia accelerations of several systems hosting brown dwarf companions, including three L dwarfs, HR~7672B \citep[L4.5;][]{2002ApJ...571..519L}, HD~33632Ab \citep[L9.5$^{+1.0}_{-3.0}$;][]{2020ApJ...904L..25C}, and HD~72946B \citep[L5.0$\pm$1.5;][]{2020A&A...633L...2M}. These objects' masses predicted by \cite{2008ApJ...689.1327S} hybrid evolution models given their ages and \Lbol\ are all consistent with their dynamical masses, although these L dwarfs have older ages ($\approx 2$~Gyr) than HD~130948BC and Gl~417BC ($\approx 800$~Myr). If the evolution-based properties of COCONUTS-3B suffer from the same systematic errors as HD~130948BC and Gl~417BC, then its true mass might be as low as $\sim 30$~M$_{\rm Jup}$. 

In addition, we note our derived physical properties of COCONUTS-3B are based on solar-metallicity evolution models while its host star has slightly super-solar bulk metallicity (Section~\ref{sec:primary_metallicity}). To estimate this metallicity effect, we re-compute the physical properties of COCONUTS-3B using two classes of the cloudless \cite{2008ApJ...689.1327S} evolution models with [Fe/H]$=0$~dex and $+0.3$~dex. Switching models from solar to a super-solar metallicity, COCONUTS-3B will have a $21$~K cooler $T_{\rm eff}$, a 0.04~dex lower $\log{(g)}$, a 0.03~$R_{\rm Jup}$ larger $R$, and a 2~M$_{\rm Jup}$ smaller $M$. These systematic differences are all much smaller than the uncertainties of our derived properties.

\subsection{Discussion}
\label{subsec:discuss}

\subsubsection{Comparison with \cite{2008ApJ...689.1327S} Atmospheric Model Spectra}
\label{subsubsec:companion_sm08spec}
To examine whether the \cite{2008ApJ...689.1327S} models can sufficiently interpret the atmospheres of mid-to-late-L dwarfs like COCONUTS-3B, we construct the \cite{2008ApJ...689.1327S} atmospheric model spectra at this companion's evolution-based physical properties. Specifically, we linearly interpolate the model spectra using the distributions of \Teff\ and \logg\ computed in Section~\ref{subsec:companion_phys} and then scale these models by the companion's evolution-based $R$ and COCONUTS-3A's parallax. As demonstrated in \cite{2020ApJ...891..171Z} and \cite{2021ApJ...916...53Z}, the difference between these evolution-based model spectra and the objects' observed spectra can reveal the shortcomings of the assumptions adopted by model atmospheres.

Figure~\ref{fig:companion_sm08_spex} compares the SpeX prism spectrum and broadband fluxes of COCONUTS-3B to the evolution-based \cite{2008ApJ...689.1327S} model spectra with \fsed$=1,2,3$. Figure~\ref{fig:companion_sm08_gnirs} further examines the moderate-resolution spectral features by comparing these models to the GNIRS XD data. The spectrophotometry of COCONUTS-3B is best described by dusty atmospheres with \fsed$=1$, although such models produce slightly brighter $z_{\rm P1}$-band and fainter $W1$-band fluxes than observed values. Also, these \fsed$=1$ models predict too shallow depths for Na~\textsc{i} and K~\textsc{i} resonance lines in $J$ band and a slightly bluer $K$-band spectral slope than the data. These differences are very likely due to the older version of the opacity database \citep[e.g.,][]{2008ApJS..174..504F} for the alkali lines (in red optical and $J$ bands), CH$_{4}$ (in $H$, $K$, and $W1$ bands), and collision-induced H$_{2}$ absorption (in $K$ band) adopted by \cite{2008ApJ...689.1327S}. New sets of cloudy models that incorporate the updated opacities \citep[e.g.,][]{2012ApJ...750...74S, 2015PhyS...90e4005R, 2016A&A...589A..21A, 2019A&A...628A.120A, 2013JMoSp.291...69Y, 2014MNRAS.440.1649Y} will likely improve the consistency between model predictions and observations, leading to more reliable estimates of these objects' physical properties via atmospheric modeling.

\subsubsection{Comparison with Previously Known Ultracool Benchmarks}
\label{subsubsec:companion_context}
\begin{figure*}[t]
\begin{center}
\includegraphics[height=6in]{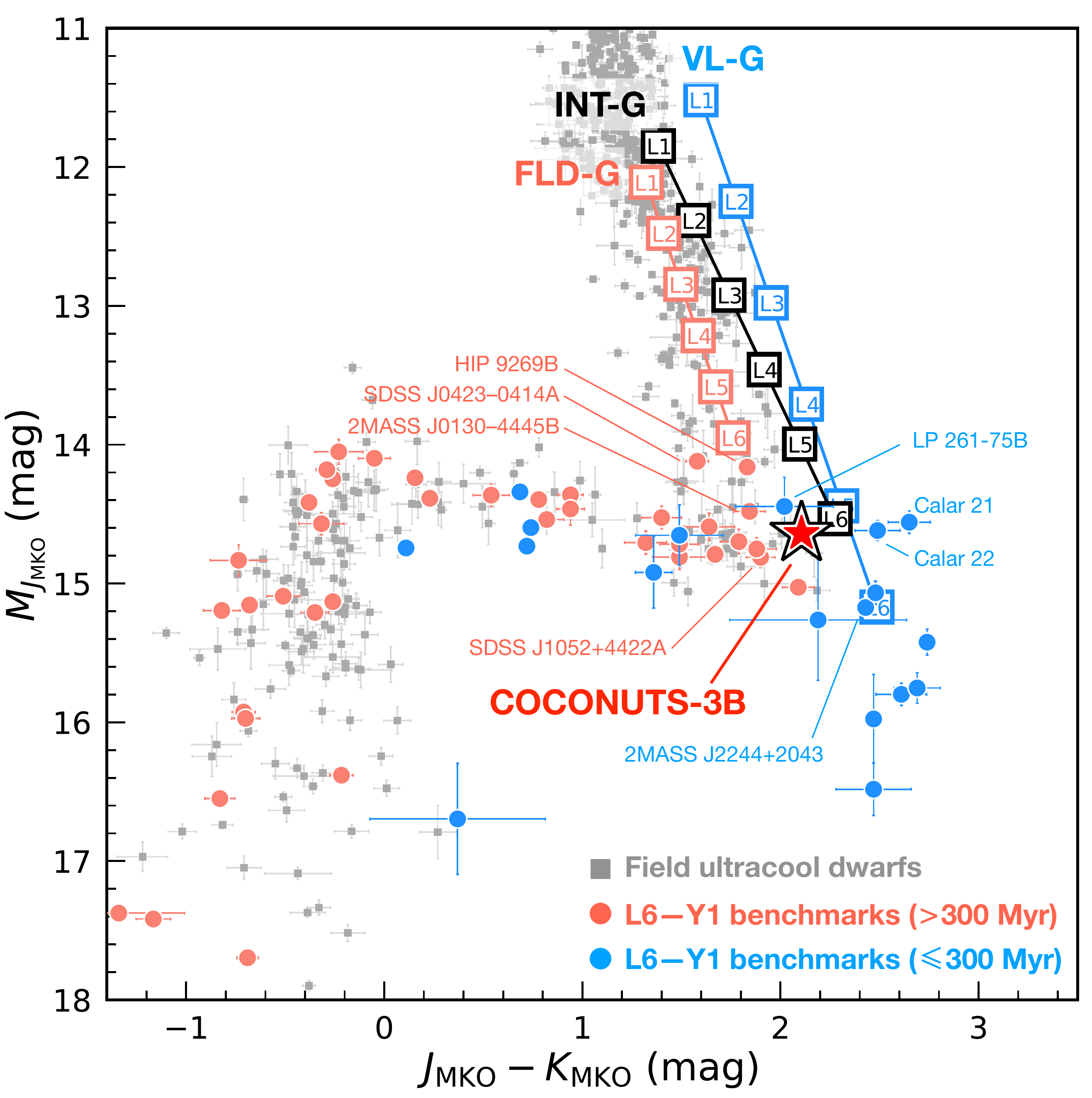} 
\caption{$J$-band absolute magnitude and $J-K$ color of COCONUTS-3B (red star) compared to those of L6$-$Y1 benchmarks compiled by \cite{2020ApJ...891..171Z} and \cite{2021ApJ...911....7Z} with ages of $>300$~Myr (orange) and $\leqslant 300$~Myr (blue), as well as the photometric sequences of L1$-$L6 dwarfs with \textsc{fld-g} (orange), \textsc{int-g} (black), and \textsc{vl-g} (blue) gravity classes. We overlay the photometry of field dwarfs (grey) with M, L, T, Y spectral types obtained from the UltracoolSheet \citep[][]{ultracoolsheet}, which have S/N$>5$ in $M_{J}$ and $J-K$ and are not young, binaries, or subdwarfs. We label all benchmarks with the same spectral type (L6) as COCONUTS-3B, including HIP~9269B \citep[][]{2014ApJ...792..119D}, SDSSp~J042348.57$-$041403.5A \citep[SDSS~J0423$-$0414A;][]{2005ApJ...634L.177B}, 2MASS~J01303563$-$4445411B \cite[2MASS~J0130$-$4445B;][]{2008AJ....136.1290R, 2011AJ....141....7D}, SDSS~J105213.51+442255.7A \citep[SDSS~J1052$+$4422A;][]{2006AJ....131.2722C, 2015ApJ...805...56D}, LP~261$-$75B \citep[][]{2000AJ....120..447K, 2006PASP..118..671R}, 2MASS~J2244$+$2043 \citep[][]{2002AJ....124.1170D}, and Calar~21 and Calar~22 \citep[][]{2014A&A...568A..77Z}. The $J-K$ color of COCONUTS-3B is among the reddest compared to all benchmarks with ages of a few 100~Myr or older. }
\label{fig:companion_context}
\end{center}
\end{figure*}

To place our discovery in the context, we compare the photometry of COCONUTS-3B to that of L6$-$Y1 benchmarks compiled by \cite{2020ApJ...891..171Z} and \cite{2021ApJ...911....7Z} in Figure~\ref{fig:companion_context}. These benchmarks include wide-orbit co-moving companions to stars, kinematic members of nearby young moving groups, and components of substellar binaries with dynamical masses. All of these objects have independently known ages or masses which are usually impossible to determine for free-floating single brown dwarfs in the field. We also compare the $J$-band absolute magnitude and $J-K$ color of COCONUTS-3B to the photometric sequences of \textsc{fld-g}, \textsc{int-g}, and \textsc{vl-g} ultracool dwarfs from \cite{2016ApJ...833...96L}. COCONUTS-3B has slightly brighter and bluer near-infrared photometry than L6 \textsc{vl-g} objects, but is much fainter and redder than their \textsc{fld-g} counterparts. Also, the $J_{\rm MKO}-K_{\rm MKO} = 2.11 \pm 0.02$~mag of COCONUTS-3B is among the reddest of all ultracool benchmarks with ages older than a few 100~Myr. COCONUTS-3B is in fact similar to the brown dwarf companion HD~206893B, which has a moderately young age ($50-700$~Myr based on its host star) and an extremely red near-infrared color \citep[][]{2017A&A...597L...2M, 2017A&A...608A..79D}.\footnote{\cite{2017A&A...608A..79D} synthesized HD~206893B's $J_{S}$-band magnitude from their VLT/SPHERE IFS spectroscopy ($R\sim30$) with S/N$\approx10$ and measured $K1$ and $K2$-band magnitudeS using IRDIS data, leading to very red colors of $J_{S}-K1 = 3.13 \pm 0.20$~mag and $J_{S}-K2 = 3.45 \pm 0.19$~mag. We synthesize the $J_{S}-K1 = 1.969 \pm 0.004$~mag and $J_{S}-K2 = 2.220 \pm 0.004$~mag for COCONUTS-3B using its SpeX prism spectrum, the dual-band $K12$ filters of SPHERE IRDIS, and a virtual $J_{S}$ filter with $100\%$ transmission spanning 1.2$-$1.3$\mu$m adopted by \cite{2017A&A...608A..79D}. COCONUTS-3B is thus bluer than HD~206893B, but still redder than field-age late-L dwarfs \citep[see Figure~7 of][]{2017A&A...608A..79D}. }

Field-age, high-gravity L6 dwarfs have a typical \logLbol\ of $-4.37 \pm 0.14$~dex and a \Teff\ of $1483 \pm 113$~K based on polynomial fits in \cite{2020ApJ...891..171Z} and \cite{2015ApJ...810..158F}, respectively. COCONUTS-3B has a 0.08~dex ($0.5\sigma$) fainter bolometric luminosity and a 121~K ($1.0\sigma$) cooler effective temperature compared to its high-gravity counterparts with similar spectral types. The companion's lower \Teff, faint absolute magnitude, and red colors are in accord with its moderately young age, intermediate surface gravity, and dusty atmospheres, given that properties of L/T transition objects are known to be surface gravity dependent \citep[e.g.,][]{2006ApJ...651.1166M, 2012ApJ...754..135M, 2016ApJS..225...10F, 2016ApJ...833...96L}.

\subsubsection{Formation Scenario}
\label{subsubsec:formation}

Formation of planetary-mass and substellar companions that reside on very wide orbits (a few 100$-$1000~au) is intriguing. These companions might form in situ like components of stellar binaries via fragmentation of the collapsing proto-stellar clouds \citep[e.g.,][]{2001MNRAS.322..231K, 2003PASP..115..763C}. Alternatively, they might form like closer-in exoplanets via core/pebble accretion or disk gravitational instability and then scatter outward to large orbital separations due to dynamical interactions with other system components \citep[e.g.,][]{2006ApJ...637L.137B, 2009ApJ...696.1600V}. For the COCONUTS-3 system, it is very likely that the companion formed in a star-like fashion, given that the companion's mass of $\approx 40$~M$_{\rm Jup}$ is $\gtrsim 10\times$ larger than the expected total mass of protoplanetary disks surrounding low-mass ($\approx 0.1$~M$_{\odot}$) M dwarfs like COCONUTS-3A \citep[e.g.,][]{2016ApJ...828...46A, 2016ApJ...831..125P, 2018A&A...618L...3M}. In addition, a companion-to-host mass ratio of $0.29 \pm 0.10$ for the COCONUTS-3 system is much higher than those of directly imaged exoplanets ($\lesssim 0.04$; e.g., \citealt{2008Sci...322.1348M, 2010Natur.468.1080M, 2015Sci...350...64M, 2020ApJ...898L..16B, 2021A&A...648A..73B}; also see Figure~3 of \citealt{2021ApJ...916L..11Z}).

\section{Conclusion}
\label{sec:conclusion}
We have reported the third discovery from our COol Companions ON Ultrawide oribiTS (COCONUTS) program. The COCONUTS-3 system is composed of a young M-dwarf primary star with strong H$\alpha$ and X-ray emission, as well as an unusually red L-dwarf companion which was previously identified as the free-floating object WISEA~J081322.19$-$152203.2 by \cite{2017AJ....153..196S}. Given the consistency in proper motions and distances, we confirm the physical association between A and B components and compute a projected separation of $61\arcsec$ ($1891$~au). Based on the primary star's optical and near-infrared spectrophotometry, we derive its physical properties (e.g., \Teffstar$= 3291 \pm 49$~K and \Mstar$= 0.123 \pm 0.006$~\Msun) and a slightly super-solar metallicity of [Fe/H]$_{\star} =0.21 \pm 0.07$~dex. We note this [Fe/H]$_{\star}$ is derived using empirical calibrations established by older and higher-gravity M dwarfs, and could be under-estimated by $0.2-0.3$~dex according to PHOENIX stellar models given COCONUTS-3A's younger age and lower surface gravity. Combining the host star's stellar activity, kinematics, HR~diagram position, lithium absorption, and surface gravity-sensitive spectroscopic features, we estimate an age of 100~Myr to 1~Gyr for the COCONUTS-3 system.

We also study the near-infrared spectra of COCONUTS-3B and derive a spectral type of L6 with an intermediate surface gravity classification ($\textsc{int-g}$). Based on the companion's observed bolometric luminosity and its host star's age, we use the hot-start \cite{2008ApJ...689.1327S} hybrid evolution models to infer the physical properties of COCONUTS-3B to be \Teff$= 1362^{+48}_{-73}$~K, \logg$= 4.96^{+0.15}_{-0.34}$~dex, $R = 1.03^{+0.12}_{-0.06}$~R$_{\rm Jup}$, and $M = 39^{+11}_{-18}$~M$_{\rm Jup}$. It has been recently suggested that the ultracool evolution models tend to over-predict the mass at a given \Lbol\ and age by $\sim 25\%$ for moderately young brown dwarfs ($<1$~Gyr) based on two substellar binary systems with directly measured dynamical masses, meaning that the true mass of COCONUTS-3B might be as low as $\sim 30$~M$_{\rm Jup}$.

We construct the \cite{2008ApJ...689.1327S} atmospheric model spectra at our derived evolution-based physical properties of COCONUTS-3B and compare them with the observed spectra. We find models with \fsed$=1$ match the companion's spectrophotometry, although these models predict slightly brighter fluxes in the red optical, shallower depths for alkali resonance lines, bluer $K$-band spectral slope, and fainter fluxes near 3.3~$\mu$m. These data-model differences are very likely due to the older version of the opacity database used by \cite{2008ApJ...689.1327S} and new sets of cloudy models that incorporate the theoretical advances over the past decade will likely improve the consistency between data and model predictions.  

Compared to field-age, high-gravity L6 dwarfs, COCONUTS-3B has much fainter absolute magnitudes, similar bolometric luminosity, and a $120$~K cooler effective temperature. This companion's $J-K$ color is among the reddest of ultracool benchmarks with ages older than a few 100~Myr. These anomalous spectrophotometric and physical properties are in accord with its intermediate surface gravity, dusty atmosphere, and moderately young age, given the surface gravity dependence of the L/T transition.

Similar to stellar binaries, the COCONUTS-3 system likely formed via fragmentation processes of the collapsing proto-stellar clouds, given the companion's mass and the large companion-to-host mass ratio ($0.29 \pm 0.10$). Under this formation scenario, COCONUTS-3B is expected to share the bulk metallicity and elemental abundances (e.g., C/O) as its host star. Modeling spectroscopy of COCONUTS-3B and comparing the resulting chemical properties to those of COCONUTS-3A can directly quantify the systematic errors of substellar and exoplanet model atmospheres. Expanding such calibration to a large ensemble of wide-orbit companions will enable thorough examination of model atmospheres over a wide parameter space, and these empirically calibrated models will lead to robust characterization of the atmospheric properties and formation of directly imaged exoplanets.

\begin{acknowledgments}
Z.Z. thanks Didier Saumon for sharing the \cite{2008ApJ...689.1327S} atmospheric model spectra; Andre-Nicolas Chene and Bin Yang for helpful discussions about the Gemini/GMOS data reduction; Eric Mamajek for discussions about typical photometric properties of dwarf stars; Andrew Mann for discussions about the empirical metallicity calibrations of young low-mass stars; Eunkyu Han for discussions about the magnetic fields in M dwarfs; Adam Kraus and Brendan Bowler for helpful comments on the manuscript. Z.P.V. acknowledges the Heising-Simons Foundation for postdoctoral scholar support at Caltech. This work has benefited from The UltracoolSheet at http://bit.ly/UltracoolSheet, maintained by Will Best, Trent Dupuy, Michael Liu, Rob Siverd, and Zhoujian Zhang, and developed from compilations by \cite{2012ApJS..201...19D}, \cite{2013Sci...341.1492D}, \cite{2016ApJ...833...96L}, \cite{2018ApJS..234....1B}, and \cite{2021AJ....161...42B}. This research has benefitted from the Ultracool RIZzo Spectral Library (\url{http://dx.doi.org/10.5281/zenodo.11313}), maintained by Jonathan Gagn\'{e} and Kelle Cruz. This research has benefitted from the Montreal Brown Dwarf and Exoplanet Spectral Library, maintained by Jonathan Gagn\'{e}. This work is based in part on observations obtained at the international Gemini Observatory, a program of NSF's NOIRLab, which is managed by the Association of Universities for Research in Astronomy (AURA) under a cooperative agreement with the National Science Foundation. on behalf of the Gemini Observatory partnership: the National Science Foundation (United States), National Research Council (Canada), Agencia Nacional de Investigaci\'{o}n y Desarrollo (Chile), Ministerio de Ciencia, Tecnolog\'{i}a e Innovaci\'{o}n (Argentina), Minist\'{e}rio da Ci\^{e}ncia, Tecnologia, Inova\c{c}\~{o}es e Comunica\c{c}\~{o}es (Brazil), and Korea Astronomy and Space Science Institute (Republic of Korea). This research has made use of the NASA/IPAC Infrared Science Archive, which is funded by the National Aeronautics and Space Administration and operated by the California Institute of Technology. This work has made use of data from the European Space Agency (ESA) mission {\it Gaia} (\url{https://www.cosmos.esa.int/gaia}), processed by the {\it Gaia} Data Processing and Analysis Consortium (DPAC, \url{https://www.cosmos.esa.int/web/gaia/dpac/consortium}). Funding for the DPAC has been provided by national institutions, in particular the institutions participating in the {\it Gaia} Multilateral Agreement.
\end{acknowledgments}

\facilities{UH~2.2m (SNIFS), Gemini (GMOS, GNIRS), IRTF (SpeX)}

\software{{\it emcee} \citep{2013PASP..125..306F}, BANYAN~$\Sigma$ \citep[version~1.2; ][]{2018ApJ...856...23G}, LACEwING \citep[][]{2017AJ....153...95R}, TOPCAT \citep[][]{2005ASPC..347...29T}, Astropy \citep{2013A&A...558A..33A, 2018AJ....156..123A}, IPython \citep{PER-GRA:2007}, Numpy \citep{numpy},  Scipy \citep{scipy}, Matplotlib \citep{Hunter:2007}.}

\appendix

\section{A Model-based Exploration of the Surface-Gravity Dependence \\of the Empirical Metallicity Calibrations for Low-mass Stars}
\label{app:logg_effect_metal}

We estimated the bulk metallicity of COCONUTS-3A (Section~\ref{sec:primary_metallicity}) using empirical calibrations established by field-age M dwarfs orbiting FGK primary stars, but these M-type calibrators have slightly higher surface gravities than the moderately young COCONUTS-3A. The metal-sensitive lines (e.g., Na \textsc{i} doublet at 2.2$\mu$m) adopted by these calibrations are also sensitive to the surface gravity, meaning that our derived [Fe/H]$_\star$ may be unreliable. Here we use stellar model atmospheres to explore the surface-gravity dependence of the empirical calibrations of low-mass stellar metallicity as developed by \citeauthor{2013AJ....145...52M} (\citeyear{2013AJ....145...52M}; M13), \citeauthor{2014AJ....147..160M} (\citeyear{2014AJ....147..160M}; M14), and \citeauthor{2014AJ....147...20N} (\citeyear{2014AJ....147...20N}; N14).

We base our analysis on the PHOENIX synthetic model spectra \citep[][]{2013A&A...553A...6H} over a parameter space of $[2300, 4500]$~K in $T_{\rm eff}$, $[3.5, 5.5]$~dex in $\log{(g)}$, and $[-1.0, +0.5]$~dex in [Fe/H] with intervals of 100~K, 0.5~dex, and 0.5~dex, respectively. We choose the range of $T_{\rm eff}$ to cover the applicable spectral types (K5--M9.5) of all empirical calibrations considered here. The $\log{(g)}$ is also a proxy of stellar age, with $\log{(g)} = 5.5$~dex corresponding to field ages ($>1$~Gyr), $5.0$~dex for moderately young ages of 100~Myr to 1~Gyr, and $4.5$~dex and $4.0$~dex for very young ages of $5-100$~Myr over the aforementioned parameter space \citep[based on the evolution models of][]{2015A&A...577A..42B}. We downgrade the spectral resolution of these models to match that of SpeX SXD and put the model wavelengths in vacuum following \cite{1996ApOpt..35.1566C}. We then apply each calibration method to compute the empirical [Fe/H]$_{\rm emp}$ for all model spectra that satisfy the method's applicable range in spectral type and metallicity. Specifically, we apply (1) the $J/H/K$-band calibration of M13 to PHOENIX models with $4500$~K $\geqslant T_{\rm eff} \geqslant 3000$~K (i.e., K5--M5) and $-1.04$~dex $\leqslant $[Fe/H]$\leqslant +0.56$~dex; (2) the M14 calibration to models with $3100$~K $\geqslant T_{\rm eff} \geqslant 2300$~K (i.e., M4.5--M9.5) and $-0.58$~dex $\leqslant $[Fe/H]$\leqslant +0.56$~dex; (3) the N14 calibration to models with $3700$~K $\geqslant T_{\rm eff} \geqslant 3000$~K (i.e., M1--M5) and $-1.00$~dex $\leqslant $[Fe/H]$\leqslant +0.35$~dex. The conversion between stellar effective temperatures and spectral types is based on the mean stellar properties compiled by E. Mamajek\footnote{\url{https://www.pas.rochester.edu/~emamajek/EEM_dwarf_UBVIJHK_colors_Teff.txt} \label{fn}} \citep[also see][]{2013ApJS..208....9P}.

\begin{figure*}[t]
\begin{center}
\includegraphics[height=3.8in]{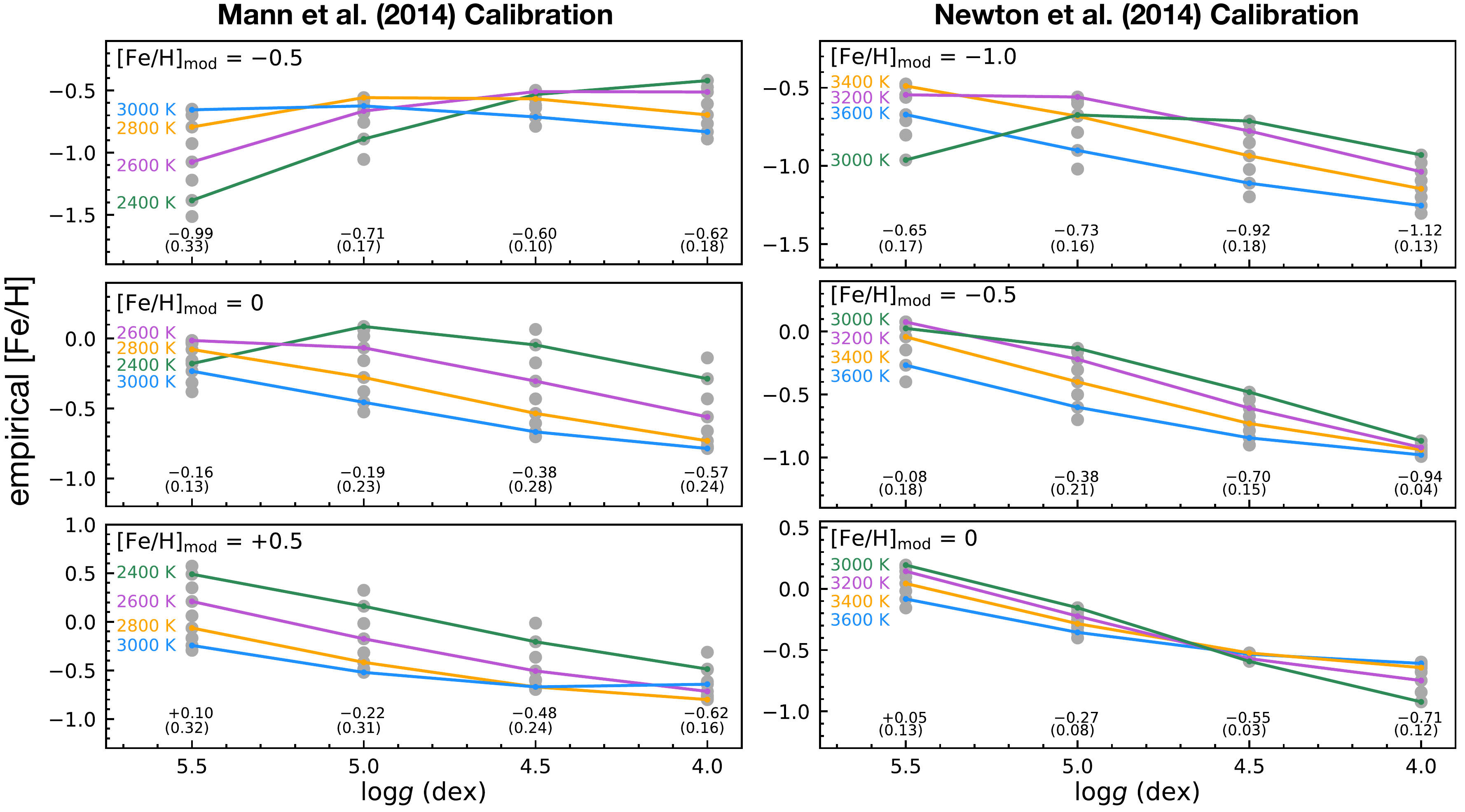} 
\caption{Computed empirical [Fe/H] (grey circles) of PHOENIX model spectra with different $T_{\rm eff}$, $\log{(g)}$, and [Fe/H] using the calibrations of \citeauthor{2014AJ....147..160M} (\citeyear{2014AJ....147..160M}; left panel) and \citeauthor{2014AJ....147...20N} (\citeyear{2014AJ....147...20N}; right panel). Typical scatter of the \cite{2014AJ....147..160M} and \cite{2014AJ....147...20N} empirical calibrations is 0.07~dex and 0.12~dex, respectively.  Surface gravity decreases from left to right in each panel, with $\log{(g)} = 5.5$~dex corresponding to field ages ($>1$~Gyr), $5.0$~dex for moderately young ages (100~Myr to 1~Gyr), and $4.5$~dex and $4.0$~dex for very young ages ($5-100$~Myr). Following the applicable range of each method, we apply the \cite{2014AJ....147..160M} calibration to models with $3100$~K $\geqslant T_{\rm eff} \geqslant 2300$~K (M4.5--M9.5) and [Fe/H]$= -0.5$~dex (top left), 0~dex (middle left), and $+0.5$~dex (bottom left), and the \cite{2014AJ....147...20N} calibration to models with $3700$~K $\geqslant T_{\rm eff} \geqslant 3000$~K (M1--M5) and [Fe/H]$= -1$~dex (top right), $-0.5$~dex (middle right), $0$~dex (bottom right). We highlight a few equal-$T_{\rm eff}$ tracks in each panel, with a sequence of 3000~K (blue), 2800~K (orange), 2600~K (purple), and 2400~K (green) for the \cite{2014AJ....147..160M} calibration, and 3600~K (blue), 3400~K (orange), 3200~K (purple), and 3000~K (green) for the \cite{2014AJ....147...20N} calibration. In each panel, we also label the mean and standard deviation (inside the parenthesis) of the computed empirical [Fe/H] at each $\log{(g)}$. }
\label{fig:feh_m14_n14}
\end{center}
\end{figure*}

\begin{figure*}[t]
\begin{center}
\includegraphics[height=3.6in]{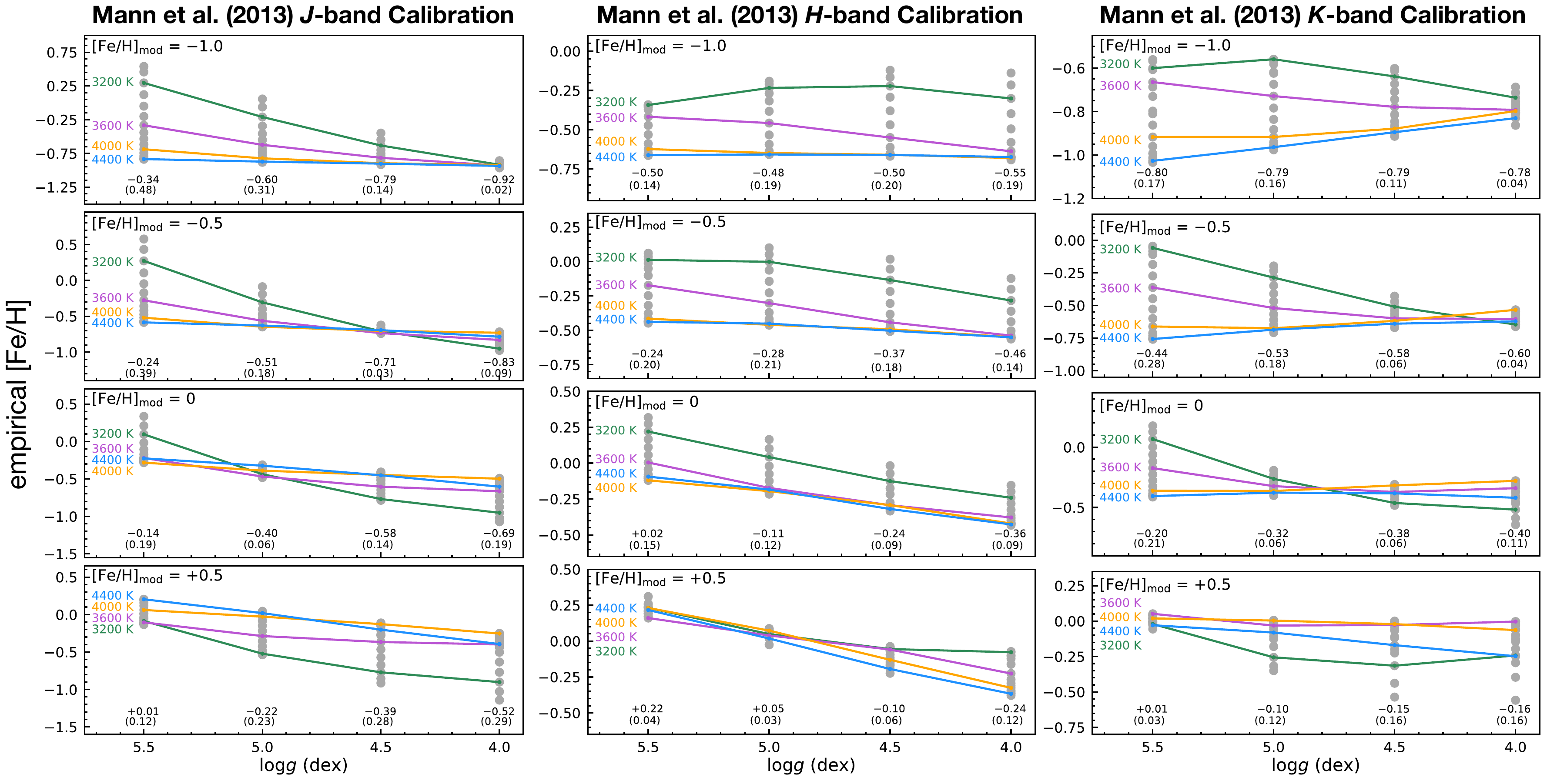} 
\caption{Computed empirical [Fe/H] (grey circles) of PHOENIX model spectra with different $T_{\rm eff}$, $\log{(g)}$, and [Fe/H] using the calibrations of \cite{2013AJ....145...52M} in $J$ band (left panel), $H$ band (middle panel), and $K$ band (right panel), with a similar format as Figure~\ref{fig:feh_m14_n14}. Typical scatter of these $J/H/K$-band empirical calibrations is 0.12~dex, 0.09~dex, and $0.08$~dex, respectively. Here we apply the \cite{2013AJ....145...52M} calibration to models with $4500$~K $\geqslant T_{\rm eff} \geqslant 3000$~K (K5--M5) and [Fe/H]$= -1$~dex (top row), $-0.5$~dex (second top row), $0$~dex (second bottom row), and $+0.5$~dex (bottom row). The colored lines in each panel are equal-$T_{\rm eff}$ tracks corresponding to $4400$~K (blue), $4000$~K (orange), $3600$~K (purple), and $3200$~K (green). }
\label{fig:feh_m13}
\end{center}
\end{figure*}

In Figures~\ref{fig:feh_m14_n14} and \ref{fig:feh_m13}, we present the computed [Fe/H]$_{\rm emp}$ by various calibrations as a function of surface gravity. While one might expect that the [Fe/H]$_{\rm emp}$ at $\log{g} = 5.5$~dex (i.e., field age of $>1$~Gyr) should all line up with the [Fe/H] of models, our analysis shows that [Fe/H]$_{\rm emp}$ exhibit systematic, $T_{\rm eff}$-dependent offsets from the model [Fe/H] \citep[e.g., also see][]{2017ApJ...851...26V}. These mismatches are likely due to the systematic uncertainties of stellar model atmospheres and the adopted opacity line lists \citep[e.g., Figure~11 of][]{2015ApJ...812..128C} and that the calibrations considered here are all tied to the SpeX SXD instrument rather than PHOENIX models. In other words, our analysis re-affirms the value of empirical metallicity calibrations as opposed to a purely model-based characterization.

Here we only focus on the trend (instead of the absolute values) of [Fe/H]$_{\rm emp}$ with respect to $\log{(g)}$. We first discuss the surface-gravity effect on the N14 and M14 calibrations (Figure~\ref{fig:feh_m14_n14}), which we use to derive [Fe/H]$_{\star}$ of COCONUTS-3A. For models with [Fe/H]$= 0$~dex and $+0.5$~dex, the M14 calibration (applied to 3100--2300~K) tends to derive a smaller [Fe/H]$_{\rm emp}$ toward lower surface gravities or younger ages. Specifically, compared to the case with $\log{(g)} = 5.5$~dex (i.e., field age), the M14-based [Fe/H]$_{\rm emp}$ is smaller by 0.2--0.3~dex for the lower $\log{(g)}$ of 5.0~dex and by 0.5--0.6~dex for $\log{(g)}$ of 4.0~dex or 4.5~dex. The N14 calibration (applied to 3700--3000~K) exhibits a similar trend with surface gravities for models with [Fe/H]$ = -1.0$~dex (except for ones with $T_{\rm eff} = 3000$~K), $-0.5$~dex, and $0$~dex. COCONUTS-3A has $T_{\rm eff} \approx 3000$~K and $\log{(g)} \approx 5.0$~dex (Section~\ref{subsec:phys}), so our derived [Fe/H]$_{\rm emp}$ might be under-estimated by $0.2-0.3$~dex according to these PHOENIX models. Qualitatively, this underestimate is expected since a lower surface gravity can cause weaker strengths of the Na~\textsc{i} doublet for early-to-mid M stars \citep[e.g.,][]{1998ApJ...493..909L, 2014AJ....147...20N} and the [Fe/H] values by both M14 and N14 calibrations monotonically decrease with decreasing equivalent widths of Na~\textsc{i} (as long as this equivalent width is below 7.5~\AA\ when applying the N14 calibration).

Interestingly, surface gravity imposes an opposite effect on the M14 calibration for models with [Fe/H]$= -0.5$~dex as compared to ones at higher metallicity (i.e., $0$~dex and $+0.5$~dex), where the M14-based [Fe/H]$_{\rm emp}$ tends to be larger with decreasing surface gravities. This means that the gravity dependence of the Na~\textsc{i} doublet strengths at [Fe/H]$= -0.5$~dex and $T_{\rm eff} < 3000$~K is different from the case with higher [Fe/H] and/or hotter $T_{\rm eff}$, according to the PHOENIX models.

The M13 calibrations (Figure~\ref{fig:feh_m13}) also tend to derive smaller [Fe/H]$_{\rm emp}$ toward lower surface gravities for models with cool $T_{\rm eff}$ ($\lesssim 3600$~K) and solar or super-solar [Fe/H]. For hot, sub-solar models with $T_{\rm eff}$ of $\gtrsim 4000$~K and [Fe/H] of $-0.5$~dex and $-1$~dex, the M13-based [Fe/H]$_{\rm emp}$ can be comparable or even larger toward lower $\log{(g)}$ compared to the case with $\log{(g)} = 5.5$~dex.

Our analysis aims to provide a qualitative perspective about the surface-gravity dependence of the empirical metallicity calibrations. We caution that the quantitative $\log{(g)}$ dependence revealed by our work should not be directly used as correction factors to the computed [Fe/H]$_{\rm emp}$, given that the systematics of the stellar model atmospheres likely change with surface gravities and effective temperatures as well. Also, compared to field-age low-mass stars, younger lower-gravity stars tend to be more active and possess stronger magnetic fields, which can alter the metal-sensitive line profiles and enhance the equivalent widths (and thus the resulting [Fe/H]) via Zeimann broadening and intensification \citep[e.g.,][]{1992ApJ...390..622B, 1994ApJ...431..844B, 2021A&ARv..29....1K, 2021ApJ...921...53L}. Estimation of the bulk metallicity for low-gravity low-mass stars would benefit from an empirical calibration customized for young stars, which is unfortunately not available to date. We note this calibration is difficult to construct with the existing sample of nearby young stars, which lacks sufficiently large variations in [Fe/H] and mostly has solar metallicity.

\section{Estimated X-ray Emission from Background Stars Near COCONUTS-3A}
\label{app:xray_bg}
Searching around COCONUTS-3A, we find a source 2RXS~J081318.4$-$152252 (2RXS~J0813$-$1522) in the Second ROSAT All-Sky Survey and its counterpart 1RXS~J081318.4$-$152250 (1RXS~J0813$-$1522) in the ROSAT All-Sky Survey Faint Source Catalog \citep{2000IAUC.7432....3V}, with the latter catalog reporting a position error of 15\arcsec. COCONUTS-3A is the closest match to both X-ray detections, with an angular separation of 13.8\arcsec and 12.2\arcsec, respectively. Seven background sources are also within 30\arcsec\ of 2RXS~J0813$-$1522 but six are outside the 15\arcsec\ position error. Here we quantify the potential X-ray fluxes of all these background stars to assess whether the ROSAT detection comes from COCONUTS-3A.

Figure~\ref{fig:xray_neighbors} and Table~\ref{tab:xray_neighbors} present coordinates and photometry of the background sources (with distances of $2-12$~kpc from Gaia~EDR3), including six FGK dwarfs and one K-type giant star based on their HR diagram positions. For the dwarf stars, we convert BP$-$RP into spectral types and then obtain typical B$-$V and bolometric luminosities, by using the mean stellar colors and properties compiled by E. Mamajek \citep[see footnote~\ref{fn} and][]{2013ApJS..208....9P}. We estimate the maximum possible X-ray fluxes of these objects by assuming their X-ray emission are in the saturated regime. We obtain the saturated \logLxbol\ for each star based on \cite{2012MNRAS.422.2024J} using the objects' B$-$V colors. We then compute the X-ray flux from each background dwarf star, with uncertainties in \logLbol, \logLxbol, and distances propagated accordingly. The remaining giant star (BG2, TYC~5996$-$1259$-$1) falls in the asymptotic giant branch, evolving across the X-ray dividing line, and thereby has faint intrinsic X-ray emission. According to \cite{1990ApJ...348..253M}, K giants have an X-ray-to-$V$-band flux ratio of \logFxv$<-6.2$, where \logFxv$=\log{(f_{X})}+0.4V+5.47$. We use this object's observed $V = 11.57 \pm 0.14$~mag to compute its maximum possible X-ray flux.

Summing up our estimated X-ray fluxes of all background stars, we obtain $(3.7 \pm1.7) \times 10^{-15}$~\ergscm. This value is only $2.2\% \pm 1.6\%$ of the observed X-ray flux $(1.7 \pm 1.0) \times 10^{-13}$~\ergscm\ of 2RXS~J0813$-$1522 (Section~\ref{subsubsec:xray}), suggesting the background sources in the neighborhood of COCONUTS-3A make a negligible contribution to our measured X-ray emission.

\end{CJK*}

\clearpage
\bibliographystyle{aasjournal}
\bibliography{ms}

\begin{thebibliography}{}
\expandafter\ifx\csname natexlab\endcsname\relax\def\natexlab#1{#1}\fi
\providecommand{\url}[1]{\href{#1}{#1}}
\providecommand{\dodoi}[1]{doi:~\href{http://doi.org/#1}{\nolinkurl{#1}}}
\providecommand{\doeprint}[1]{\href{http://ascl.net/#1}{\nolinkurl{http://ascl.net/#1}}}
\providecommand{\doarXiv}[1]{\href{https://arxiv.org/abs/#1}{\nolinkurl{https://arxiv.org/abs/#1}}}

\bibitem[{{Ackerman} \& {Marley}(2001)}]{2001ApJ...556..872A}
{Ackerman}, A.~S., \& {Marley}, M.~S. 2001, \apj, 556, 872,
  \dodoi{10.1086/321540}

\bibitem[{{Aldering} {et~al.}(2002){Aldering}, {Adam}, {Antilogus}, {Astier},
  {Bacon}, {Bongard}, {Bonnaud}, {Copin}, {Hardin}, \&
  {Henault}}]{2002SPIE.4836...61A}
{Aldering}, G., {Adam}, G., {Antilogus}, P., {et~al.} 2002, in Society of
  Photo-Optical Instrumentation Engineers (SPIE) Conference Series, Vol. 4836,
  Survey and Other Telescope Technologies and Discoveries, ed. J.~A. {Tyson} \&
  S.~{Wolff}, 61--72

\bibitem[{{Allard} {et~al.}(2016){Allard}, {Spiegelman}, \&
  {Kielkopf}}]{2016A&A...589A..21A}
{Allard}, N.~F., {Spiegelman}, F., \& {Kielkopf}, J.~F. 2016, \aap, 589, A21,
  \dodoi{10.1051/0004-6361/201628270}

\bibitem[{{Allard} {et~al.}(2019){Allard}, {Spiegelman}, {Leininger}, \&
  {Molliere}}]{2019A&A...628A.120A}
{Allard}, N.~F., {Spiegelman}, F., {Leininger}, T., \& {Molliere}, P. 2019,
  \aap, 628, A120, \dodoi{10.1051/0004-6361/201935593}

\bibitem[{{Allers} \& {Liu}(2013)}]{2013ApJ...772...79A}
{Allers}, K.~N., \& {Liu}, M.~C. 2013, \apj, 772, 79,
  \dodoi{10.1088/0004-637X/772/2/79}

\bibitem[{{Ansdell} {et~al.}(2016){Ansdell}, {Williams}, {van der Marel},
  {Carpenter}, {Guidi}, {Hogerheijde}, {Mathews}, {Manara}, {Miotello},
  {Natta}, {Oliveira}, {Tazzari}, {Testi}, {van Dishoeck}, \& {van
  Terwisga}}]{2016ApJ...828...46A}
{Ansdell}, M., {Williams}, J.~P., {van der Marel}, N., {et~al.} 2016, \apj,
  828, 46, \dodoi{10.3847/0004-637X/828/1/46}

\bibitem[{{Artigau} {et~al.}(2015){Artigau}, {Gagn{\'e}}, {Faherty}, {Malo},
  {Naud}, {Doyon}, {Lafreni{\`e}re}, \& {Beletsky}}]{2015ApJ...806..254A}
{Artigau}, {\'E}., {Gagn{\'e}}, J., {Faherty}, J., {et~al.} 2015, \apj, 806,
  254, \dodoi{10.1088/0004-637X/806/2/254}

\bibitem[{{Astropy Collaboration} {et~al.}(2013){Astropy Collaboration},
  {Robitaille}, {Tollerud}, {Greenfield}, {Droettboom}, {Bray}, {Aldcroft},
  {Davis}, {Ginsburg}, {Price-Whelan}, {Kerzendorf}, {Conley}, {Crighton},
  {Barbary}, {Muna}, {Ferguson}, {Grollier}, {Parikh}, {Nair}, {Unther},
  {Deil}, {Woillez}, {Conseil}, {Kramer}, {Turner}, {Singer}, {Fox}, {Weaver},
  {Zabalza}, {Edwards}, {Azalee Bostroem}, {Burke}, {Casey}, {Crawford},
  {Dencheva}, {Ely}, {Jenness}, {Labrie}, {Lim}, {Pierfederici}, {Pontzen},
  {Ptak}, {Refsdal}, {Servillat}, \& {Streicher}}]{2013A&A...558A..33A}
{Astropy Collaboration}, {Robitaille}, T.~P., {Tollerud}, E.~J., {et~al.} 2013,
  \aap, 558, A33, \dodoi{10.1051/0004-6361/201322068}

\bibitem[{{Astropy Collaboration} {et~al.}(2018){Astropy Collaboration},
  {Price-Whelan}, {Sip{\H o}cz}, {G{\"u}nther}, {Lim}, {Crawford}, {Conseil},
  {Shupe}, {Craig}, {Dencheva}, {Ginsburg}, {VanderPlas}, {Bradley},
  {P{\'e}rez-Su{\'a}rez}, {de Val-Borro}, {Aldcroft}, {Cruz}, {Robitaille},
  {Tollerud}, {Ardelean}, {Babej}, {Bach}, {Bachetti}, {Bakanov}, {Bamford},
  {Barentsen}, {Barmby}, {Baumbach}, {Berry}, {Biscani}, {Boquien}, {Bostroem},
  {Bouma}, {Brammer}, {Bray}, {Breytenbach}, {Buddelmeijer}, {Burke},
  {Calderone}, {Cano Rodr{\'{\i}}guez}, {Cara}, {Cardoso}, {Cheedella},
  {Copin}, {Corrales}, {Crichton}, {D'Avella}, {Deil}, {Depagne}, {Dietrich},
  {Donath}, {Droettboom}, {Earl}, {Erben}, {Fabbro}, {Ferreira}, {Finethy},
  {Fox}, {Garrison}, {Gibbons}, {Goldstein}, {Gommers}, {Greco}, {Greenfield},
  {Groener}, {Grollier}, {Hagen}, {Hirst}, {Homeier}, {Horton}, {Hosseinzadeh},
  {Hu}, {Hunkeler}, {Ivezi{\'c}}, {Jain}, {Jenness}, {Kanarek}, {Kendrew},
  {Kern}, {Kerzendorf}, {Khvalko}, {King}, {Kirkby}, {Kulkarni}, {Kumar},
  {Lee}, {Lenz}, {Littlefair}, {Ma}, {Macleod}, {Mastropietro}, {McCully},
  {Montagnac}, {Morris}, {Mueller}, {Mumford}, {Muna}, {Murphy}, {Nelson},
  {Nguyen}, {Ninan}, {N{\"o}the}, {Ogaz}, {Oh}, {Parejko}, {Parley}, {Pascual},
  {Patil}, {Patil}, {Plunkett}, {Prochaska}, {Rastogi}, {Reddy Janga},
  {Sabater}, {Sakurikar}, {Seifert}, {Sherbert}, {Sherwood-Taylor}, {Shih},
  {Sick}, {Silbiger}, {Singanamalla}, {Singer}, {Sladen}, {Sooley},
  {Sornarajah}, {Streicher}, {Teuben}, {Thomas}, {Tremblay}, {Turner},
  {Terr{\'o}n}, {van Kerkwijk}, {de la Vega}, {Watkins}, {Weaver}, {Whitmore},
  {Woillez}, {Zabalza}, \& {Astropy Contributors}}]{2018AJ....156..123A}
{Astropy Collaboration}, {Price-Whelan}, A.~M., {Sip{\H o}cz}, B.~M., {et~al.}
  2018, \aj, 156, 123, \dodoi{10.3847/1538-3881/aabc4f}

\bibitem[{{Bacon} {et~al.}(2001){Bacon}, {Copin}, {Monnet}, {Miller},
  {Allington-Smith}, {Bureau}, {Carollo}, {Davies}, {Emsellem}, \&
  {Kuntschner}}]{2001MNRAS.326...23B}
{Bacon}, R., {Copin}, Y., {Monnet}, G., {et~al.} 2001, \mnras, 326, 23,
  \dodoi{10.1046/j.1365-8711.2001.04612.x}

\bibitem[{{Bailer-Jones} {et~al.}(2021){Bailer-Jones}, {Rybizki}, {Fouesneau},
  {Demleitner}, \& {Andrae}}]{2021AJ....161..147B}
{Bailer-Jones}, C.~A.~L., {Rybizki}, J., {Fouesneau}, M., {Demleitner}, M., \&
  {Andrae}, R. 2021, \aj, 161, 147, \dodoi{10.3847/1538-3881/abd806}

\bibitem[{{Baraffe} {et~al.}(2015){Baraffe}, {Homeier}, {Allard}, \&
  {Chabrier}}]{2015A&A...577A..42B}
{Baraffe}, I., {Homeier}, D., {Allard}, F., \& {Chabrier}, G. 2015, \aap, 577,
  A42, \dodoi{10.1051/0004-6361/201425481}

\bibitem[{{Barrado y Navascu{\'e}s} \&
  {Mart{\'\i}n}(2003)}]{2003AJ....126.2997B}
{Barrado y Navascu{\'e}s}, D., \& {Mart{\'\i}n}, E.~L. 2003, \aj, 126, 2997,
  \dodoi{10.1086/379673}

\bibitem[{{Barrado y Navascu{\'e}s} {et~al.}(2004){Barrado y Navascu{\'e}s},
  {Stauffer}, \& {Jayawardhana}}]{2004ApJ...614..386B}
{Barrado y Navascu{\'e}s}, D., {Stauffer}, J.~R., \& {Jayawardhana}, R. 2004,
  \apj, 614, 386, \dodoi{10.1086/423485}

\bibitem[{{Basri} \& {Marcy}(1994)}]{1994ApJ...431..844B}
{Basri}, G., \& {Marcy}, G.~W. 1994, \apj, 431, 844, \dodoi{10.1086/174535}

\bibitem[{{Basri} {et~al.}(1992){Basri}, {Marcy}, \&
  {Valenti}}]{1992ApJ...390..622B}
{Basri}, G., {Marcy}, G.~W., \& {Valenti}, J.~A. 1992, \apj, 390, 622,
  \dodoi{10.1086/171312}

\bibitem[{{Bell} {et~al.}(2015){Bell}, {Mamajek}, \&
  {Naylor}}]{2015MNRAS.454..593B}
{Bell}, C. P.~M., {Mamajek}, E.~E., \& {Naylor}, T. 2015, \mnras, 454, 593,
  \dodoi{10.1093/mnras/stv1981}

\bibitem[{{Bellm} {et~al.}(2019){Bellm}, {Kulkarni}, {Graham}, {Dekany},
  {Smith}, {Riddle}, {Masci}, {Helou}, {Prince}, {Adams}, {Barbarino},
  {Barlow}, {Bauer}, {Beck}, {Belicki}, {Biswas}, {Blagorodnova}, {Bodewits},
  {Bolin}, {Brinnel}, {Brooke}, {Bue}, {Bulla}, {Burruss}, {Cenko}, {Chang},
  {Connolly}, {Coughlin}, {Cromer}, {Cunningham}, {De}, {Delacroix}, {Desai},
  {Duev}, {Eadie}, {Farnham}, {Feeney}, {Feindt}, {Flynn}, {Franckowiak},
  {Frederick}, {Fremling}, {Gal-Yam}, {Gezari}, {Giomi}, {Goldstein},
  {Golkhou}, {Goobar}, {Groom}, {Hacopians}, {Hale}, {Henning}, {Ho}, {Hover},
  {Howell}, {Hung}, {Huppenkothen}, {Imel}, {Ip}, {Ivezi{\'c}}, {Jackson},
  {Jones}, {Juric}, {Kasliwal}, {Kaspi}, {Kaye}, {Kelley}, {Kowalski},
  {Kramer}, {Kupfer}, {Landry}, {Laher}, {Lee}, {Lin}, {Lin}, {Lunnan},
  {Giomi}, {Mahabal}, {Mao}, {Miller}, {Monkewitz}, {Murphy}, {Ngeow},
  {Nordin}, {Nugent}, {Ofek}, {Patterson}, {Penprase}, {Porter}, {Rauch},
  {Rebbapragada}, {Reiley}, {Rigault}, {Rodriguez}, {van Roestel}, {Rusholme},
  {van Santen}, {Schulze}, {Shupe}, {Singer}, {Soumagnac}, {Stein}, {Surace},
  {Sollerman}, {Szkody}, {Taddia}, {Terek}, {Van Sistine}, {van Velzen},
  {Vestrand}, {Walters}, {Ward}, {Ye}, {Yu}, {Yan}, \&
  {Zolkower}}]{2019PASP..131a8002B}
{Bellm}, E.~C., {Kulkarni}, S.~R., {Graham}, M.~J., {et~al.} 2019, \pasp, 131,
  018002, \dodoi{10.1088/1538-3873/aaecbe}

\bibitem[{Best {et~al.}(2020b)Best, Dupuy, Liu, Siverd, \&
  Zhang}]{ultracoolsheet}
Best, W. M.~J., Dupuy, T.~J., Liu, M.~C., Siverd, R.~J., \& Zhang, Z. 2020b,
  The UltracoolSheet: Photometry, Astrometry, Spectroscopy, and Multiplicity
  for 3000+ Ultracool Dwarfs and Imaged Exoplanets,  Zenodo,
  \dodoi{10.5281/zenodo.4169085}

\bibitem[{{Best} {et~al.}(2021){Best}, {Liu}, {Magnier}, \&
  {Dupuy}}]{2021AJ....161...42B}
{Best}, W. M.~J., {Liu}, M.~C., {Magnier}, E.~A., \& {Dupuy}, T.~J. 2021, \aj,
  161, 42, \dodoi{10.3847/1538-3881/abc893}

\bibitem[{{Best} {et~al.}(2018){Best}, {Magnier}, {Liu}, {Aller}, {Zhang},
  {Burgett}, {Chambers}, {Draper}, {Flewelling}, {Kaiser}, {Kudritzki},
  {Metcalfe}, {Tonry}, {Wainscoat}, \& {Waters}}]{2018ApJS..234....1B}
{Best}, W.~M.~J., {Magnier}, E.~A., {Liu}, M.~C., {et~al.} 2018, \apjs, 234, 1,
  \dodoi{10.3847/1538-4365/aa9982}

\bibitem[{{Binks} \& {Jeffries}(2014)}]{2014MNRAS.438L..11B}
{Binks}, A.~S., \& {Jeffries}, R.~D. 2014, \mnras, 438, L11,
  \dodoi{10.1093/mnrasl/slt141}

\bibitem[{{Binks} \& {Jeffries}(2016)}]{2016MNRAS.455.3345B}
---. 2016, \mnras, 455, 3345, \dodoi{10.1093/mnras/stv2431}

\bibitem[{{Bohn} {et~al.}(2020){Bohn}, {Kenworthy}, {Ginski}, {Rieder},
  {Mamajek}, {Meshkat}, {Pecaut}, {Reggiani}, {de Boer}, {Keller}, {Snik}, \&
  {Southworth}}]{2020ApJ...898L..16B}
{Bohn}, A.~J., {Kenworthy}, M.~A., {Ginski}, C., {et~al.} 2020, \apjl, 898,
  L16, \dodoi{10.3847/2041-8213/aba27e}

\bibitem[{{Bohn} {et~al.}(2021){Bohn}, {Ginski}, {Kenworthy}, {Mamajek},
  {Pecaut}, {Mugrauer}, {Vogt}, {Adam}, {Meshkat}, {Reggiani}, \&
  {Snik}}]{2021A&A...648A..73B}
{Bohn}, A.~J., {Ginski}, C., {Kenworthy}, M.~A., {et~al.} 2021, \aap, 648, A73,
  \dodoi{10.1051/0004-6361/202140508}

\bibitem[{{Boller} {et~al.}(2016){Boller}, {Freyberg}, {Tr{\"u}mper}, {Haberl},
  {Voges}, \& {Nandra}}]{2016A&A...588A.103B}
{Boller}, T., {Freyberg}, M.~J., {Tr{\"u}mper}, J., {et~al.} 2016, \aap, 588,
  A103, \dodoi{10.1051/0004-6361/201525648}

\bibitem[{{Bonfils} {et~al.}(2005){Bonfils}, {Delfosse}, {Udry}, {Santos},
  {Forveille}, \& {S{\'e}gransan}}]{2005A&A...442..635B}
{Bonfils}, X., {Delfosse}, X., {Udry}, S., {et~al.} 2005, \aap, 442, 635,
  \dodoi{10.1051/0004-6361:20053046}

\bibitem[{{Booth} {et~al.}(2017){Booth}, {Poppenhaeger}, {Watson}, {Silva
  Aguirre}, \& {Wolk}}]{2017MNRAS.471.1012B}
{Booth}, R.~S., {Poppenhaeger}, K., {Watson}, C.~A., {Silva Aguirre}, V., \&
  {Wolk}, S.~J. 2017, \mnras, 471, 1012, \dodoi{10.1093/mnras/stx1630}

\bibitem[{{Boss}(2006)}]{2006ApJ...637L.137B}
{Boss}, A.~P. 2006, \apjl, 637, L137, \dodoi{10.1086/500613}

\bibitem[{{Brandt} {et~al.}(2021){Brandt}, {Dupuy}, {Li}, {Chen}, {Brandt},
  {Wong}, {Currie}, {Bowler}, {Liu}, {Best}, \&
  {Phillips}}]{2021arXiv210907525B}
{Brandt}, G.~M., {Dupuy}, T.~J., {Li}, Y., {et~al.} 2021, arXiv e-prints,
  arXiv:2109.07525.
\newblock \doarXiv{2109.07525}

\bibitem[{{Brandt} {et~al.}(2019){Brandt}, {Dupuy}, \&
  {Bowler}}]{2019AJ....158..140B}
{Brandt}, T.~D., {Dupuy}, T.~J., \& {Bowler}, B.~P. 2019, \aj, 158, 140,
  \dodoi{10.3847/1538-3881/ab04a8}

\bibitem[{{Brandt} \& {Huang}(2015)}]{2015ApJ...807...58B}
{Brandt}, T.~D., \& {Huang}, C.~X. 2015, \apj, 807, 58,
  \dodoi{10.1088/0004-637X/807/1/58}

\bibitem[{{Burgasser}(2007)}]{2007ApJ...659..655B}
{Burgasser}, A.~J. 2007, \apj, 659, 655, \dodoi{10.1086/511027}

\bibitem[{{Burgasser}(2014)}]{2014ASInC..11....7B}
{Burgasser}, A.~J. 2014, in Astronomical Society of India Conference Series,
  Vol.~11, Astronomical Society of India Conference Series, 7--16

\bibitem[{{Burgasser} {et~al.}(2010){Burgasser}, {Cruz}, {Cushing}, {Gelino},
  {Looper}, {Faherty}, {Kirkpatrick}, \& {Reid}}]{2010ApJ...710.1142B}
{Burgasser}, A.~J., {Cruz}, K.~L., {Cushing}, M., {et~al.} 2010, \apj, 710,
  1142, \dodoi{10.1088/0004-637X/710/2/1142}

\bibitem[{{Burgasser} {et~al.}(2006){Burgasser}, {Geballe}, {Leggett},
  {Kirkpatrick}, \& {Golimowski}}]{2006ApJ...637.1067B}
{Burgasser}, A.~J., {Geballe}, T.~R., {Leggett}, S.~K., {Kirkpatrick}, J.~D.,
  \& {Golimowski}, D.~A. 2006, \apj, 637, 1067, \dodoi{10.1086/498563}

\bibitem[{{Burgasser} {et~al.}(2005){Burgasser}, {Reid}, {Leggett},
  {Kirkpatrick}, {Liebert}, \& {Burrows}}]{2005ApJ...634L.177B}
{Burgasser}, A.~J., {Reid}, I.~N., {Leggett}, S.~K., {et~al.} 2005, \apjl, 634,
  L177, \dodoi{10.1086/498866}

\bibitem[{{Chabrier}(2003)}]{2003PASP..115..763C}
{Chabrier}, G. 2003, \pasp, 115, 763, \dodoi{10.1086/376392}

\bibitem[{{Chabrier} \& {Baraffe}(1997)}]{1997A&A...327.1039C}
{Chabrier}, G., \& {Baraffe}, I. 1997, \aap, 327, 1039.
\newblock \doarXiv{astro-ph/9704118}

\bibitem[{{Chambers} {et~al.}(2016){Chambers}, {Magnier}, {Metcalfe},
  {Flewelling}, {Huber}, {Waters}, {Denneau}, {Draper}, {Farrow}, {Finkbeiner},
  {Holmberg}, {Koppenhoefer}, {Price}, {Rest}, {Saglia}, {Schlafly}, {Smartt},
  {Sweeney}, {Wainscoat}, {Burgett}, {Chastel}, {Grav}, {Heasley}, {Hodapp},
  {Jedicke}, {Kaiser}, {Kudritzki}, {Luppino}, {Lupton}, {Monet}, {Morgan},
  {Onaka}, {Shiao}, {Stubbs}, {Tonry}, {White}, {Ba{\~n}ados}, {Bell},
  {Bender}, {Bernard}, {Boegner}, {Boffi}, {Botticella}, {Calamida},
  {Casertano}, {Chen}, {Chen}, {Cole}, {Deacon}, {Frenk}, {Fitzsimmons},
  {Gezari}, {Gibbs}, {Goessl}, {Goggia}, {Gourgue}, {Goldman}, {Grant},
  {Grebel}, {Hambly}, {Hasinger}, {Heavens}, {Heckman}, {Henderson}, {Henning},
  {Holman}, {Hopp}, {Ip}, {Isani}, {Jackson}, {Keyes}, {Koekemoer}, {Kotak},
  {Le}, {Liska}, {Long}, {Lucey}, {Liu}, {Martin}, {Masci}, {McLean}, {Mindel},
  {Misra}, {Morganson}, {Murphy}, {Obaika}, {Narayan}, {Nieto-Santisteban},
  {Norberg}, {Peacock}, {Pier}, {Postman}, {Primak}, {Rae}, {Rai}, {Riess},
  {Riffeser}, {Rix}, {R{\"o}ser}, {Russel}, {Rutz}, {Schilbach}, {Schultz},
  {Scolnic}, {Strolger}, {Szalay}, {Seitz}, {Small}, {Smith}, {Soderblom},
  {Taylor}, {Thomson}, {Taylor}, {Thakar}, {Thiel}, {Thilker}, {Unger},
  {Urata}, {Valenti}, {Wagner}, {Walder}, {Walter}, {Watters}, {Werner},
  {Wood-Vasey}, \& {Wyse}}]{2016arXiv161205560C}
{Chambers}, K.~C., {Magnier}, E.~A., {Metcalfe}, N., {et~al.} 2016, arXiv
  e-prints.
\newblock \doarXiv{1612.05560}

\bibitem[{{Chiu} {et~al.}(2006){Chiu}, {Fan}, {Leggett}, {Golimowski}, {Zheng},
  {Geballe}, {Schneider}, \& {Brinkmann}}]{2006AJ....131.2722C}
{Chiu}, K., {Fan}, X., {Leggett}, S.~K., {et~al.} 2006, \aj, 131, 2722,
  \dodoi{10.1086/501431}

\bibitem[{{Choi} {et~al.}(2016){Choi}, {Dotter}, {Conroy}, {Cantiello},
  {Paxton}, \& {Johnson}}]{2016ApJ...823..102C}
{Choi}, J., {Dotter}, A., {Conroy}, C., {et~al.} 2016, \apj, 823, 102,
  \dodoi{10.3847/0004-637X/823/2/102}

\bibitem[{{Ciddor}(1996)}]{1996ApOpt..35.1566C}
{Ciddor}, P.~E. 1996, \ao, 35, 1566, \dodoi{10.1364/AO.35.001566}

\bibitem[{{Cifuentes} {et~al.}(2020){Cifuentes}, {Caballero},
  {Cort{\'e}s-Contreras}, {Montes}, {Abell{\'a}n}, {Dorda}, {Holgado},
  {Zapatero Osorio}, {Morales}, {Amado}, {Passegger}, {Quirrenbach}, {Reiners},
  {Ribas}, {Sanz-Forcada}, {Schweitzer}, {Seifert}, \&
  {Solano}}]{2020A&A...642A.115C}
{Cifuentes}, C., {Caballero}, J.~A., {Cort{\'e}s-Contreras}, M., {et~al.} 2020,
  \aap, 642, A115, \dodoi{10.1051/0004-6361/202038295}

\bibitem[{{Cohen} {et~al.}(2003){Cohen}, {Wheaton}, \&
  {Megeath}}]{2003AJ....126.1090C}
{Cohen}, M., {Wheaton}, W.~A., \& {Megeath}, S.~T. 2003, \aj, 126, 1090,
  \dodoi{10.1086/376474}

\bibitem[{{Covey} {et~al.}(2007){Covey}, {Ivezi{\'c}}, {Schlegel},
  {Finkbeiner}, {Padmanabhan}, {Lupton}, {Ag{\"u}eros}, {Bochanski}, {Hawley},
  {West}, {Seth}, {Kimball}, {Gogarten}, {Claire}, {Haggard}, {Kaib},
  {Schneider}, \& {Sesar}}]{2007AJ....134.2398C}
{Covey}, K.~R., {Ivezi{\'c}}, {\v{Z}}., {Schlegel}, D., {et~al.} 2007, \aj,
  134, 2398, \dodoi{10.1086/522052}

\bibitem[{Cruz \& Gagn{\'e}(2014)}]{cruz_gagne_2014}
Cruz, K., \& Gagn{\'e}, J. 2014, The Ultracool RIZzo Spectral Library,  Zenodo,
  \dodoi{10.5281/zenodo.11313}

\bibitem[{{Cruz} {et~al.}(2004){Cruz}, {Burgasser}, {Reid}, \&
  {Liebert}}]{2004ApJ...604L..61C}
{Cruz}, K.~L., {Burgasser}, A.~J., {Reid}, I.~N., \& {Liebert}, J. 2004, \apjl,
  604, L61, \dodoi{10.1086/383415}

\bibitem[{{Cruz} {et~al.}(2009){Cruz}, {Kirkpatrick}, \&
  {Burgasser}}]{2009AJ....137.3345C}
{Cruz}, K.~L., {Kirkpatrick}, J.~D., \& {Burgasser}, A.~J. 2009, \aj, 137,
  3345, \dodoi{10.1088/0004-6256/137/2/3345}

\bibitem[{{Cruz} {et~al.}(2018){Cruz}, {N{\'u}{\~n}ez}, {Burgasser},
  {Abrahams}, {Rice}, {Reid}, \& {Looper}}]{2018AJ....155...34C}
{Cruz}, K.~L., {N{\'u}{\~n}ez}, A., {Burgasser}, A.~J., {et~al.} 2018, \aj,
  155, 34, \dodoi{10.3847/1538-3881/aa9d8a}

\bibitem[{{Cruz} {et~al.}(2003){Cruz}, {Reid}, {Liebert}, {Kirkpatrick}, \&
  {Lowrance}}]{2003AJ....126.2421C}
{Cruz}, K.~L., {Reid}, I.~N., {Liebert}, J., {Kirkpatrick}, J.~D., \&
  {Lowrance}, P.~J. 2003, \aj, 126, 2421, \dodoi{10.1086/378607}

\bibitem[{{Currie} {et~al.}(2020){Currie}, {Brandt}, {Kuzuhara}, {Chilcote},
  {Guyon}, {Marois}, {Groff}, {Lozi}, {Vievard}, {Sahoo}, {Deo}, {Jovanovic},
  {Martinache}, {Wagner}, {Dupuy}, {Wahl}, {Letawsky}, {Li}, {Zeng}, {Brandt},
  {Michalik}, {Grady}, {Janson}, {Knapp}, {Kwon}, {Lawson}, {McElwain},
  {Uyama}, {Wisniewski}, \& {Tamura}}]{2020ApJ...904L..25C}
{Currie}, T., {Brandt}, T.~D., {Kuzuhara}, M., {et~al.} 2020, \apjl, 904, L25,
  \dodoi{10.3847/2041-8213/abc631}

\bibitem[{{Cushing} {et~al.}(2005){Cushing}, {Rayner}, \&
  {Vacca}}]{2005ApJ...623.1115C}
{Cushing}, M.~C., {Rayner}, J.~T., \& {Vacca}, W.~D. 2005, \apj, 623, 1115,
  \dodoi{10.1086/428040}

\bibitem[{{Cushing} {et~al.}(2004){Cushing}, {Vacca}, \&
  {Rayner}}]{2004PASP..116..362C}
{Cushing}, M.~C., {Vacca}, W.~D., \& {Rayner}, J.~T. 2004, \pasp, 116, 362,
  \dodoi{10.1086/382907}

\bibitem[{{Cushing} {et~al.}(2008){Cushing}, {Marley}, {Saumon}, {Kelly},
  {Vacca}, {Rayner}, {Freedman}, {Lodders}, \& {Roellig}}]{2008ApJ...678.1372C}
{Cushing}, M.~C., {Marley}, M.~S., {Saumon}, D., {et~al.} 2008, \apj, 678,
  1372, \dodoi{10.1086/526489}

\bibitem[{{Cutri} {et~al.}(2003){Cutri}, {Skrutskie}, {van Dyk}, {Beichman},
  {Carpenter}, {Chester}, {Cambresy}, {Evans}, {Fowler}, {Gizis}, {Howard},
  {Huchra}, {Jarrett}, {Kopan}, {Kirkpatrick}, {Light}, {Marsh}, {McCallon},
  {Schneider}, {Stiening}, {Sykes}, {Weinberg}, {Wheaton}, {Wheelock}, \&
  {Zacarias}}]{2003yCat.2246....0C}
{Cutri}, R.~M., {Skrutskie}, M.~F., {van Dyk}, S., {et~al.} 2003, VizieR Online
  Data Catalog, II/246

\bibitem[{{Cutri}(2014)}]{2014yCat.2328....0C}
{Cutri}, R.~M.~e. 2014, VizieR Online Data Catalog, 2328

\bibitem[{{Czekala} {et~al.}(2015){Czekala}, {Andrews}, {Mandel}, {Hogg}, \&
  {Green}}]{2015ApJ...812..128C}
{Czekala}, I., {Andrews}, S.~M., {Mandel}, K.~S., {Hogg}, D.~W., \& {Green},
  G.~M. 2015, \apj, 812, 128, \dodoi{10.1088/0004-637X/812/2/128}

\bibitem[{{Dahm}(2015)}]{2015ApJ...813..108D}
{Dahm}, S.~E. 2015, \apj, 813, 108, \dodoi{10.1088/0004-637X/813/2/108}

\bibitem[{{Dahn} {et~al.}(2002){Dahn}, {Harris}, {Vrba}, {Guetter}, {Canzian},
  {Henden}, {Levine}, {Luginbuhl}, {Monet}, {Monet}, {Pier}, {Stone}, {Walker},
  {Burgasser}, {Gizis}, {Kirkpatrick}, {Liebert}, \&
  {Reid}}]{2002AJ....124.1170D}
{Dahn}, C.~C., {Harris}, H.~C., {Vrba}, F.~J., {et~al.} 2002, \aj, 124, 1170,
  \dodoi{10.1086/341646}

\bibitem[{{Dawson} {et~al.}(2013){Dawson}, {Schlegel}, {Ahn}, {Anderson},
  {Aubourg}, {Bailey}, {Barkhouser}, {Bautista}, {Beifiori}, {Berlind},
  {Bhardwaj}, {Bizyaev}, {Blake}, {Blanton}, {Blomqvist}, {Bolton}, {Borde},
  {Bovy}, {Brandt}, {Brewington}, {Brinkmann}, {Brown}, {Brownstein}, {Bundy},
  {Busca}, {Carithers}, {Carnero}, {Carr}, {Chen}, {Comparat}, {Connolly},
  {Cope}, {Croft}, {Cuesta}, {da Costa}, {Davenport}, {Delubac}, {de Putter},
  {Dhital}, {Ealet}, {Ebelke}, {Eisenstein}, {Escoffier}, {Fan}, {Filiz Ak},
  {Finley}, {Font-Ribera}, {G{\'e}nova-Santos}, {Gunn}, {Guo}, {Haggard},
  {Hall}, {Hamilton}, {Harris}, {Harris}, {Ho}, {Hogg}, {Holder}, {Honscheid},
  {Huehnerhoff}, {Jordan}, {Jordan}, {Kauffmann}, {Kazin}, {Kirkby}, {Klaene},
  {Kneib}, {Le Goff}, {Lee}, {Long}, {Loomis}, {Lundgren}, {Lupton}, {Maia},
  {Makler}, {Malanushenko}, {Malanushenko}, {Mandelbaum}, {Manera}, {Maraston},
  {Margala}, {Masters}, {McBride}, {McDonald}, {McGreer}, {McMahon}, {Mena},
  {Miralda-Escud{\'e}}, {Montero-Dorta}, {Montesano}, {Muna}, {Myers},
  {Naugle}, {Nichol}, {Noterdaeme}, {Nuza}, {Olmstead}, {Oravetz}, {Oravetz},
  {Owen}, {Padmanabhan}, {Palanque-Delabrouille}, {Pan}, {Parejko},
  {P{\^a}ris}, {Percival}, {P{\'e}rez-Fournon}, {P{\'e}rez-R{\`a}fols},
  {Petitjean}, {Pfaffenberger}, {Pforr}, {Pieri}, {Prada}, {Price-Whelan},
  {Raddick}, {Rebolo}, {Rich}, {Richards}, {Rockosi}, {Roe}, {Ross}, {Ross},
  {Rossi}, {Rubi{\~n}o-Martin}, {Samushia}, {S{\'a}nchez}, {Sayres}, {Schmidt},
  {Schneider}, {Sc{\'o}ccola}, {Seo}, {Shelden}, {Sheldon}, {Shen}, {Shu},
  {Slosar}, {Smee}, {Snedden}, {Stauffer}, {Steele}, {Strauss}, {Streblyanska},
  {Suzuki}, {Swanson}, {Tal}, {Tanaka}, {Thomas}, {Tinker}, {Tojeiro},
  {Tremonti}, {Vargas Maga{\~n}a}, {Verde}, {Viel}, {Wake}, {Watson}, {Weaver},
  {Weinberg}, {Weiner}, {West}, {White}, {Wood-Vasey}, {Yeche}, {Zehavi},
  {Zhao}, \& {Zheng}}]{2013AJ....145...10D}
{Dawson}, K.~S., {Schlegel}, D.~J., {Ahn}, C.~P., {et~al.} 2013, \aj, 145, 10,
  \dodoi{10.1088/0004-6256/145/1/10}

\bibitem[{{Deacon} {et~al.}(2014){Deacon}, {Liu}, {Magnier}, {Aller}, {Best},
  {Dupuy}, {Bowler}, {Mann}, {Redstone}, {Burgett}, {Chambers}, {Draper},
  {Flewelling}, {Hodapp}, {Kaiser}, {Kudritzki}, {Morgan}, {Metcalfe}, {Price},
  {Tonry}, \& {Wainscoat}}]{2014ApJ...792..119D}
{Deacon}, N.~R., {Liu}, M.~C., {Magnier}, E.~A., {et~al.} 2014, \apj, 792, 119,
  \dodoi{10.1088/0004-637X/792/2/119}

\bibitem[{{Delorme} {et~al.}(2017){Delorme}, {Schmidt}, {Bonnefoy}, {Desidera},
  {Ginski}, {Charnay}, {Lazzoni}, {Christiaens}, {Messina}, {D'Orazi}, {Milli},
  {Schlieder}, {Gratton}, {Rodet}, {Lagrange}, {Absil}, {Vigan}, {Galicher},
  {Hagelberg}, {Bonavita}, {Lavie}, {Zurlo}, {Olofsson}, {Boccaletti},
  {Cantalloube}, {Mouillet}, {Chauvin}, {Hambsch}, {Langlois}, {Udry},
  {Henning}, {Beuzit}, {Mordasini}, {Lucas}, {Marocco}, {Biller}, {Carson},
  {Cheetham}, {Covino}, {De Caprio}, {Delboulbe}, {Feldt}, {Girard}, {Hubin},
  {Maire}, {Pavlov}, {Petit}, {Rouan}, {Roelfsema}, \&
  {Wildi}}]{2017A&A...608A..79D}
{Delorme}, P., {Schmidt}, T., {Bonnefoy}, M., {et~al.} 2017, \aap, 608, A79,
  \dodoi{10.1051/0004-6361/201731145}

\bibitem[{{Dhital} {et~al.}(2011){Dhital}, {Burgasser}, {Looper}, \&
  {Stassun}}]{2011AJ....141....7D}
{Dhital}, S., {Burgasser}, A.~J., {Looper}, D.~L., \& {Stassun}, K.~G. 2011,
  \aj, 141, 7, \dodoi{10.1088/0004-6256/141/1/7}

\bibitem[{{Dhital} {et~al.}(2012){Dhital}, {West}, {Stassun}, {Bochanski},
  {Massey}, \& {Bastien}}]{2012AJ....143...67D}
{Dhital}, S., {West}, A.~A., {Stassun}, K.~G., {et~al.} 2012, \aj, 143, 67,
  \dodoi{10.1088/0004-6256/143/3/67}

\bibitem[{{Dotter}(2016)}]{2016ApJS..222....8D}
{Dotter}, A. 2016, \apjs, 222, 8, \dodoi{10.3847/0067-0049/222/1/8}

\bibitem[{{Douglas} {et~al.}(2019){Douglas}, {Curtis}, {Ag{\"u}eros},
  {Cargile}, {Brewer}, {Meibom}, \& {Jansen}}]{2019ApJ...879..100D}
{Douglas}, S.~T., {Curtis}, J.~L., {Ag{\"u}eros}, M.~A., {et~al.} 2019, \apj,
  879, 100, \dodoi{10.3847/1538-4357/ab2468}

\bibitem[{{Douglas} {et~al.}(2014){Douglas}, {Ag{\"u}eros}, {Covey}, {Bowsher},
  {Bochanski}, {Cargile}, {Kraus}, {Law}, {Lemonias}, {Arce}, {Fierroz}, \&
  {Kundert}}]{2014ApJ...795..161D}
{Douglas}, S.~T., {Ag{\"u}eros}, M.~A., {Covey}, K.~R., {et~al.} 2014, \apj,
  795, 161, \dodoi{10.1088/0004-637X/795/2/161}

\bibitem[{{Dupuy} \& {Kraus}(2013)}]{2013Sci...341.1492D}
{Dupuy}, T.~J., \& {Kraus}, A.~L. 2013, Science, 341, 1492,
  \dodoi{10.1126/science.1241917}

\bibitem[{{Dupuy} \& {Liu}(2012)}]{2012ApJS..201...19D}
{Dupuy}, T.~J., \& {Liu}, M.~C. 2012, \apjs, 201, 19,
  \dodoi{10.1088/0067-0049/201/2/19}

\bibitem[{{Dupuy} {et~al.}(2009){Dupuy}, {Liu}, \&
  {Ireland}}]{2009ApJ...692..729D}
{Dupuy}, T.~J., {Liu}, M.~C., \& {Ireland}, M.~J. 2009, \apj, 692, 729,
  \dodoi{10.1088/0004-637X/692/1/729}

\bibitem[{{Dupuy} {et~al.}(2014){Dupuy}, {Liu}, \&
  {Ireland}}]{2014ApJ...790..133D}
---. 2014, \apj, 790, 133, \dodoi{10.1088/0004-637X/790/2/133}

\bibitem[{{Dupuy} {et~al.}(2015){Dupuy}, {Liu}, {Leggett}, {Ireland}, {Chiu},
  \& {Golimowski}}]{2015ApJ...805...56D}
{Dupuy}, T.~J., {Liu}, M.~C., {Leggett}, S.~K., {et~al.} 2015, \apj, 805, 56,
  \dodoi{10.1088/0004-637X/805/1/56}

\bibitem[{{Elias} {et~al.}(2006){Elias}, {Joyce}, {Liang}, {Muller}, {Hileman},
  \& {George}}]{2006SPIE.6269E..4CE}
{Elias}, J.~H., {Joyce}, R.~R., {Liang}, M., {et~al.} 2006, in Society of
  Photo-Optical Instrumentation Engineers (SPIE) Conference Series, Vol. 6269,
  Society of Photo-Optical Instrumentation Engineers (SPIE) Conference Series,
  ed. I.~S. {McLean} \& M.~{Iye}, 62694C

\bibitem[{{Faherty} {et~al.}(2016){Faherty}, {Riedel}, {Cruz}, {Gagne},
  {Filippazzo}, {Lambrides}, {Fica}, {Weinberger}, {Thorstensen}, {Tinney},
  {Baldassare}, {Lemonier}, \& {Rice}}]{2016ApJS..225...10F}
{Faherty}, J.~K., {Riedel}, A.~R., {Cruz}, K.~L., {et~al.} 2016, \apjs, 225,
  10, \dodoi{10.3847/0067-0049/225/1/10}

\bibitem[{{Filippazzo} {et~al.}(2015){Filippazzo}, {Rice}, {Faherty}, {Cruz},
  {Van Gordon}, \& {Looper}}]{2015ApJ...810..158F}
{Filippazzo}, J.~C., {Rice}, E.~L., {Faherty}, J., {et~al.} 2015, \apj, 810,
  158, \dodoi{10.1088/0004-637X/810/2/158}

\bibitem[{{Findeisen} {et~al.}(2011){Findeisen}, {Hillenbrand}, \&
  {Soderblom}}]{2011AJ....142...23F}
{Findeisen}, K., {Hillenbrand}, L., \& {Soderblom}, D. 2011, \aj, 142, 23,
  \dodoi{10.1088/0004-6256/142/1/23}

\bibitem[{{Fleming} {et~al.}(1995){Fleming}, {Molendi}, {Maccacaro}, \&
  {Wolter}}]{1995ApJS...99..701F}
{Fleming}, T.~A., {Molendi}, S., {Maccacaro}, T., \& {Wolter}, A. 1995, \apjs,
  99, 701, \dodoi{10.1086/192203}

\bibitem[{{Foreman-Mackey} {et~al.}(2013){Foreman-Mackey}, {Hogg}, {Lang}, \&
  {Goodman}}]{2013PASP..125..306F}
{Foreman-Mackey}, D., {Hogg}, D.~W., {Lang}, D., \& {Goodman}, J. 2013, \pasp,
  125, 306, \dodoi{10.1086/670067}

\bibitem[{{Freedman} {et~al.}(2008){Freedman}, {Marley}, \&
  {Lodders}}]{2008ApJS..174..504F}
{Freedman}, R.~S., {Marley}, M.~S., \& {Lodders}, K. 2008, \apjs, 174, 504,
  \dodoi{10.1086/521793}

\bibitem[{{Gagn{\'e}} {et~al.}(2020){Gagn{\'e}}, {David}, {Mamajek}, {Mann},
  {Faherty}, \& {B{\'e}dard}}]{2020ApJ...903...96G}
{Gagn{\'e}}, J., {David}, T.~J., {Mamajek}, E.~E., {et~al.} 2020, \apj, 903,
  96, \dodoi{10.3847/1538-4357/abb77e}

\bibitem[{{Gagn{\'e}} \& {Faherty}(2018)}]{2018ApJ...862..138G}
{Gagn{\'e}}, J., \& {Faherty}, J.~K. 2018, \apj, 862, 138,
  \dodoi{10.3847/1538-4357/aaca2e}

\bibitem[{{Gagn{\'e}} {et~al.}(2014){Gagn{\'e}}, {Lafreni{\`e}re}, {Doyon},
  {Malo}, \& {Artigau}}]{2014ApJ...783..121G}
{Gagn{\'e}}, J., {Lafreni{\`e}re}, D., {Doyon}, R., {Malo}, L., \& {Artigau},
  {\'E}. 2014, \apj, 783, 121, \dodoi{10.1088/0004-637X/783/2/121}

\bibitem[{{Gagn{\'e}} {et~al.}(2018){Gagn{\'e}}, {Mamajek}, {Malo}, {Riedel},
  {Rodriguez}, {Lafreni{\`e}re}, {Faherty}, {Roy-Loubier}, {Pueyo}, {Robin}, \&
  {Doyon}}]{2018ApJ...856...23G}
{Gagn{\'e}}, J., {Mamajek}, E.~E., {Malo}, L., {et~al.} 2018, \apj, 856, 23,
  \dodoi{10.3847/1538-4357/aaae09}

\bibitem[{{Gaia Collaboration} {et~al.}(2020){Gaia Collaboration}, {Brown},
  {Vallenari}, {Prusti}, {de Bruijne}, {Babusiaux}, \&
  {Biermann}}]{2020arXiv201201533G}
{Gaia Collaboration}, {Brown}, A.~G.~A., {Vallenari}, A., {et~al.} 2020, arXiv
  e-prints, arXiv:2012.01533.
\newblock \doarXiv{2012.01533}

\bibitem[{{Gaia Collaboration} {et~al.}(2016){Gaia Collaboration}, {Prusti},
  {de Bruijne}, {Brown}, {Vallenari}, {Babusiaux}, {Bailer-Jones}, {Bastian},
  {Biermann}, {Evans}, {Eyer}, {Jansen}, {Jordi}, {Klioner}, {Lammers},
  {Lindegren}, {Luri}, {Mignard}, {Milligan}, {Panem}, {Poinsignon},
  {Pourbaix}, {Randich}, {Sarri}, {Sartoretti}, {Siddiqui}, {Soubiran},
  {Valette}, {van Leeuwen}, {Walton}, {Aerts}, {Arenou}, {Cropper}, {Drimmel},
  {H{\o}g}, {Katz}, {Lattanzi}, {O'Mullane}, {Grebel}, {Holland}, {Huc},
  {Passot}, {Bramante}, {Cacciari}, {Casta{\~n}eda}, {Chaoul}, {Cheek}, {De
  Angeli}, {Fabricius}, {Guerra}, {Hern{\'a}ndez}, {Jean-Antoine-Piccolo},
  {Masana}, {Messineo}, {Mowlavi}, {Nienartowicz}, {Ord{\'o}{\~n}ez-Blanco},
  {Panuzzo}, {Portell}, {Richards}, {Riello}, {Seabroke}, {Tanga},
  {Th{\'e}venin}, {Torra}, {Els}, {Gracia-Abril}, {Comoretto},
  {Garcia-Reinaldos}, {Lock}, {Mercier}, {Altmann}, {Andrae}, {Astraatmadja},
  {Bellas-Velidis}, {Benson}, {Berthier}, {Blomme}, {Busso}, {Carry},
  {Cellino}, {Clementini}, {Cowell}, {Creevey}, {Cuypers}, {Davidson}, {De
  Ridder}, {de Torres}, {Delchambre}, {Dell'Oro}, {Ducourant}, {Fr{\'e}mat},
  {Garc{\'\i}a-Torres}, {Gosset}, {Halbwachs}, {Hambly}, {Harrison}, {Hauser},
  {Hestroffer}, {Hodgkin}, {Huckle}, {Hutton}, {Jasniewicz}, {Jordan},
  {Kontizas}, {Korn}, {Lanzafame}, {Manteiga}, {Moitinho}, {Muinonen},
  {Osinde}, {Pancino}, {Pauwels}, {Petit}, {Recio-Blanco}, {Robin}, {Sarro},
  {Siopis}, {Smith}, {Smith}, {Sozzetti}, {Thuillot}, {van Reeven}, {Viala},
  {Abbas}, {Abreu Aramburu}, {Accart}, {Aguado}, {Allan}, {Allasia},
  {Altavilla}, {{\'A}lvarez}, {Alves}, {Anderson}, {Andrei}, {Anglada Varela},
  {Antiche}, {Antoja}, {Ant{\'o}n}, {Arcay}, {Atzei}, {Ayache}, {Bach},
  {Baker}, {Balaguer-N{\'u}{\~n}ez}, {Barache}, {Barata}, {Barbier}, {Barblan},
  {Baroni}, {Barrado y Navascu{\'e}s}, {Barros}, {Barstow}, {Becciani},
  {Bellazzini}, {Bellei}, {Bello Garc{\'\i}a}, {Belokurov}, {Bendjoya},
  {Berihuete}, {Bianchi}, {Bienaym{\'e}}, {Billebaud}, {Blagorodnova},
  {Blanco-Cuaresma}, {Boch}, {Bombrun}, {Borrachero}, {Bouquillon}, {Bourda},
  {Bouy}, {Bragaglia}, {Breddels}, {Brouillet}, {Br{\"u}semeister},
  {Bucciarelli}, {Budnik}, {Burgess}, {Burgon}, {Burlacu}, {Busonero}, {Buzzi},
  {Caffau}, {Cambras}, {Campbell}, {Cancelliere}, {Cantat-Gaudin}, {Carlucci},
  {Carrasco}, {Castellani}, {Charlot}, {Charnas}, {Charvet}, {Chassat},
  {Chiavassa}, {Clotet}, {Cocozza}, {Collins}, {Collins}, {Costigan}, {Crifo},
  {Cross}, {Crosta}, {Crowley}, {Dafonte}, {Damerdji}, {Dapergolas}, {David},
  {David}, {De Cat}, {de Felice}, {de Laverny}, {De Luise}, {De March}, {de
  Martino}, {de Souza}, {Debosscher}, {del Pozo}, {Delbo}, {Delgado},
  {Delgado}, {di Marco}, {Di Matteo}, {Diakite}, {Distefano}, {Dolding}, {Dos
  Anjos}, {Drazinos}, {Dur{\'a}n}, {Dzigan}, {Ecale}, {Edvardsson}, {Enke},
  {Erdmann}, {Escolar}, {Espina}, {Evans}, {Eynard Bontemps}, {Fabre},
  {Fabrizio}, {Faigler}, {Falc{\~a}o}, {Farr{\`a}s Casas}, {Faye}, {Federici},
  {Fedorets}, {Fern{\'a}ndez-Hern{\'a}ndez}, {Fernique}, {Fienga}, {Figueras},
  {Filippi}, {Findeisen}, {Fonti}, {Fouesneau}, {Fraile}, {Fraser}, {Fuchs},
  {Furnell}, {Gai}, {Galleti}, {Galluccio}, {Garabato}, {Garc{\'\i}a-Sedano},
  {Gar{\'e}}, {Garofalo}, {Garralda}, {Gavras}, {Gerssen}, {Geyer}, {Gilmore},
  {Girona}, {Giuffrida}, {Gomes}, {Gonz{\'a}lez-Marcos},
  {Gonz{\'a}lez-N{\'u}{\~n}ez}, {Gonz{\'a}lez-Vidal}, {Granvik}, {Guerrier},
  {Guillout}, {Guiraud}, {G{\'u}rpide}, {Guti{\'e}rrez-S{\'a}nchez}, {Guy},
  {Haigron}, {Hatzidimitriou}, {Haywood}, {Heiter}, {Helmi}, {Hobbs},
  {Hofmann}, {Holl}, {Holland }, {Hunt}, {Hypki}, {Icardi}, {Irwin}, {Jevardat
  de Fombelle}, {Jofr{\'e}}, {Jonker}, {Jorissen}, {Julbe}, {Karampelas},
  {Kochoska}, {Kohley}, {Kolenberg}, {Kontizas}, {Koposov}, {Kordopatis},
  {Koubsky}, {Kowalczyk}, {Krone-Martins}, {Kudryashova}, {Kull}, {Bachchan},
  {Lacoste-Seris}, {Lanza}, {Lavigne}, {Le Poncin-Lafitte}, {Lebreton},
  {Lebzelter}, {Leccia}, {Leclerc}, {Lecoeur-Taibi}, {Lemaitre}, {Lenhardt},
  {Leroux}, {Liao}, {Licata}, {Lindstr{\o}m}, {Lister}, {Livanou}, {Lobel},
  {L{\"o}ffler}, {L{\'o}pez}, {Lopez-Lozano}, {Lorenz}, {Loureiro},
  {MacDonald}, {Magalh{\~a}es Fernandes}, {Managau}, {Mann}, {Mantelet},
  {Marchal}, {Marchant}, {Marconi}, {Marie}, {Marinoni}, {Marrese},
  {Marschalk{\'o}}, {Marshall}, {Mart{\'\i}n-Fleitas}, {Martino}, {Mary},
  {Matijevi{\v{c}}}, {Mazeh}, {McMillan}, {Messina}, {Mestre}, {Michalik},
  {Millar}, {Miranda}, {Molina}, {Molinaro}, {Molinaro}, {Moln{\'a}r},
  {Moniez}, {Montegriffo}, {Monteiro}, {Mor}, {Mora}, {Morbidelli}, {Morel},
  {Morgenthaler}, {Morley}, {Morris}, {Mulone}, {Muraveva}, {Musella},
  {Narbonne}, {Nelemans}, {Nicastro}, {Noval}, {Ord{\'e}novic},
  {Ordieres-Mer{\'e}}, {Osborne}, {Pagani}, {Pagano}, {Pailler}, {Palacin},
  {Palaversa}, {Parsons}, {Paulsen}, {Pecoraro}, {Pedrosa}, {Pentik{\"a}inen},
  {Pereira}, {Pichon}, {Piersimoni}, {Pineau}, {Plachy}, {Plum}, {Poujoulet},
  {Pr{\v{s}}a}, {Pulone}, {Ragaini}, {Rago}, {Rambaux}, {Ramos-Lerate},
  {Ranalli}, {Rauw}, {Read}, {Regibo}, {Renk}, {Reyl{\'e}}, {Ribeiro},
  {Rimoldini}, {Ripepi}, {Riva}, {Rixon}, {Roelens}, {Romero-G{\'o}mez},
  {Rowell}, {Royer}, {Rudolph}, {Ruiz-Dern}, {Sadowski}, {Sagrist{\`a}
  Sell{\'e}s}, {Sahlmann}, {Salgado}, {Salguero}, {Sarasso}, {Savietto},
  {Schnorhk}, {Schultheis}, {Sciacca}, {Segol}, {Segovia}, {Segransan},
  {Serpell}, {Shih}, {Smareglia}, {Smart}, {Smith}, {Solano}, {Solitro},
  {Sordo}, {Soria Nieto}, {Souchay}, {Spagna}, {Spoto}, {Stampa}, {Steele},
  {Steidelm{\"u}ller}, {Stephenson}, {Stoev}, {Suess}, {S{\"u}veges}, {Surdej},
  {Szabados}, {Szegedi-Elek}, {Tapiador}, {Taris}, {Tauran}, {Taylor},
  {Teixeira}, {Terrett}, {Tingley}, {Trager}, {Turon}, {Ulla}, {Utrilla},
  {Valentini}, {van Elteren}, {Van Hemelryck}, {van Leeuwen}, {Varadi},
  {Vecchiato}, {Veljanoski}, {Via}, {Vicente}, {Vogt}, {Voss}, {Votruba},
  {Voutsinas}, {Walmsley}, {Weiler}, {Weingrill}, {Werner}, {Wevers},
  {Whitehead}, {Wyrzykowski}, {Yoldas}, {{\v{Z}}erjal}, {Zucker}, {Zurbach},
  {Zwitter}, {Alecu}, {Allen}, {Allende Prieto}, {Amorim},
  {Anglada-Escud{\'e}}, {Arsenijevic}, {Azaz}, {Balm}, {Beck}, {Bernstein},
  {Bigot}, {Bijaoui}, {Blasco}, {Bonfigli}, {Bono}, {Boudreault}, {Bressan},
  {Brown}, {Brunet}, {Bunclark}, {Buonanno}, {Butkevich}, {Carret}, {Carrion},
  {Chemin}, {Ch{\'e}reau}, {Corcione}, {Darmigny}, {de Boer}, {de Teodoro}, {de
  Zeeuw}, {Delle Luche}, {Domingues}, {Dubath}, {Fodor}, {Fr{\'e}zouls},
  {Fries}, {Fustes}, {Fyfe}, {Gallardo}, {Gallegos}, {Gardiol}, {Gebran},
  {Gomboc}, {G{\'o}mez}, {Grux}, {Gueguen}, {Heyrovsky}, {Hoar}, {Iannicola},
  {Isasi Parache}, {Janotto}, {Joliet}, {Jonckheere}, {Keil}, {Kim},
  {Klagyivik}, {Klar}, {Knude}, {Kochukhov}, {Kolka}, {Kos}, {Kutka}, {Lainey},
  {LeBouquin}, {Liu}, {Loreggia}, {Makarov}, {Marseille}, {Martayan},
  {Martinez-Rubi}, {Massart}, {Meynadier}, {Mignot}, {Munari}, {Nguyen},
  {Nordlander}, {Ocvirk}, {O'Flaherty}, {Olias Sanz}, {Ortiz}, {Osorio},
  {Oszkiewicz}, {Ouzounis}, {Palmer}, {Park}, {Pasquato}, {Peltzer}, {Peralta},
  {P{\'e}turaud}, {Pieniluoma}, {Pigozzi}, {Poels}, {Prat}, {Prod'homme},
  {Raison}, {Rebordao}, {Risquez}, {Rocca-Volmerange}, {Rosen}, {Ruiz-Fuertes},
  {Russo}, {Sembay}, {Serraller Vizcaino}, {Short}, {Siebert}, {Silva},
  {Sinachopoulos}, {Slezak}, {Soffel}, {Sosnowska}, {Strai{\v{z}}ys}, {ter
  Linden}, {Terrell}, {Theil}, {Tiede}, {Troisi}, {Tsalmantza}, {Tur},
  {Vaccari}, {Vachier}, {Valles}, {Van Hamme}, {Veltz}, {Virtanen}, {Wallut},
  {Wichmann}, {Wilkinson}, {Ziaeepour}, \& {Zschocke}}]{2016AandA...595A...1G}
{Gaia Collaboration}, {Prusti}, T., {de Bruijne}, J.~H.~J., {et~al.} 2016,
  \aap, 595, A1, \dodoi{10.1051/0004-6361/201629272}

\bibitem[{{Gaia Collaboration} {et~al.}(2018{\natexlab{a}}){Gaia
  Collaboration}, {Babusiaux}, {van Leeuwen}, {Barstow}, {Jordi}, {Vallenari},
  {Bossini}, {Bressan}, {Cantat-Gaudin}, {van Leeuwen}, {Brown}, {Prusti}, {de
  Bruijne}, {Bailer-Jones}, {Biermann}, {Evans}, {Eyer}, {Jansen}, {Klioner},
  {Lammers}, {Lindegren}, {Luri}, {Mignard}, {Panem}, {Pourbaix}, {Randich},
  {Sartoretti}, {Siddiqui}, {Soubiran}, {Walton}, {Arenou}, {Bastian},
  {Cropper}, {Drimmel}, {Katz}, {Lattanzi}, {Bakker}, {Cacciari},
  {Casta{\~n}eda}, {Chaoul}, {Cheek}, {De Angeli}, {Fabricius}, {Guerra},
  {Holl}, {Masana}, {Messineo}, {Mowlavi}, {Nienartowicz}, {Panuzzo},
  {Portell}, {Riello}, {Seabroke}, {Tanga}, {Th{\'e}venin}, {Gracia-Abril},
  {Comoretto}, {Garcia-Reinaldos}, {Teyssier}, {Altmann}, {Andrae}, {Audard},
  {Bellas-Velidis}, {Benson}, {Berthier}, {Blomme}, {Burgess}, {Busso},
  {Carry}, {Cellino}, {Clementini}, {Clotet}, {Creevey}, {Davidson}, {De
  Ridder}, {Delchambre}, {Dell'Oro}, {Ducourant},
  {Fern{\'a}ndez-Hern{\'a}ndez}, {Fouesneau}, {Fr{\'e}mat}, {Galluccio},
  {Garc{\'\i}a-Torres}, {Gonz{\'a}lez-N{\'u}{\~n}ez}, {Gonz{\'a}lez-Vidal},
  {Gosset}, {Guy}, {Halbwachs}, {Hambly}, {Harrison}, {Hern{\'a}ndez},
  {Hestroffer}, {Hodgkin}, {Hutton}, {Jasniewicz}, {Jean-Antoine-Piccolo},
  {Jordan}, {Korn}, {Krone-Martins}, {Lanzafame}, {Lebzelter}, {L{\"o}ffler},
  {Manteiga}, {Marrese}, {Mart{\'\i}n-Fleitas}, {Moitinho}, {Mora}, {Muinonen},
  {Osinde}, {Pancino}, {Pauwels}, {Petit}, {Recio-Blanco}, {Richards},
  {Rimoldini}, {Robin}, {Sarro}, {Siopis}, {Smith}, {Sozzetti}, {S{\"u}veges},
  {Torra}, {van Reeven}, {Abbas}, {Abreu Aramburu}, {Accart}, {Aerts},
  {Altavilla}, {{\'A}lvarez}, {Alvarez}, {Alves}, {Anderson}, {Andrei},
  {Anglada Varela}, {Antiche}, {Antoja}, {Arcay}, {Astraatmadja}, {Bach},
  {Baker}, {Balaguer-N{\'u}{\~n}ez}, {Balm}, {Barache}, {Barata}, {Barbato},
  {Barblan}, {Barklem}, {Barrado}, {Barros}, {Bartholom{\'e} Mu{\~n}oz},
  {Bassilana}, {Becciani}, {Bellazzini}, {Berihuete}, {Bertone}, {Bianchi},
  {Bienaym{\'e}}, {Blanco-Cuaresma}, {Boch}, {Boeche}, {Bombrun}, {Borrachero},
  {Bouquillon}, {Bourda}, {Bragaglia}, {Bramante}, {Breddels}, {Brouillet},
  {Br{\"u}semeister}, {Brugaletta}, {Bucciarelli}, {Burlacu}, {Busonero},
  {Butkevich}, {Buzzi}, {Caffau}, {Cancelliere}, {Cannizzaro}, {Carballo},
  {Carlucci}, {Carrasco}, {Casamiquela}, {Castellani}, {Castro-Ginard},
  {Charlot}, {Chemin}, {Chiavassa}, {Cocozza}, {Costigan}, {Cowell}, {Crifo},
  {Crosta}, {Crowley}, {Cuypers}, {Dafonte}, {Damerdji}, {Dapergolas}, {David},
  {David}, {de Laverny}, {De Luise}, {De March}, {de Martino}, {de Souza}, {de
  Torres}, {Debosscher}, {del Pozo}, {Delbo}, {Delgado}, {Delgado}, {Diakite},
  {Diener}, {Distefano}, {Dolding}, {Drazinos}, {Dur{\'a}n}, {Edvardsson},
  {Enke}, {Eriksson}, {Esquej}, {Eynard Bontemps}, {Fabre}, {Fabrizio},
  {Faigler}, {Falc{\~a}o}, {Farr{\`a}s Casas}, {Federici}, {Fedorets},
  {Fernique}, {Figueras}, {Filippi}, {Findeisen}, {Fonti}, {Fraile}, {Fraser},
  {Fr{\'e}zouls}, {Gai}, {Galleti}, {Garabato}, {Garc{\'\i}a-Sedano},
  {Garofalo}, {Garralda}, {Gavel}, {Gavras}, {Gerssen}, {Geyer}, {Giacobbe},
  {Gilmore}, {Girona}, {Giuffrida}, {Glass}, {Gomes}, {Granvik}, {Gueguen},
  {Guerrier}, {Guiraud}, {Guti{\'e}}, {Haigron}, {Hatzidimitriou}, {Hauser},
  {Haywood}, {Heiter}, {Helmi}, {Heu}, {Hilger}, {Hobbs}, {Hofmann}, {Holland},
  {Huckle}, {Hypki}, {Icardi}, {Jan{\ss}en}, {Jevardat de Fombelle}, {Jonker},
  {Juh{\'a}sz}, {Julbe}, {Karampelas}, {Kewley}, {Klar}, {Kochoska}, {Kohley},
  {Kolenberg}, {Kontizas}, {Kontizas}, {Koposov}, {Kordopatis},
  {Kostrzewa-Rutkowska}, {Koubsky}, {Lambert}, {Lanza}, {Lasne}, {Lavigne}, {Le
  Fustec}, {Le Poncin-Lafitte}, {Lebreton}, {Leccia}, {Leclerc},
  {Lecoeur-Taibi}, {Lenhardt}, {Leroux}, {Liao}, {Licata}, {Lindstr{\o}m},
  {Lister}, {Livanou}, {Lobel}, {L{\'o}pez}, {Managau}, {Mann}, {Mantelet},
  {Marchal}, {Marchant}, {Marconi}, {Marinoni}, {Marschalk{\'o}}, {Marshall},
  {Martino}, {Marton}, {Mary}, {Massari}, {Matijevi{\v{c}}}, {Mazeh},
  {McMillan}, {Messina}, {Michalik}, {Millar}, {Molina}, {Molinaro},
  {Moln{\'a}r}, {Montegriffo}, {Mor}, {Morbidelli}, {Morel}, {Morris},
  {Mulone}, {Muraveva}, {Musella}, {Nelemans}, {Nicastro}, {Noval},
  {O'Mullane}, {Ord{\'e}novic}, {Ord{\'o}{\~n}ez-Blanco}, {Osborne}, {Pagani},
  {Pagano}, {Pailler}, {Palacin}, {Palaversa}, {Panahi}, {Pawlak},
  {Piersimoni}, {Pineau}, {Plachy}, {Plum}, {Poggio}, {Poujoulet},
  {Pr{\v{s}}a}, {Pulone}, {Racero}, {Ragaini}, {Rambaux}, {Ramos-Lerate},
  {Regibo}, {Reyl{\'e}}, {Riclet}, {Ripepi}, {Riva}, {Rivard}, {Rixon},
  {Roegiers}, {Roelens}, {Romero-G{\'o}mez}, {Rowell}, {Royer}, {Ruiz-Dern},
  {Sadowski}, {Sagrist{\`a} Sell{\'e}s}, {Sahlmann}, {Salgado}, {Salguero},
  {Sanna}, {Santana-Ros}, {Sarasso}, {Savietto}, {Schultheis}, {Sciacca},
  {Segol}, {Segovia}, {S{\'e}gransan}, {Shih}, {Siltala}, {Silva}, {Smart},
  {Smith}, {Solano}, {Solitro}, {Sordo}, {Soria Nieto}, {Souchay}, {Spagna},
  {Spoto}, {Stampa}, {Steele}, {Steidelm{\"u}ller}, {Stephenson}, {Stoev},
  {Suess}, {Surdej}, {Szabados}, {Szegedi-Elek}, {Tapiador}, {Taris}, {Tauran},
  {Taylor}, {Teixeira}, {Terrett}, {Teyssandier}, {Thuillot}, {Titarenko},
  {Torra Clotet}, {Turon}, {Ulla}, {Utrilla}, {Uzzi}, {Vaillant}, {Valentini},
  {Valette}, {van Elteren}, {Van Hemelryck}, {Vaschetto}, {Vecchiato},
  {Veljanoski}, {Viala}, {Vicente}, {Vogt}, {von Essen}, {Voss}, {Votruba},
  {Voutsinas}, {Walmsley}, {Weiler}, {Wertz}, {Wevers}, {Wyrzykowski},
  {Yoldas}, {{\v{Z}}erjal}, {Ziaeepour}, {Zorec}, {Zschocke}, {Zucker},
  {Zurbach}, \& {Zwitter}}]{2018A&A...616A..10G}
{Gaia Collaboration}, {Babusiaux}, C., {van Leeuwen}, F., {et~al.}
  2018{\natexlab{a}}, \aap, 616, A10, \dodoi{10.1051/0004-6361/201832843}

\bibitem[{{Gaia Collaboration} {et~al.}(2018{\natexlab{b}}){Gaia
  Collaboration}, {Brown}, {Vallenari}, {Prusti}, {de Bruijne}, {Babusiaux},
  {Bailer-Jones}, {Biermann}, {Evans}, {Eyer}, \&
  et~al.}]{2018AandA...616A...1G}
{Gaia Collaboration}, {Brown}, A.~G.~A., {Vallenari}, A., {et~al.}
  2018{\natexlab{b}}, \aap, 616, A1, \dodoi{10.1051/0004-6361/201833051}

\bibitem[{{Gizis} {et~al.}(2015){Gizis}, {Allers}, {Liu}, {Harris}, {Faherty},
  {Burgasser}, \& {Kirkpatrick}}]{2015ApJ...799..203G}
{Gizis}, J.~E., {Allers}, K.~N., {Liu}, M.~C., {et~al.} 2015, \apj, 799, 203,
  \dodoi{10.1088/0004-637X/799/2/203}

\bibitem[{{Gizis} {et~al.}(2012){Gizis}, {Faherty}, {Liu}, {Castro}, {Shaw},
  {Vrba}, {Harris}, {Aller}, \& {Deacon}}]{2012AJ....144...94G}
{Gizis}, J.~E., {Faherty}, J.~K., {Liu}, M.~C., {et~al.} 2012, \aj, 144, 94,
  \dodoi{10.1088/0004-6256/144/4/94}

\bibitem[{{Hewett} {et~al.}(2006){Hewett}, {Warren}, {Leggett}, \&
  {Hodgkin}}]{2006MNRAS.367..454H}
{Hewett}, P.~C., {Warren}, S.~J., {Leggett}, S.~K., \& {Hodgkin}, S.~T. 2006,
  \mnras, 367, 454, \dodoi{10.1111/j.1365-2966.2005.09969.x}

\bibitem[{{Hook} {et~al.}(2004){Hook}, {J{\o}rgensen}, {Allington-Smith},
  {Davies}, {Metcalfe}, {Murowinski}, \& {Crampton}}]{2004PASP..116..425H}
{Hook}, I.~M., {J{\o}rgensen}, I., {Allington-Smith}, J.~R., {et~al.} 2004,
  \pasp, 116, 425, \dodoi{10.1086/383624}

\bibitem[{Hunter(2007)}]{Hunter:2007}
Hunter, J.~D. 2007, Computing in Science \& Engineering, 9, 90,
  \dodoi{10.1109/MCSE.2007.55}

\bibitem[{{Husser} {et~al.}(2013){Husser}, {Wende-von Berg}, {Dreizler},
  {Homeier}, {Reiners}, {Barman}, \& {Hauschildt}}]{2013A&A...553A...6H}
{Husser}, T.~O., {Wende-von Berg}, S., {Dreizler}, S., {et~al.} 2013, \aap,
  553, A6, \dodoi{10.1051/0004-6361/201219058}

\bibitem[{{Jackson} {et~al.}(2012){Jackson}, {Davis}, \&
  {Wheatley}}]{2012MNRAS.422.2024J}
{Jackson}, A.~P., {Davis}, T.~A., \& {Wheatley}, P.~J. 2012, \mnras, 422, 2024,
  \dodoi{10.1111/j.1365-2966.2012.20657.x}

\bibitem[{{Johnson} {et~al.}(2012){Johnson}, {Gazak}, {Apps}, {Muirhead},
  {Crepp}, {Crossfield}, {Boyajian}, {von Braun}, {Rojas-Ayala}, {Howard},
  {Covey}, {Schlawin}, {Hamren}, {Morton}, {Marcy}, \&
  {Lloyd}}]{2012AJ....143..111J}
{Johnson}, J.~A., {Gazak}, J.~Z., {Apps}, K., {et~al.} 2012, \aj, 143, 111,
  \dodoi{10.1088/0004-6256/143/5/111}

\bibitem[{Jones {et~al.}(2001)Jones, Oliphant, Peterson, {et~al.}}]{scipy}
Jones, E., Oliphant, T., Peterson, P., {et~al.} 2001, {SciPy}: Open source
  scientific tools for {Python}.
\newblock \url{http://www.scipy.org/}

\bibitem[{{Kellogg} {et~al.}(2015){Kellogg}, {Metchev}, {Gei{\ss}ler}, {Hicks},
  {Kirkpatrick}, \& {Kurtev}}]{2015AJ....150..182K}
{Kellogg}, K., {Metchev}, S., {Gei{\ss}ler}, K., {et~al.} 2015, \aj, 150, 182,
  \dodoi{10.1088/0004-6256/150/6/182}

\bibitem[{{Kenyon} \& {Hartmann}(1995)}]{1995ApJS..101..117K}
{Kenyon}, S.~J., \& {Hartmann}, L. 1995, \apjs, 101, 117,
  \dodoi{10.1086/192235}

\bibitem[{{Kesseli} {et~al.}(2017){Kesseli}, {West}, {Veyette}, {Harrison},
  {Feldman}, \& {Bochanski}}]{2017ApJS..230...16K}
{Kesseli}, A.~Y., {West}, A.~A., {Veyette}, M., {et~al.} 2017, \apjs, 230, 16,
  \dodoi{10.3847/1538-4365/aa656d}

\bibitem[{{Kiman} {et~al.}(2021){Kiman}, {Faherty}, {Cruz}, {Gagn{\'e}},
  {Angus}, {Schmidt}, {Mann}, {Bardalez Gagliuffi}, \&
  {Rice}}]{2021AJ....161..277K}
{Kiman}, R., {Faherty}, J.~K., {Cruz}, K.~L., {et~al.} 2021, \aj, 161, 277,
  \dodoi{10.3847/1538-3881/abf561}

\bibitem[{{Kirkpatrick} {et~al.}(1999){Kirkpatrick}, {Reid}, {Liebert},
  {Cutri}, {Nelson}, {Beichman}, {Dahn}, {Monet}, {Gizis}, \&
  {Skrutskie}}]{1999ApJ...519..802K}
{Kirkpatrick}, J.~D., {Reid}, I.~N., {Liebert}, J., {et~al.} 1999, \apj, 519,
  802, \dodoi{10.1086/307414}

\bibitem[{{Kirkpatrick} {et~al.}(2000){Kirkpatrick}, {Reid}, {Liebert},
  {Gizis}, {Burgasser}, {Monet}, {Dahn}, {Nelson}, \&
  {Williams}}]{2000AJ....120..447K}
---. 2000, \aj, 120, 447, \dodoi{10.1086/301427}

\bibitem[{{Kirkpatrick} {et~al.}(2010){Kirkpatrick}, {Looper}, {Burgasser},
  {Schurr}, {Cutri}, {Cushing}, {Cruz}, {Sweet}, {Knapp}, {Barman},
  {Bochanski}, {Roellig}, {McLean}, {McGovern}, \&
  {Rice}}]{2010ApJS..190..100K}
{Kirkpatrick}, J.~D., {Looper}, D.~L., {Burgasser}, A.~J., {et~al.} 2010,
  \apjs, 190, 100, \dodoi{10.1088/0067-0049/190/1/100}

\bibitem[{{Kirkpatrick} {et~al.}(2011){Kirkpatrick}, {Cushing}, {Gelino},
  {Griffith}, {Skrutskie}, {Marsh}, {Wright}, {Mainzer}, {Eisenhardt},
  {McLean}, {Thompson}, {Bauer}, {Benford}, {Bridge}, {Lake}, {Petty},
  {Stanford}, {Tsai}, {Bailey}, {Beichman}, {Bloom}, {Bochanski}, {Burgasser},
  {Capak}, {Cruz}, {Hinz}, {Kartaltepe}, {Knox}, {Manohar}, {Masters},
  {Morales-Calder{\'o}n}, {Prato}, {Rodigas}, {Salvato}, {Schurr}, {Scoville},
  {Simcoe}, {Stapelfeldt}, {Stern}, {Stock}, \& {Vacca}}]{2011ApJS..197...19K}
{Kirkpatrick}, J.~D., {Cushing}, M.~C., {Gelino}, C.~R., {et~al.} 2011, \apjs,
  197, 19, \dodoi{10.1088/0067-0049/197/2/19}

\bibitem[{{Kirkpatrick} {et~al.}(2021){Kirkpatrick}, {Gelino}, {Faherty},
  {Meisner}, {Caselden}, {Schneider}, {Marocco}, {Cayago}, {Smart},
  {Eisenhardt}, {Kuchner}, {Wright}, {Cushing}, {Allers}, {Bardalez Gagliuffi},
  {Burgasser}, {Gagn{\'e}}, {Logsdon}, {Martin}, {Ingalls}, {Lowrance},
  {Abrahams}, {Aganze}, {Gerasimov}, {Gonzales}, {Hsu}, {Kamraj}, {Kiman},
  {Rees}, {Theissen}, {Ammar}, {Andersen}, {Beaulieu}, {Colin}, {Elachi},
  {Goodman}, {Gramaize}, {Hamlet}, {Hong}, {Jonkeren}, {Khalil}, {Martin},
  {Pendrill}, {Pumphrey}, {Rothermich}, {Sainio}, {Stenner}, {Tanner},
  {Th{\'e}venot}, {Voloshin}, {Walla}, {W{\k{e}}dracki}, \& {Backyard Worlds:
  Planet 9 Collaboration}}]{2021ApJS..253....7K}
{Kirkpatrick}, J.~D., {Gelino}, C.~R., {Faherty}, J.~K., {et~al.} 2021, \apjs,
  253, 7, \dodoi{10.3847/1538-4365/abd107}

\bibitem[{{Kochukhov}(2021)}]{2021A&ARv..29....1K}
{Kochukhov}, O. 2021, \aapr, 29, 1, \dodoi{10.1007/s00159-020-00130-3}

\bibitem[{{Kraus} {et~al.}(2017){Kraus}, {Herczeg}, {Rizzuto}, {Mann},
  {Slesnick}, {Carpenter}, {Hillenbrand}, \& {Mamajek}}]{2017ApJ...838..150K}
{Kraus}, A.~L., {Herczeg}, G.~J., {Rizzuto}, A.~C., {et~al.} 2017, \apj, 838,
  150, \dodoi{10.3847/1538-4357/aa62a0}

\bibitem[{{Kroupa}(2001)}]{2001MNRAS.322..231K}
{Kroupa}, P. 2001, \mnras, 322, 231, \dodoi{10.1046/j.1365-8711.2001.04022.x}

\bibitem[{{Kupfer} {et~al.}(2021){Kupfer}, {Prince}, {van Roestel}, {Bellm},
  {Bildsten}, {Coughlin}, {Drake}, {Graham}, {Klein}, {Kulkarni}, {Masci},
  {Walters}, {Andreoni}, {Biswas}, {Bradshaw}, {Duev}, {Dekany}, {Guidry},
  {Hermes}, {Laher}, \& {Riddle}}]{2021MNRAS.505.1254K}
{Kupfer}, T., {Prince}, T.~A., {van Roestel}, J., {et~al.} 2021, \mnras, 505,
  1254, \dodoi{10.1093/mnras/stab1344}

\bibitem[{{Lantz} {et~al.}(2004){Lantz}, {Aldering}, {Antilogus}, {Bonnaud},
  {Capoani}, {Castera}, {Copin}, {Dubet}, {Gangler}, \&
  {Henault}}]{2004SPIE.5249..146L}
{Lantz}, B., {Aldering}, G., {Antilogus}, P., {et~al.} 2004, in Society of
  Photo-Optical Instrumentation Engineers (SPIE) Conference Series, Vol. 5249,
  Optical Design and Engineering, ed. L.~{Mazuray}, P.~J. {Rogers}, \&
  R.~{Wartmann}, 146--155

\bibitem[{{L{\'e}pine} {et~al.}(2007){L{\'e}pine}, {Rich}, \&
  {Shara}}]{2007ApJ...669.1235L}
{L{\'e}pine}, S., {Rich}, R.~M., \& {Shara}, M.~M. 2007, \apj, 669, 1235,
  \dodoi{10.1086/521614}

\bibitem[{{Lindegren}(2018)}]{Lindegren2018}
{Lindegren}, L. 2018, Gaia Technical Note: GAIA-C3-TN-LU-LL-124-01

\bibitem[{{Liu} {et~al.}(2016){Liu}, {Dupuy}, \&
  {Allers}}]{2016ApJ...833...96L}
{Liu}, M.~C., {Dupuy}, T.~J., \& {Allers}, K.~N. 2016, \apj, 833, 96,
  \dodoi{10.3847/1538-4357/833/1/96}

\bibitem[{{Liu} {et~al.}(2002){Liu}, {Fischer}, {Graham}, {Lloyd}, {Marcy}, \&
  {Butler}}]{2002ApJ...571..519L}
{Liu}, M.~C., {Fischer}, D.~A., {Graham}, J.~R., {et~al.} 2002, \apj, 571, 519,
  \dodoi{10.1086/339845}

\bibitem[{{Liu} {et~al.}(2013){Liu}, {Magnier}, {Deacon}, {Allers}, {Dupuy},
  {Kotson}, {Aller}, {Burgett}, {Chambers}, {Draper}, {Hodapp}, {Jedicke},
  {Kaiser}, {Kudritzki}, {Metcalfe}, {Morgan}, {Price}, {Tonry}, \&
  {Wainscoat}}]{2013ApJ...777L..20L}
{Liu}, M.~C., {Magnier}, E.~A., {Deacon}, N.~R., {et~al.} 2013, \apjl, 777,
  L20, \dodoi{10.1088/2041-8205/777/2/L20}

\bibitem[{{Looper} {et~al.}(2008){Looper}, {Kirkpatrick}, {Cutri}, {Barman},
  {Burgasser}, {Cushing}, {Roellig}, {McGovern}, {McLean}, {Rice}, {Swift}, \&
  {Schurr}}]{2008ApJ...686..528L}
{Looper}, D.~L., {Kirkpatrick}, J.~D., {Cutri}, R.~M., {et~al.} 2008, \apj,
  686, 528, \dodoi{10.1086/591025}

\bibitem[{{L{\'o}pez-Valdivia} {et~al.}(2021){L{\'o}pez-Valdivia}, {Sokal},
  {Mace}, {Kidder}, {Hussaini}, {Nofi}, {Prato}, {Johns-Krull}, {Oh}, {Lee},
  {Park}, {Oh}, {Kraus}, {Kaplan}, {Llama}, {Mann}, {Kim}, {Gully-Santiago},
  {Lee}, {Pak}, {Hwang}, \& {Jaffe}}]{2021ApJ...921...53L}
{L{\'o}pez-Valdivia}, R., {Sokal}, K.~R., {Mace}, G.~N., {et~al.} 2021, \apj,
  921, 53, \dodoi{10.3847/1538-4357/ac1a7b}

\bibitem[{{Luhman} {et~al.}(1998){Luhman}, {Brice{\~n}o}, {Rieke}, \&
  {Hartmann}}]{1998ApJ...493..909L}
{Luhman}, K.~L., {Brice{\~n}o}, C., {Rieke}, G.~H., \& {Hartmann}, L. 1998,
  \apj, 493, 909, \dodoi{10.1086/305171}

\bibitem[{{Lyo} {et~al.}(2004){Lyo}, {Lawson}, \&
  {Bessell}}]{2004MNRAS.355..363L}
{Lyo}, A.~R., {Lawson}, W.~A., \& {Bessell}, M.~S. 2004, \mnras, 355, 363,
  \dodoi{10.1111/j.1365-2966.2004.08318.x}

\bibitem[{{Macintosh} {et~al.}(2015){Macintosh}, {Graham}, {Barman}, {De Rosa},
  {Konopacky}, {Marley}, {Marois}, {Nielsen}, {Pueyo}, {Rajan}, {Rameau},
  {Saumon}, {Wang}, {Patience}, {Ammons}, {Arriaga}, {Artigau}, {Beckwith},
  {Brewster}, {Bruzzone}, {Bulger}, {Burningham}, {Burrows}, {Chen}, {Chiang},
  {Chilcote}, {Dawson}, {Dong}, {Doyon}, {Draper}, {Duch{\^e}ne}, {Esposito},
  {Fabrycky}, {Fitzgerald}, {Follette}, {Fortney}, {Gerard}, {Goodsell},
  {Greenbaum}, {Hibon}, {Hinkley}, {Cotten}, {Hung}, {Ingraham},
  {Johnson-Groh}, {Kalas}, {Lafreniere}, {Larkin}, {Lee}, {Line}, {Long},
  {Maire}, {Marchis}, {Matthews}, {Max}, {Metchev}, {Millar-Blanchaer},
  {Mittal}, {Morley}, {Morzinski}, {Murray-Clay}, {Oppenheimer}, {Palmer},
  {Patel}, {Perrin}, {Poyneer}, {Rafikov}, {Rantakyr{\"o}}, {Rice}, {Rojo},
  {Rudy}, {Ruffio}, {Ruiz}, {Sadakuni}, {Saddlemyer}, {Salama}, {Savransky},
  {Schneider}, {Sivaramakrishnan}, {Song}, {Soummer}, {Thomas}, {Vasisht},
  {Wallace}, {Ward-Duong}, {Wiktorowicz}, {Wolff}, \&
  {Zuckerman}}]{2015Sci...350...64M}
{Macintosh}, B., {Graham}, J.~R., {Barman}, T., {et~al.} 2015, Science, 350,
  64, \dodoi{10.1126/science.aac5891}

\bibitem[{{Maggio} {et~al.}(1990){Maggio}, {Vaiana}, {Haisch}, {Stern},
  {Bookbinder}, {Harnden}, \& {Rosner}}]{1990ApJ...348..253M}
{Maggio}, A., {Vaiana}, G.~S., {Haisch}, B.~M., {et~al.} 1990, \apj, 348, 253,
  \dodoi{10.1086/168234}

\bibitem[{{Magnier} {et~al.}(2020){Magnier}, {Schlafly}, {Finkbeiner}, {Tonry},
  {Goldman}, {R{\"o}ser}, {Schilbach}, {Casertano}, {Chambers}, {Flewelling},
  {Huber}, {Price}, {Sweeney}, {Waters}, {Denneau}, {Draper}, {Hodapp},
  {Jedicke}, {Kaiser}, {Kudritzki}, {Metcalfe}, {Stubbs}, \&
  {Wainscoat}}]{2020ApJS..251....6M}
{Magnier}, E.~A., {Schlafly}, E.~F., {Finkbeiner}, D.~P., {et~al.} 2020, \apjs,
  251, 6, \dodoi{10.3847/1538-4365/abb82a}

\bibitem[{{Maire} {et~al.}(2020){Maire}, {Baudino}, {Desidera}, {Messina},
  {Brandner}, {Godoy}, {Cantalloube}, {Galicher}, {Bonnefoy}, {Hagelberg},
  {Olofsson}, {Absil}, {Chauvin}, {Henning}, \&
  {Langlois}}]{2020A&A...633L...2M}
{Maire}, A.~L., {Baudino}, J.~L., {Desidera}, S., {et~al.} 2020, \aap, 633, L2,
  \dodoi{10.1051/0004-6361/201937134}

\bibitem[{{Malo} {et~al.}(2014){Malo}, {Artigau}, {Doyon}, {Lafreni{\`e}re},
  {Albert}, \& {Gagn{\'e}}}]{2014ApJ...788...81M}
{Malo}, L., {Artigau}, {\'E}., {Doyon}, R., {et~al.} 2014, \apj, 788, 81,
  \dodoi{10.1088/0004-637X/788/1/81}

\bibitem[{{Malo} {et~al.}(2013){Malo}, {Doyon}, {Lafreni{\`e}re}, {Artigau},
  {Gagn{\'e}}, {Baron}, \& {Riedel}}]{2013ApJ...762...88M}
{Malo}, L., {Doyon}, R., {Lafreni{\`e}re}, D., {et~al.} 2013, \apj, 762, 88,
  \dodoi{10.1088/0004-637X/762/2/88}

\bibitem[{{Manara} {et~al.}(2018){Manara}, {Morbidelli}, \&
  {Guillot}}]{2018A&A...618L...3M}
{Manara}, C.~F., {Morbidelli}, A., \& {Guillot}, T. 2018, \aap, 618, L3,
  \dodoi{10.1051/0004-6361/201834076}

\bibitem[{{Mann} {et~al.}(2013){Mann}, {Brewer}, {Gaidos}, {L{\'e}pine}, \&
  {Hilton}}]{2013AJ....145...52M}
{Mann}, A.~W., {Brewer}, J.~M., {Gaidos}, E., {L{\'e}pine}, S., \& {Hilton},
  E.~J. 2013, \aj, 145, 52, \dodoi{10.1088/0004-6256/145/2/52}

\bibitem[{{Mann} {et~al.}(2014){Mann}, {Deacon}, {Gaidos}, {Ansdell}, {Brewer},
  {Liu}, {Magnier}, \& {Aller}}]{2014AJ....147..160M}
{Mann}, A.~W., {Deacon}, N.~R., {Gaidos}, E., {et~al.} 2014, \aj, 147, 160,
  \dodoi{10.1088/0004-6256/147/6/160}

\bibitem[{{Mann} {et~al.}(2015){Mann}, {Feiden}, {Gaidos}, {Boyajian}, \& {von
  Braun}}]{2015ApJ...804...64M}
{Mann}, A.~W., {Feiden}, G.~A., {Gaidos}, E., {Boyajian}, T., \& {von Braun},
  K. 2015, \apj, 804, 64, \dodoi{10.1088/0004-637X/804/1/64}

\bibitem[{{Mann} {et~al.}(2019){Mann}, {Dupuy}, {Kraus}, {Gaidos}, {Ansdell},
  {Ireland}, {Rizzuto}, {Hung}, {Dittmann}, {Factor}, {Feiden}, {Martinez},
  {Ru{\'\i}z-Rodr{\'\i}guez}, \& {Thao}}]{2019ApJ...871...63M}
{Mann}, A.~W., {Dupuy}, T., {Kraus}, A.~L., {et~al.} 2019, \apj, 871, 63,
  \dodoi{10.3847/1538-4357/aaf3bc}

\bibitem[{{Marley} \& {Robinson}(2015)}]{2015ARA&A..53..279M}
{Marley}, M.~S., \& {Robinson}, T.~D. 2015, \araa, 53, 279,
  \dodoi{10.1146/annurev-astro-082214-122522}

\bibitem[{{Marley} {et~al.}(2012){Marley}, {Saumon}, {Cushing}, {Ackerman},
  {Fortney}, \& {Freedman}}]{2012ApJ...754..135M}
{Marley}, M.~S., {Saumon}, D., {Cushing}, M., {et~al.} 2012, \apj, 754, 135,
  \dodoi{10.1088/0004-637X/754/2/135}

\bibitem[{{Marley} {et~al.}(2021){Marley}, {Saumon}, {Visscher}, {Lupu},
  {Freedman}, {Morley}, {Fortney}, {Seay}, {Smith}, {Teal}, \&
  {Wang}}]{2021arXiv210707434M}
{Marley}, M.~S., {Saumon}, D., {Visscher}, C., {et~al.} 2021, arXiv e-prints,
  arXiv:2107.07434.
\newblock \doarXiv{2107.07434}

\bibitem[{{Marocco} {et~al.}(2021){Marocco}, {Eisenhardt}, {Fowler},
  {Kirkpatrick}, {Meisner}, {Schlafly}, {Stanford}, {Garcia}, {Caselden},
  {Cushing}, {Cutri}, {Faherty}, {Gelino}, {Gonzalez}, {Jarrett}, {Koontz},
  {Mainzer}, {Marchese}, {Mobasher}, {Schlegel}, {Stern}, {Teplitz}, \&
  {Wright}}]{2021ApJS..253....8M}
{Marocco}, F., {Eisenhardt}, P. R.~M., {Fowler}, J.~W., {et~al.} 2021, \apjs,
  253, 8, \dodoi{10.3847/1538-4365/abd805}

\bibitem[{{Marois} {et~al.}(2008){Marois}, {Macintosh}, {Barman}, {Zuckerman},
  {Song}, {Patience}, {Lafreni{\`e}re}, \& {Doyon}}]{2008Sci...322.1348M}
{Marois}, C., {Macintosh}, B., {Barman}, T., {et~al.} 2008, Science, 322, 1348,
  \dodoi{10.1126/science.1166585}

\bibitem[{{Marois} {et~al.}(2010){Marois}, {Zuckerman}, {Konopacky},
  {Macintosh}, \& {Barman}}]{2010Natur.468.1080M}
{Marois}, C., {Zuckerman}, B., {Konopacky}, Q.~M., {Macintosh}, B., \&
  {Barman}, T. 2010, \nat, 468, 1080, \dodoi{10.1038/nature09684}

\bibitem[{{Martin} {et~al.}(2005){Martin}, {Fanson}, {Schiminovich},
  {Morrissey}, {Friedman}, {Barlow}, {Conrow}, {Grange}, {Jelinsky},
  {Milliard}, {Siegmund}, {Bianchi}, {Byun}, {Donas}, {Forster}, {Heckman},
  {Lee}, {Madore}, {Malina}, {Neff}, {Rich}, {Small}, {Surber}, {Szalay},
  {Welsh}, \& {Wyder}}]{2005ApJ...619L...1M}
{Martin}, D.~C., {Fanson}, J., {Schiminovich}, D., {et~al.} 2005, \apjl, 619,
  L1, \dodoi{10.1086/426387}

\bibitem[{{Masci} {et~al.}(2019){Masci}, {Laher}, {Rusholme}, {Shupe}, {Groom},
  {Surace}, {Jackson}, {Monkewitz}, {Beck}, {Flynn}, {Terek}, {Landry},
  {Hacopians}, {Desai}, {Howell}, {Brooke}, {Imel}, {Wachter}, {Ye}, {Lin},
  {Cenko}, {Cunningham}, {Rebbapragada}, {Bue}, {Miller}, {Mahabal}, {Bellm},
  {Patterson}, {Juri{\'c}}, {Golkhou}, {Ofek}, {Walters}, {Graham}, {Kasliwal},
  {Dekany}, {Kupfer}, {Burdge}, {Cannella}, {Barlow}, {Van Sistine}, {Giomi},
  {Fremling}, {Blagorodnova}, {Levitan}, {Riddle}, {Smith}, {Helou}, {Prince},
  \& {Kulkarni}}]{2019PASP..131a8003M}
{Masci}, F.~J., {Laher}, R.~R., {Rusholme}, B., {et~al.} 2019, \pasp, 131,
  018003, \dodoi{10.1088/1538-3873/aae8ac}

\bibitem[{{McLean} {et~al.}(2003){McLean}, {McGovern}, {Burgasser},
  {Kirkpatrick}, {Prato}, \& {Kim}}]{2003ApJ...596..561M}
{McLean}, I.~S., {McGovern}, M.~R., {Burgasser}, A.~J., {et~al.} 2003, \apj,
  596, 561, \dodoi{10.1086/377636}

\bibitem[{{McMahon} {et~al.}(2013){McMahon}, {Banerji}, {Gonzalez}, {Koposov},
  {Bejar}, {Lodieu}, {Rebolo}, \& {VHS Collaboration}}]{2013Msngr.154...35M}
{McMahon}, R.~G., {Banerji}, M., {Gonzalez}, E., {et~al.} 2013, The Messenger,
  154, 35

\bibitem[{{Mentuch} {et~al.}(2008){Mentuch}, {Brandeker}, {van Kerkwijk},
  {Jayawardhana}, \& {Hauschildt}}]{2008ApJ...689.1127M}
{Mentuch}, E., {Brandeker}, A., {van Kerkwijk}, M.~H., {Jayawardhana}, R., \&
  {Hauschildt}, P.~H. 2008, \apj, 689, 1127, \dodoi{10.1086/592764}

\bibitem[{{Metchev} \& {Hillenbrand}(2006)}]{2006ApJ...651.1166M}
{Metchev}, S.~A., \& {Hillenbrand}, L.~A. 2006, \apj, 651, 1166,
  \dodoi{10.1086/507836}

\bibitem[{{Milli} {et~al.}(2017){Milli}, {Hibon}, {Christiaens}, {Choquet},
  {Bonnefoy}, {Kennedy}, {Wyatt}, {Absil}, {G{\'o}mez Gonz{\'a}lez}, {del
  Burgo}, {Matr{\`a}}, {Augereau}, {Boccaletti}, {Delacroix}, {Ertel}, {Dent},
  {Forsberg}, {Fusco}, {Girard}, {Habraken}, {Huby}, {Karlsson}, {Lagrange},
  {Mawet}, {Mouillet}, {Perrin}, {Pinte}, {Pueyo}, {Reyes}, {Soummer},
  {Surdej}, {Tarricq}, \& {Wahhaj}}]{2017A&A...597L...2M}
{Milli}, J., {Hibon}, P., {Christiaens}, V., {et~al.} 2017, \aap, 597, L2,
  \dodoi{10.1051/0004-6361/201629908}

\bibitem[{{Murphy} \& {Lawson}(2015)}]{2015MNRAS.447.1267M}
{Murphy}, S.~J., \& {Lawson}, W.~A. 2015, \mnras, 447, 1267,
  \dodoi{10.1093/mnras/stu2450}

\bibitem[{{Neves} {et~al.}(2012){Neves}, {Bonfils}, {Santos}, {Delfosse},
  {Forveille}, {Allard}, {Nat{\'a}rio}, {Fernandes}, \&
  {Udry}}]{2012A&A...538A..25N}
{Neves}, V., {Bonfils}, X., {Santos}, N.~C., {et~al.} 2012, \aap, 538, A25,
  \dodoi{10.1051/0004-6361/201118115}

\bibitem[{{Newton} {et~al.}(2014){Newton}, {Charbonneau}, {Irwin},
  {Berta-Thompson}, {Rojas-Ayala}, {Covey}, \& {Lloyd}}]{2014AJ....147...20N}
{Newton}, E.~R., {Charbonneau}, D., {Irwin}, J., {et~al.} 2014, \aj, 147, 20,
  \dodoi{10.1088/0004-6256/147/1/20}

\bibitem[{{Newton} {et~al.}(2017){Newton}, {Irwin}, {Charbonneau}, {Berlind},
  {Calkins}, \& {Mink}}]{2017ApJ...834...85N}
{Newton}, E.~R., {Irwin}, J., {Charbonneau}, D., {et~al.} 2017, \apj, 834, 85,
  \dodoi{10.3847/1538-4357/834/1/85}

\bibitem[{Oliphant(2006)}]{numpy}
Oliphant, T. 2006, {NumPy}: A guide to {NumPy}, USA: Trelgol Publishing.
\newblock \url{http://www.numpy.org/}

\bibitem[{{Pascucci} {et~al.}(2016){Pascucci}, {Testi}, {Herczeg}, {Long},
  {Manara}, {Hendler}, {Mulders}, {Krijt}, {Ciesla}, {Henning}, {Mohanty},
  {Drabek-Maunder}, {Apai}, {Sz{\H{u}}cs}, {Sacco}, \&
  {Olofsson}}]{2016ApJ...831..125P}
{Pascucci}, I., {Testi}, L., {Herczeg}, G.~J., {et~al.} 2016, \apj, 831, 125,
  \dodoi{10.3847/0004-637X/831/2/125}

\bibitem[{{Paxton} {et~al.}(2011){Paxton}, {Bildsten}, {Dotter}, {Herwig},
  {Lesaffre}, \& {Timmes}}]{2011ApJS..192....3P}
{Paxton}, B., {Bildsten}, L., {Dotter}, A., {et~al.} 2011, \apjs, 192, 3,
  \dodoi{10.1088/0067-0049/192/1/3}

\bibitem[{{Pecaut} \& {Mamajek}(2013)}]{2013ApJS..208....9P}
{Pecaut}, M.~J., \& {Mamajek}, E.~E. 2013, \apjs, 208, 9,
  \dodoi{10.1088/0067-0049/208/1/9}

\bibitem[{P\'erez \& Granger(2007)}]{PER-GRA:2007}
P\'erez, F., \& Granger, B.~E. 2007, Computing in Science and Engineering, 9,
  21, \dodoi{10.1109/MCSE.2007.53}

\bibitem[{{Preibisch} \& {Feigelson}(2005)}]{2005ApJS..160..390P}
{Preibisch}, T., \& {Feigelson}, E.~D. 2005, \apjs, 160, 390,
  \dodoi{10.1086/432094}

\bibitem[{{Rayner} {et~al.}(2009){Rayner}, {Cushing}, \&
  {Vacca}}]{2009ApJS..185..289R}
{Rayner}, J.~T., {Cushing}, M.~C., \& {Vacca}, W.~D. 2009, \apjs, 185, 289,
  \dodoi{10.1088/0067-0049/185/2/289}

\bibitem[{{Rayner} {et~al.}(2003){Rayner}, {Toomey}, {Onaka}, {Denault},
  {Stahlberger}, {Vacca}, {Cushing}, \& {Wang}}]{2003PASP..115..362R}
{Rayner}, J.~T., {Toomey}, D.~W., {Onaka}, P.~M., {et~al.} 2003, \pasp, 115,
  362, \dodoi{10.1086/367745}

\bibitem[{{Rebull} {et~al.}(2018){Rebull}, {Stauffer}, {Cody}, {Hillenbrand},
  {David}, \& {Pinsonneault}}]{2018AJ....155..196R}
{Rebull}, L.~M., {Stauffer}, J.~R., {Cody}, A.~M., {et~al.} 2018, \aj, 155,
  196, \dodoi{10.3847/1538-3881/aab605}

\bibitem[{{Reid} {et~al.}(2008){Reid}, {Cruz}, {Kirkpatrick}, {Allen},
  {Mungall}, {Liebert}, {Lowrance}, \& {Sweet}}]{2008AJ....136.1290R}
{Reid}, I.~N., {Cruz}, K.~L., {Kirkpatrick}, J.~D., {et~al.} 2008, \aj, 136,
  1290, \dodoi{10.1088/0004-6256/136/3/1290}

\bibitem[{{Reid} \& {Walkowicz}(2006)}]{2006PASP..118..671R}
{Reid}, I.~N., \& {Walkowicz}, L.~M. 2006, \pasp, 118, 671,
  \dodoi{10.1086/503446}

\bibitem[{{Riedel} {et~al.}(2017){Riedel}, {Blunt}, {Lambrides}, {Rice},
  {Cruz}, \& {Faherty}}]{2017AJ....153...95R}
{Riedel}, A.~R., {Blunt}, S.~C., {Lambrides}, E.~L., {et~al.} 2017, \aj, 153,
  95, \dodoi{10.3847/1538-3881/153/3/95}

\bibitem[{{Riedel} {et~al.}(2014){Riedel}, {Finch}, {Henry}, {Subasavage},
  {Jao}, {Malo}, {Rodriguez}, {White}, {Gies}, {Dieterich}, {Winters},
  {Davison}, {Nelan}, {Blunt}, {Cruz}, {Rice}, \&
  {Ianna}}]{2014AJ....147...85R}
{Riedel}, A.~R., {Finch}, C.~T., {Henry}, T.~J., {et~al.} 2014, \aj, 147, 85,
  \dodoi{10.1088/0004-6256/147/4/85}

\bibitem[{{Rodriguez} {et~al.}(2013){Rodriguez}, {Zuckerman}, {Kastner},
  {Bessell}, {Faherty}, \& {Murphy}}]{2013ApJ...774..101R}
{Rodriguez}, D.~R., {Zuckerman}, B., {Kastner}, J.~H., {et~al.} 2013, \apj,
  774, 101, \dodoi{10.1088/0004-637X/774/2/101}

\bibitem[{{Rojas-Ayala} {et~al.}(2010){Rojas-Ayala}, {Covey}, {Muirhead}, \&
  {Lloyd}}]{2010ApJ...720L.113R}
{Rojas-Ayala}, B., {Covey}, K.~R., {Muirhead}, P.~S., \& {Lloyd}, J.~P. 2010,
  \apjl, 720, L113, \dodoi{10.1088/2041-8205/720/1/L113}

\bibitem[{{Rojas-Ayala} {et~al.}(2012){Rojas-Ayala}, {Covey}, {Muirhead}, \&
  {Lloyd}}]{2012ApJ...748...93R}
---. 2012, \apj, 748, 93, \dodoi{10.1088/0004-637X/748/2/93}

\bibitem[{{Ryabchikova} {et~al.}(2015){Ryabchikova}, {Piskunov}, {Kurucz},
  {Stempels}, {Heiter}, {Pakhomov}, \& {Barklem}}]{2015PhyS...90e4005R}
{Ryabchikova}, T., {Piskunov}, N., {Kurucz}, R.~L., {et~al.} 2015, \physscr,
  90, 054005, \dodoi{10.1088/0031-8949/90/5/054005}

\bibitem[{{Saumon} \& {Marley}(2008)}]{2008ApJ...689.1327S}
{Saumon}, D., \& {Marley}, M.~S. 2008, \apj, 689, 1327, \dodoi{10.1086/592734}

\bibitem[{{Saumon} {et~al.}(2012){Saumon}, {Marley}, {Abel}, {Frommhold}, \&
  {Freedman}}]{2012ApJ...750...74S}
{Saumon}, D., {Marley}, M.~S., {Abel}, M., {Frommhold}, L., \& {Freedman},
  R.~S. 2012, \apj, 750, 74, \dodoi{10.1088/0004-637X/750/1/74}

\bibitem[{{Schlaufman} \& {Laughlin}(2010)}]{2010A&A...519A.105S}
{Schlaufman}, K.~C., \& {Laughlin}, G. 2010, \aap, 519, A105,
  \dodoi{10.1051/0004-6361/201015016}

\bibitem[{{Schlieder} {et~al.}(2012{\natexlab{a}}){Schlieder}, {L{\'e}pine},
  {Rice}, {Simon}, {Fielding}, \& {Tomasino}}]{2012AJ....143..114S}
{Schlieder}, J.~E., {L{\'e}pine}, S., {Rice}, E., {et~al.} 2012{\natexlab{a}},
  \aj, 143, 114, \dodoi{10.1088/0004-6256/143/5/114}

\bibitem[{{Schlieder} {et~al.}(2012{\natexlab{b}}){Schlieder}, {L{\'e}pine}, \&
  {Simon}}]{2012AJ....143...80S}
{Schlieder}, J.~E., {L{\'e}pine}, S., \& {Simon}, M. 2012{\natexlab{b}}, \aj,
  143, 80, \dodoi{10.1088/0004-6256/143/4/80}

\bibitem[{{Schmidt} {et~al.}(2015){Schmidt}, {Hawley}, {West}, {Bochanski},
  {Davenport}, {Ge}, \& {Schneider}}]{2015AJ....149..158S}
{Schmidt}, S.~J., {Hawley}, S.~L., {West}, A.~A., {et~al.} 2015, \aj, 149, 158,
  \dodoi{10.1088/0004-6256/149/5/158}

\bibitem[{{Schneider} {et~al.}(2014){Schneider}, {Cushing}, {Kirkpatrick},
  {Mace}, {Gelino}, {Faherty}, {Fajardo-Acosta}, \&
  {Sheppard}}]{2014AJ....147...34S}
{Schneider}, A.~C., {Cushing}, M.~C., {Kirkpatrick}, J.~D., {et~al.} 2014, \aj,
  147, 34, \dodoi{10.1088/0004-6256/147/2/34}

\bibitem[{{Schneider} \& {Shkolnik}(2018)}]{2018AJ....155..122S}
{Schneider}, A.~C., \& {Shkolnik}, E.~L. 2018, \aj, 155, 122,
  \dodoi{10.3847/1538-3881/aaaa24}

\bibitem[{{Schneider} {et~al.}(2017){Schneider}, {Windsor}, {Cushing},
  {Kirkpatrick}, \& {Shkolnik}}]{2017AJ....153..196S}
{Schneider}, A.~C., {Windsor}, J., {Cushing}, M.~C., {Kirkpatrick}, J.~D., \&
  {Shkolnik}, E.~L. 2017, \aj, 153, 196, \dodoi{10.3847/1538-3881/aa6624}

\bibitem[{{Shappee} {et~al.}(2014){Shappee}, {Prieto}, {Grupe}, {Kochanek},
  {Stanek}, {De Rosa}, {Mathur}, {Zu}, {Peterson}, {Pogge}, {Komossa}, {Im},
  {Jencson}, {Holoien}, {Basu}, {Beacom}, {Szczygie{\l}}, {Brimacombe},
  {Adams}, {Campillay}, {Choi}, {Contreras}, {Dietrich}, {Dubberley},
  {Elphick}, {Foale}, {Giustini}, {Gonzalez}, {Hawkins}, {Howell}, {Hsiao},
  {Koss}, {Leighly}, {Morrell}, {Mudd}, {Mullins}, {Nugent}, {Parrent},
  {Phillips}, {Pojmanski}, {Rosing}, {Ross}, {Sand}, {Terndrup}, {Valenti},
  {Walker}, \& {Yoon}}]{2014ApJ...788...48S}
{Shappee}, B.~J., {Prieto}, J.~L., {Grupe}, D., {et~al.} 2014, \apj, 788, 48,
  \dodoi{10.1088/0004-637X/788/1/48}

\bibitem[{{Shkolnik} {et~al.}(2009){Shkolnik}, {Liu}, \&
  {Reid}}]{2009ApJ...699..649S}
{Shkolnik}, E., {Liu}, M.~C., \& {Reid}, I.~N. 2009, \apj, 699, 649,
  \dodoi{10.1088/0004-637X/699/1/649}

\bibitem[{{Showman} {et~al.}(2020){Showman}, {Tan}, \&
  {Parmentier}}]{2020SSRv..216..139S}
{Showman}, A.~P., {Tan}, X., \& {Parmentier}, V. 2020, \ssr, 216, 139,
  \dodoi{10.1007/s11214-020-00758-8}

\bibitem[{{Silaj} \& {Landstreet}(2014)}]{2014A&A...566A.132S}
{Silaj}, J., \& {Landstreet}, J.~D. 2014, \aap, 566, A132,
  \dodoi{10.1051/0004-6361/201321468}

\bibitem[{{Slesnick} {et~al.}(2006{\natexlab{a}}){Slesnick}, {Carpenter}, \&
  {Hillenbrand}}]{2006AJ....131.3016S}
{Slesnick}, C.~L., {Carpenter}, J.~M., \& {Hillenbrand}, L.~A.
  2006{\natexlab{a}}, \aj, 131, 3016, \dodoi{10.1086/503560}

\bibitem[{{Slesnick} {et~al.}(2006{\natexlab{b}}){Slesnick}, {Carpenter},
  {Hillenbrand}, \& {Mamajek}}]{2006AJ....132.2665S}
{Slesnick}, C.~L., {Carpenter}, J.~M., {Hillenbrand}, L.~A., \& {Mamajek},
  E.~E. 2006{\natexlab{b}}, \aj, 132, 2665, \dodoi{10.1086/508937}

\bibitem[{{Soderblom}(2010)}]{2010ARA&A..48..581S}
{Soderblom}, D.~R. 2010, \araa, 48, 581,
  \dodoi{10.1146/annurev-astro-081309-130806}

\bibitem[{{Stassun} {et~al.}(2019){Stassun}, {Oelkers}, {Paegert}, {Torres},
  {Pepper}, {De Lee}, {Collins}, {Latham}, {Muirhead}, {Chittidi},
  {Rojas-Ayala}, {Fleming}, {Rose}, {Tenenbaum}, {Ting}, {Kane}, {Barclay},
  {Bean}, {Brassuer}, {Charbonneau}, {Ge}, {Lissauer}, {Mann}, {McLean},
  {Mullally}, {Narita}, {Plavchan}, {Ricker}, {Sasselov}, {Seager}, {Sharma},
  {Shiao}, {Sozzetti}, {Stello}, {Vanderspek}, {Wallace}, \&
  {Winn}}]{2019AJ....158..138S}
{Stassun}, K.~G., {Oelkers}, R.~J., {Paegert}, M., {et~al.} 2019, \aj, 158,
  138, \dodoi{10.3847/1538-3881/ab3467}

\bibitem[{{Taylor}(2005)}]{2005ASPC..347...29T}
{Taylor}, M.~B. 2005, in Astronomical Society of the Pacific Conference Series,
  Vol. 347, Astronomical Data Analysis Software and Systems XIV, ed.
  P.~{Shopbell}, M.~{Britton}, \& R.~{Ebert}, 29

\bibitem[{{Terrien} {et~al.}(2012){Terrien}, {Mahadevan}, {Bender},
  {Deshpande}, {Ramsey}, \& {Bochanski}}]{2012ApJ...747L..38T}
{Terrien}, R.~C., {Mahadevan}, S., {Bender}, C.~F., {et~al.} 2012, \apjl, 747,
  L38, \dodoi{10.1088/2041-8205/747/2/L38}

\bibitem[{{Tremblin} {et~al.}(2016){Tremblin}, {Amundsen}, {Chabrier},
  {Baraffe}, {Drummond}, {Hinkley}, {Mourier}, \&
  {Venot}}]{2016ApJ...817L..19T}
{Tremblin}, P., {Amundsen}, D.~S., {Chabrier}, G., {et~al.} 2016, \apjl, 817,
  L19, \dodoi{10.3847/2041-8205/817/2/L19}

\bibitem[{{Tremblin} {et~al.}(2017){Tremblin}, {Chabrier}, {Baraffe}, {Liu},
  {Magnier}, {Lagage}, {Alves de Oliveira}, {Burgasser}, {Amundsen}, \&
  {Drummond}}]{2017ApJ...850...46T}
{Tremblin}, P., {Chabrier}, G., {Baraffe}, I., {et~al.} 2017, \apj, 850, 46,
  \dodoi{10.3847/1538-4357/aa9214}

\bibitem[{{Vacca} {et~al.}(2003){Vacca}, {Cushing}, \&
  {Rayner}}]{2003PASP..115..389V}
{Vacca}, W.~D., {Cushing}, M.~C., \& {Rayner}, J.~T. 2003, \pasp, 115, 389,
  \dodoi{10.1086/346193}

\bibitem[{{Veras} {et~al.}(2009){Veras}, {Crepp}, \&
  {Ford}}]{2009ApJ...696.1600V}
{Veras}, D., {Crepp}, J.~R., \& {Ford}, E.~B. 2009, \apj, 696, 1600,
  \dodoi{10.1088/0004-637X/696/2/1600}

\bibitem[{{Veyette} {et~al.}(2017){Veyette}, {Muirhead}, {Mann}, {Brewer},
  {Allard}, \& {Homeier}}]{2017ApJ...851...26V}
{Veyette}, M.~J., {Muirhead}, P.~S., {Mann}, A.~W., {et~al.} 2017, \apj, 851,
  26, \dodoi{10.3847/1538-4357/aa96aa}

\bibitem[{{Voges} {et~al.}(2000){Voges}, {Aschenbach}, {Boller}, {Brauninger},
  {Briel}, {Burkert}, {Dennerl}, {Englhauser}, {Gruber}, {Haberl}, {Hartner},
  {Hasinger}, {Pfeffermann}, {Pietsch}, {Predehl}, {Schmitt}, {Trumper}, \&
  {Zimmermann}}]{2000IAUC.7432....3V}
{Voges}, W., {Aschenbach}, B., {Boller}, T., {et~al.} 2000, \iaucirc, 7432, 3

\bibitem[{{Vrba} {et~al.}(2004){Vrba}, {Henden}, {Luginbuhl}, {Guetter},
  {Munn}, {Canzian}, {Burgasser}, {Kirkpatrick}, {Fan}, {Geballe},
  {Golimowski}, {Knapp}, {Leggett}, {Schneider}, \&
  {Brinkmann}}]{2004AJ....127.2948V}
{Vrba}, F.~J., {Henden}, A.~A., {Luginbuhl}, C.~B., {et~al.} 2004, \aj, 127,
  2948, \dodoi{10.1086/383554}

\bibitem[{{Walkowicz} {et~al.}(2004){Walkowicz}, {Hawley}, \&
  {West}}]{2004PASP..116.1105W}
{Walkowicz}, L.~M., {Hawley}, S.~L., \& {West}, A.~A. 2004, \pasp, 116, 1105,
  \dodoi{10.1086/426792}

\bibitem[{{West} \& {Hawley}(2008)}]{2008PASP..120.1161W}
{West}, A.~A., \& {Hawley}, S.~L. 2008, \pasp, 120, 1161,
  \dodoi{10.1086/593024}

\bibitem[{{West} {et~al.}(2008){West}, {Hawley}, {Bochanski}, {Covey}, {Reid},
  {Dhital}, {Hilton}, \& {Masuda}}]{2008AJ....135..785W}
{West}, A.~A., {Hawley}, S.~L., {Bochanski}, J.~J., {et~al.} 2008, \aj, 135,
  785, \dodoi{10.1088/0004-6256/135/3/785}

\bibitem[{{West} {et~al.}(2011){West}, {Morgan}, {Bochanski}, {Andersen},
  {Bell}, {Kowalski}, {Davenport}, {Hawley}, {Schmidt}, {Bernat}, {Hilton},
  {Muirhead}, {Covey}, {Rojas-Ayala}, {Schlawin}, {Gooding}, {Schluns},
  {Dhital}, {Pineda}, \& {Jones}}]{2011AJ....141...97W}
{West}, A.~A., {Morgan}, D.~P., {Bochanski}, J.~J., {et~al.} 2011, \aj, 141,
  97, \dodoi{10.1088/0004-6256/141/3/97}

\bibitem[{{Wright} {et~al.}(2010){Wright}, {Eisenhardt}, {Mainzer}, {Ressler},
  {Cutri}, {Jarrett}, {Kirkpatrick}, {Padgett}, {McMillan}, {Skrutskie},
  {Stanford}, {Cohen}, {Walker}, {Mather}, {Leisawitz}, {Gautier}, {McLean},
  {Benford}, {Lonsdale}, {Blain}, {Mendez}, {Irace}, {Duval}, {Liu}, {Royer},
  {Heinrichsen}, {Howard}, {Shannon}, {Kendall}, {Walsh}, {Larsen}, {Cardon},
  {Schick}, {Schwalm}, {Abid}, {Fabinsky}, {Naes}, \&
  {Tsai}}]{2010AJ....140.1868W}
{Wright}, E.~L., {Eisenhardt}, P.~R.~M., {Mainzer}, A.~K., {et~al.} 2010, \aj,
  140, 1868, \dodoi{10.1088/0004-6256/140/6/1868}

\bibitem[{{Yurchenko} \& {Tennyson}(2014)}]{2014MNRAS.440.1649Y}
{Yurchenko}, S.~N., \& {Tennyson}, J. 2014, \mnras, 440, 1649,
  \dodoi{10.1093/mnras/stu326}

\bibitem[{{Yurchenko} {et~al.}(2013){Yurchenko}, {Tennyson}, {Barber}, \&
  {Thiel}}]{2013JMoSp.291...69Y}
{Yurchenko}, S.~N., {Tennyson}, J., {Barber}, R.~J., \& {Thiel}, W. 2013,
  Journal of Molecular Spectroscopy, 291, 69, \dodoi{10.1016/j.jms.2013.05.014}

\bibitem[{{Zapatero Osorio} {et~al.}(2014){Zapatero Osorio}, {G{\'a}lvez
  Ortiz}, {Bihain}, {Bailer-Jones}, {Rebolo}, {Henning}, {Boudreault},
  {B{\'e}jar}, {Goldman}, {Mundt}, \& {Caballero}}]{2014A&A...568A..77Z}
{Zapatero Osorio}, M.~R., {G{\'a}lvez Ortiz}, M.~C., {Bihain}, G., {et~al.}
  2014, \aap, 568, A77, \dodoi{10.1051/0004-6361/201423848}

\bibitem[{{Zhang}(2020)}]{2020RAA....20...99Z}
{Zhang}, X. 2020, Research in Astronomy and Astrophysics, 20, 099,
  \dodoi{10.1088/1674-4527/20/7/99}

\bibitem[{{Zhang} {et~al.}(2021{\natexlab{a}}){Zhang}, {Liu}, {Best}, {Dupuy},
  \& {Siverd}}]{2021ApJ...911....7Z}
{Zhang}, Z., {Liu}, M.~C., {Best}, W. M.~J., {Dupuy}, T.~J., \& {Siverd}, R.~J.
  2021{\natexlab{a}}, \apj, 911, 7, \dodoi{10.3847/1538-4357/abe3fa}

\bibitem[{{Zhang} {et~al.}(2021{\natexlab{b}}){Zhang}, {Liu}, {Claytor},
  {Best}, {Dupuy}, \& {Siverd}}]{2021ApJ...916L..11Z}
{Zhang}, Z., {Liu}, M.~C., {Claytor}, Z.~R., {et~al.} 2021{\natexlab{b}},
  \apjl, 916, L11, \dodoi{10.3847/2041-8213/ac1123}

\bibitem[{{Zhang} {et~al.}(2021{\natexlab{c}}){Zhang}, {Liu}, {Marley}, {Line},
  \& {Best}}]{2021ApJ...916...53Z}
{Zhang}, Z., {Liu}, M.~C., {Marley}, M.~S., {Line}, M.~R., \& {Best}, W. M.~J.
  2021{\natexlab{c}}, \apj, 916, 53, \dodoi{10.3847/1538-4357/abf8b2}

\bibitem[{{Zhang} {et~al.}(2018){Zhang}, {Liu}, {Best}, {Magnier}, {Aller},
  {Chambers}, {Draper}, {Flewelling}, {Hodapp}, {Kaiser}, {Kudritzki},
  {Metcalfe}, {Wainscoat}, \& {Waters}}]{2018ApJ...858...41Z}
{Zhang}, Z., {Liu}, M.~C., {Best}, W.~M.~J., {et~al.} 2018, \apj, 858, 41,
  \dodoi{10.3847/1538-4357/aab269}

\bibitem[{{Zhang} {et~al.}(2020){Zhang}, {Liu}, {Hermes}, {Magnier}, {Marley},
  {Tremblay}, {Tucker}, {Do}, {Payne}, \& {Shappee}}]{2020ApJ...891..171Z}
{Zhang}, Z., {Liu}, M.~C., {Hermes}, J.~J., {et~al.} 2020, \apj, 891, 171,
  \dodoi{10.3847/1538-4357/ab765c}

\bibitem[{{Zuckerman} \& {Song}(2004)}]{2004ARA&A..42..685Z}
{Zuckerman}, B., \& {Song}, I. 2004, \araa, 42, 685,
  \dodoi{10.1146/annurev.astro.42.053102.134111}

\end{thebibliography}

\clearpage
\tabletypesize{\scriptsize}

{\tabletypesize{\scriptsize}
\begin{deluxetable}{lcllcll} 
\tablewidth{0pc} 
\setlength{\tabcolsep}{0.05in} 
\tablecaption{Properties of COCONUTS-3 \label{tab:info}} 
\tablehead{ \multicolumn{1}{l}{}  &  \multicolumn{1}{c}{}  &  \multicolumn{2}{c}{COCONUTS-3A}  &  \multicolumn{1}{c}{}  &  \multicolumn{2}{c}{COCONUTS-3B}  \\ 
\cline{3-4} \cline{6-7}  
\multicolumn{1}{l}{Properties}  &  \multicolumn{1}{c}{}  &  \multicolumn{1}{l}{Value}  &  \multicolumn{1}{l}{Ref.}  &  \multicolumn{1}{c}{}  &  \multicolumn{1}{l}{Value}  &  \multicolumn{1}{l}{Ref.}  }
\startdata
Spectral Type                          & &    M5.5$\pm0.5$ (opt), M4.5$\pm0.5$ (NIR)               & This Work   & &    L6$\pm1$ \textsc{int-g}                         & This Work  \\
Age (Myr)                                & &   $100-1000$                   & This Work   & &     --                          & $\dots$  \\
$[$Fe/H$]$ (dex)                     & &   $0.21 \pm 0.07$        & This Work   & &     --                          & $\dots$  \\
\hline
\multicolumn{7}{c}{Astrometry and Kinematics} \\
\hline
R.A., Decl. (epoch~J2000; hms, dms)       & &   08:13:18.84, $-$15:22:39.6               & Gaia16,20   & &    08:13:22.30, $-$15:22:04.4          & Cham16, Magn20    \\
Gaia EDR3 $\mu_{\alpha}\cos{\delta}$, $\mu_{\delta}$ (mas/yr)   & &   $-131.02 \pm 0.02$, $93.39 \pm 0.02$   & Gaia16,20   & &      &     \\
CatWISE2020 $\mu_{\alpha}\cos{\delta}$, $\mu_{\delta}$ (mas/yr)   & &   $-120.9 \pm 4.4$, $92.1 \pm 5.3$   & Maro21   & &    $-120.8 \pm 6.9$, $92.5 \pm 7.6$  & Maro21    \\
PS1 $\mu_{\alpha}\cos{\delta}$, $\mu_{\delta}$ (mas/yr)   & &   $-126.4 \pm 4.4$, $84.3 \pm 3.3$   & Cham16,Magn20   & &    $-112.6 \pm 21.0$, $101.1 \pm 21.0$  & Cham16,Magn20    \\
Gaia EDR3 Parallax (mas)                         & &   $32.33 \pm 0.02$   & Gaia16,20   & &   --    & $\dots$    \\
Distance (pc)                           & &   $30.88 \pm 0.02$   & Bail21   & &   $32 \pm 7$\tablenotemark{\scriptsize a}    & This Work    \\
Tangential Velocity (km/s)        & &   $23.57 \pm 0.02$   & This Work   & &   $22.3 \pm 1.1$    & This Work    \\
Radial Velocity (km/s)              & &   $41 \pm 60$   & This Work   & &   --    & $\dots$    \\
Position Angle (East of North; deg)  & &   --                & $\dots$   & &   $54.75 \pm 0.03$   & This Work    \\
Projected Separation  & &   --                & $\dots$   & &   $61.22\arcsec\pm0.03\arcsec$ ($1891\pm1.6$~au)    & This Work    \\
\hline
\multicolumn{7}{c}{Spectrophotometric Properties} \\
\hline
Gaia DR2 $G$ (mag)        & &   $14.0348 \pm 0.0005$   & Gaia16,18   & &   --    & $\dots$    \\
Gaia DR2 $BP$ (mag)        & &   $15.835 \pm 0.004$   & Gaia16,18   & &   --    & $\dots$    \\
Gaia DR2 $RP$ (mag)        & &   $12.748 \pm 0.002$   & Gaia16,18   & &   --    & $\dots$    \\
Pan-STARRS1 $g$ (mag)        & &   $16.194 \pm 0.003$   & Cham16   & &   --    & $\dots$    \\
Pan-STARRS1 $r$ (mag)        & &   $14.953 \pm 0.003$   & Cham16   & &   --    & $\dots$    \\
Pan-STARRS1 $i$ (mag)        & &   $13.311 \pm 0.003$\tablenotemark{\scriptsize b}   & Cham16   & &   $22.02 \pm 0.25$    & Cham16    \\
Pan-STARRS1 $z$ (mag)        & &   $12.594 \pm 0.035$\tablenotemark{\scriptsize b}   & Cham16   & &   $20.60 \pm 0.07$    & Cham16    \\
Pan-STARRS1 $y$ (mag)        & &   $12.196 \pm 0.005$\tablenotemark{\scriptsize b}   & Cham16   & &   $19.53 \pm 0.05$    & Cham16    \\
VHS $J$ (mag)        & &   $12.1354 \pm 0.0007$   & McMa13   & &   $17.124 \pm 0.013$    & McMa13    \\
VHS \Ks (mag)        & &   $11.3509 \pm 0.0009$   & McMa13   & &   $15.050 \pm 0.012$    & McMa13    \\
2MASS $J$ (mag)        & &   $12.18 \pm 0.01$\tablenotemark{\scriptsize c}   & This Work   & &   $17.19 \pm 0.03$\tablenotemark{\scriptsize c}    & This Work    \\
2MASS $H$ (mag)        & &   $11.59 \pm 0.05$\tablenotemark{\scriptsize c}   & This Work   & &   $15.89 \pm 0.06$\tablenotemark{\scriptsize c}    & This Work    \\
2MASS \Ks (mag)        & &   $11.32 \pm 0.01$\tablenotemark{\scriptsize c}   & This Work   & &   $15.02 \pm 0.03$\tablenotemark{\scriptsize c}    & This Work    \\
MKO $Y$ (mag)        & &   $12.71 \pm 0.05$\tablenotemark{\scriptsize c}   & This Work   & &   $18.39 \pm 0.07$\tablenotemark{\scriptsize c}    & This Work    \\
MKO $J$ (mag)        & &   $12.14 \pm 0.01$\tablenotemark{\scriptsize c}   & This Work   & &   $17.09 \pm 0.03$\tablenotemark{\scriptsize c}    & This Work    \\
MKO $H$ (mag)        & &   $11.61 \pm 0.05$\tablenotemark{\scriptsize c}   & This Work   & &   $15.95 \pm 0.06$\tablenotemark{\scriptsize c}    & This Work    \\
MKO $K$ (mag)        & &   $11.30 \pm 0.01$\tablenotemark{\scriptsize c}   & This Work   & &   $14.99 \pm 0.03$\tablenotemark{\scriptsize c}    & This Work    \\
$W1$ (mag)              & &   $9.825 \pm 0.013$   & Maro21   & &   $14.037 \pm 0.014$    & Maro21    \\
$W2$ (mag)              & &   $9.645 \pm 0.009$   & Maro21   & &   $13.534 \pm 0.012$    & Maro21    \\
EW(H$\alpha$)  (\r{A})     & &   $-2.9 \pm 0.2$, $-7.18 \pm 0.02$   & This Work   & &   --    & \dots    \\
\hline
\multicolumn{7}{c}{Physical Properties} \\
\hline
$\log{(L_{X})}$ (dex)     & &   $28.3 \pm 0.2$   & This Work   & &   --    & \dots    \\
$\log{(L_{\rm bol}/L_{\odot})}$ (dex)    & &   $-2.80 \pm 0.04$  & This Work    & &    $-4.45 \pm 0.03$                                 & This Work    \\
$P_{\rm rot}$ (hours)    & &   $> 2.7$  & This Work    & &    --                                 & \dots    \\
\Teff\ (K)                                               & &   $2966 \pm 85$         & This Work    & &    $1362^{+48}_{-73}$         & This Work    \\
\logg\ (dex)                                           & &   $5.17 \pm 0.03$       & This Work    & &    $4.96^{+0.15}_{-0.34}$        & This Work    \\
\R\ (A: R$_{\odot}$; B: R$_{\rm Jup}$)  & &   $0.151 \pm 0.004$   & This Work    & &    $1.03^{+0.12}_{-0.06}$         & This Work    \\
\M\ (A: M$_{\odot}$; B: M$_{\rm Jup}$) & &   $0.123 \pm 0.006$      & This Work    & &    $39^{+11}_{-18}$         & This Work    \\
\enddata 
\tablenotetext{a}{Photometric distance derived using the object's \JMKO magnitude, the L6 \textsc{int-g} spectral classification, and the  \cite{2016ApJ...833...96L} empirical relations. } 
\tablenotetext{b}{The $i_{\rm P1}/z_{\rm P1}/y_{\rm P1}$ photometry of the primary star is saturated. } 
\tablenotetext{c}{These photometry are synthesized using the objects' near-infrared spectra and VHS broadband photometry (Sections~\ref{subsec:phys} and \ref{subsec:companion_lbol}). } 
\tablerefs{Bail21: \cite{2021AJ....161..147B}, Cutr03: \cite{2003yCat.2246....0C}, Cham16: \cite{2016arXiv161205560C}, Gaia16: \cite{2016AandA...595A...1G}, Gaia18: \cite{2018AandA...616A...1G}, Gaia20: \cite{2020arXiv201201533G}, McMa13: \cite{2013Msngr.154...35M}, Magn20: \cite{2020ApJS..251....6M}, Maro21: \cite{2021ApJS..253....8M}, Schn17: \cite{2017AJ....153..196S}.} 
\end{deluxetable} 
}

\begin{deluxetable}{l|c|l} 
\tablewidth{0pc} 
\tablecaption{Metallicity and Age of COCONUTS-3A \label{tab:metal_age_primary}} 
\tablehead{ \multicolumn{1}{l}{}  &  \multicolumn{1}{c}{Value\tablenotemark{\scriptsize a}}  &  \multicolumn{1}{l}{Notes}   } 
\startdata 
Metallicity &  $0.16 \pm 0.12$ dex  &  \cite{2014AJ....147...20N} calibration (SXD; $R \sim 2000$)  \\ 
  &  $0.23 \pm 0.08$ dex &  \cite{2014AJ....147..160M} $K$-band calibration (SXD)  \\ 
  &  ($-0.15 \pm 0.17$ dex)  &  \cite{2013AJ....145...52M} optical-band calibration (SNIFS; $R \sim 1200$)  \\ 
  &  ($0.09 \pm 0.13$ dex)  &  \cite{2013AJ....145...52M} $J$-band calibration (SXD)  \\ 
  &  ($-0.17 \pm 0.11$ dex)  &  \cite{2013AJ....145...52M} $H$-band calibration (SXD)  \\ 
  &  ($0.12 \pm 0.09$ dex)  &  \cite{2013AJ....145...52M} $K$-band calibration (SXD)  \\ 
  &  ($0.11 \pm 0.15$ dex)  &  \cite{2012ApJ...747L..38T} $H$-band calibration as updated by \cite{2013AJ....145...52M} (SXD)  \\ 
  &  ($0.15 \pm 0.14$ dex)  &  \cite{2012ApJ...747L..38T} $K$-band calibration as updated by \cite{2013AJ....145...52M} (SXD)  \\ 
\hline 
\textbf{Adopted Metallicity} &  $\mathbf{0.21 \pm 0.07}$ {\bf dex} &   \\ 
\hline \hline 
Stellar-Activity Age  &  $870^{+940}_{-570}$ Myr  &  EW$_{\rm H\alpha} = -7.18 \pm 0.02$~\r{A} (GMOS; $R \sim 4400$)  \\
  &  ($2.9^{+2.4}_{-1.4}$ Gyr)  &  EW$_{\rm H\alpha} = -2.9 \pm 0.2$~\r{A} (SNIFS)  \\
  &  $\lesssim 850$ Myr  &  $\log{(L_{X})} = 28.3 \pm 0.2$~dex, $\log{(R_{X})} = -2.7 \pm 0.3$ dex  \\
  &  $10-750$ Myr or older  & $\log{(F_{NUV}/F_{J})} < -3.7$ dex, $NUV-G > 8.3$ mag, $NUV-J > 10.1$ mag   \\
Kinematics Age  &  $\gtrsim 150$ Myr  &  $UVW = (-34.3 \pm 32.6, -12.1 \pm 48.8, -3.9 \pm 10.7)$ km s$^{-1}$  \\
HR Diagram Age  &  $100$ Myr $-1.5$ Gyr  &  $M_{G} = 11.5826 \pm 0.0015$ mag, BP$-$RP$= 3.088 \pm 0.004$ mag  \\
Lithium Age  &  $\gtrsim 100$ Myr  & EW$_{\rm Li} < 4$ mA (GMOS)   \\
Spectral-Index Age  &  a few 100 Myr  & Na I index$= 1.23$, Na-8189 index$= 0.84$, EW$_{\rm Na8200} = 4.50$ \r{A} (GMOS)  \\
  &  a few $10-100$ Myr  & EW$_{\rm Na1.138} = 5.6 \pm 0.2$ \r{A} (SXD) \\
Rotation-based Age  &  a few Myr to Gyr  & $P_{\rm rot} \gtrsim  2.7$~hours \\
\hline 
\textbf{Adopted Age} &  $\mathbf{100}$ {\bf Myr} $\mathbf{-1}$ {\bf Gyr}  &      \\ 
\enddata 
\tablenotetext{a}{Metallicities inside the parenthesis are derived from empirical calibrations that are not applicable to COCONUTS-3A's spectral type, and have been therefore excluded when computing the final metallicity. Ages inside the parenthesis is not well-constrained and thereby excluded for determining the final age.   } 
\end{deluxetable}

\clearpage

\begin{deluxetable}{lccccccccc} 
\tablewidth{0pc} 
\tablecaption{Gravity Classification of COCONUTS-3A and 3B based on the \cite{2013ApJ...772...79A} System \label{tab:al13_index}} 
\tablehead{ \multicolumn{1}{l}{Spectral Indices}  &  \multicolumn{1}{c}{}  &  \multicolumn{2}{c}{COCONUTS-3A}  &  \multicolumn{1}{c}{}  &  \multicolumn{5}{c}{COCONUTS-3B}   \\
\cline{3-4} \cline{6-10}
\multicolumn{1}{l}{or Equivalent Widths (\r{A})}  &  \multicolumn{1}{c}{}  &  \multicolumn{2}{c}{SpeX SXD}  &  \multicolumn{1}{c}{}  &  \multicolumn{2}{c}{SpeX prism}  &  \multicolumn{1}{c}{}  &  \multicolumn{2}{c}{GNIRS XD}} 
\startdata 
FeH$_{z}$    &    &    \multicolumn{2}{c}{$1.053 \pm 0.002$}    &    &    \multicolumn{2}{c}{$1.126^{+0.021}_{-0.023}$}   &    &    \multicolumn{2}{c}{$1.170^{+0.016}_{-0.015}$}     \\
VO$_{z}$    &    &    \multicolumn{2}{c}{$0.9902^{+0.0013}_{-0.0012}$}    &    &    \multicolumn{2}{c}{$0.986^{+0.011}_{-0.012}$}   &    &    \multicolumn{2}{c}{$1.004 \pm 0.007$}     \\
K~\textsc{i}$_{J}$    &    &    \multicolumn{2}{c}{$1.0231^{+0.0011}_{-0.0012}$}    &    &    \multicolumn{2}{c}{$1.030^{+0.009}_{-0.008}$}   &    &    \multicolumn{2}{c}{$1.061 \pm 0.004$}     \\
H-cont    &    &    \multicolumn{2}{c}{$0.9909^{+0.0010}_{-0.0009}$}    &    &    \multicolumn{2}{c}{$0.895 \pm 0.008$}   &    &    \multicolumn{2}{c}{$0.906 \pm 0.004$}     \\
Na~\textsc{i} $1.138$ $\mu$m    &    &    \multicolumn{2}{c}{$5.611^{+0.174}_{-0.165}$}    &    &    \multicolumn{2}{c}{$\dots$}   &    &    \multicolumn{2}{c}{$7.464 \pm 0.746$}     \\
K~\textsc{i} $1.169$ $\mu$m    &    &    \multicolumn{2}{c}{$\dots$}    &    &    \multicolumn{2}{c}{$\dots$}    &    &    \multicolumn{2}{c}{$6.616 \pm 0.693$}    \\
K~\textsc{i} $1.177$ $\mu$m    &    &    \multicolumn{2}{c}{$\dots$}    &    &    \multicolumn{2}{c}{$\dots$}    &    &    \multicolumn{2}{c}{$8.042 \pm 0.533$}    \\
K~\textsc{i} $1.253$ $\mu$m    &    &    \multicolumn{2}{c}{$1.435^{+0.124}_{-0.135}$}    &    &    \multicolumn{2}{c}{$\dots$}    &    &    \multicolumn{2}{c}{$4.814 \pm 0.397$}    \\
\hline
Gravity Class    &    &    \multicolumn{2}{c}{\textsc{int-g}}    &    &    \multicolumn{2}{c}{\textsc{int-g}}    &    &    \multicolumn{2}{c}{\textsc{int-g}}    \\
\enddata 
\end{deluxetable}

\begin{deluxetable}{lcccccc} 
\tablewidth{0pc} 
\tablecaption{Spectral Types of COCONUTS-3B based on the \cite{2006ApJ...637.1067B} System\label{tab:b06_index}} 
\tablehead{ \multicolumn{1}{l}{Spectroscopy}  &  \multicolumn{1}{c}{H$_{2}$O-$J$}  &  \multicolumn{1}{c}{CH$_{4}$-$J$}  &  \multicolumn{1}{c}{H$_{2}$O-$H$}  &  \multicolumn{1}{c}{CH$_{4}$-$H$}  &  \multicolumn{1}{c}{CH$_{4}$-$K$}  &  \multicolumn{1}{c}{Averaged Spectral Type}   } 
\startdata 
SpeX prism ($R \approx 75$) &  $0.682$ (L7.8) &  $0.902$ ($<$T0) &  $0.709$ (L6.2) &  $1.125$ ($<$T1) &  $1.006$ (L4.2) &  L6.1$\pm$1.8  \\ 
GNIRS XD ($R \approx 750$) &  $0.710$ (L7.0) &  $0.908$ ($<$T0) &  $0.703$ (L6.4) &  $1.118$ ($<$T1) &  $0.994$ (L4.6) &  L6.0$\pm$1.3  \\ 
\enddata 
\end{deluxetable}

\begin{longrotatetable} 
\begin{deluxetable}{lccccccccccccccccc} 
\tablewidth{0pc} 
\setlength{\tabcolsep}{0.03in} 
\tablecaption{Properties of Background Stars Near COCONUTS-3A \label{tab:xray_neighbors}} 
\tablehead{ \multicolumn{1}{l}{}  &  \multicolumn{1}{c}{}  &  \multicolumn{4}{c}{Gaia EDR3\tablenotemark{\scriptsize a}}  &  \multicolumn{1}{c}{}  &  \multicolumn{2}{c}{Gaia DR2}  &  \multicolumn{1}{c}{}  &  \multicolumn{1}{c}{}  &  \multicolumn{1}{c}{}  &  \multicolumn{1}{c}{}  &  \multicolumn{5}{c}{Estimated Properties}  \\ 
\cline{3-6} \cline{8-9} \cline{14-18} 
\multicolumn{1}{l}{Object}  &  \multicolumn{1}{c}{}  &  \multicolumn{1}{c}{R.A.}  &  \multicolumn{1}{c}{Decl.}  &  \multicolumn{1}{c}{parallax}  &  \multicolumn{1}{c}{distance}  &  \multicolumn{1}{c}{}  &  \multicolumn{1}{c}{$G$}  &  \multicolumn{1}{c}{BP$-$RP}  &  \multicolumn{1}{c}{}  &  \multicolumn{1}{c}{$M_{G}$\tablenotemark{\scriptsize b}}  &  \multicolumn{1}{c}{Separation\tablenotemark{\scriptsize c}}  &  \multicolumn{1}{c}{}  &  \multicolumn{1}{c}{SpT}  &  \multicolumn{1}{c}{$B-V$}  &  \multicolumn{1}{c}{$\log{(L_{\rm bol}/L_{\odot})}$}  &  \multicolumn{1}{c}{$\log{(L_{X}/L_{\rm bol})}$}  &  \multicolumn{1}{c}{$F_{X}$}     \\ 
\multicolumn{1}{l}{}  &  \multicolumn{1}{c}{}  &  \multicolumn{1}{c}{(hh:mm:ss.ss)}  &  \multicolumn{1}{c}{(dd:mm:ss.s)}  &  \multicolumn{1}{c}{(mas)}  &  \multicolumn{1}{c}{(pc)}  &  \multicolumn{1}{c}{}  &  \multicolumn{1}{c}{(mag)}  &  \multicolumn{1}{c}{(mag)}  &  \multicolumn{1}{c}{}  &  \multicolumn{1}{c}{(mag)}  &  \multicolumn{1}{c}{(\arcsec)}  &  \multicolumn{1}{c}{}  &  \multicolumn{1}{c}{}  &  \multicolumn{1}{c}{(mag)}  &  \multicolumn{1}{c}{(dex)}  &  \multicolumn{1}{c}{(dex)}  &  \multicolumn{1}{c}{($10^{-15}$ erg~s$^{-1}$~cm$^{-2}$)}  } 
\startdata 
 BG1 & & 08:13:20.87 & -15:25:40.2 & $0.34 \pm 0.04$ & $2596 \pm 276$ & & $16.045 \pm 0.001$ & $0.775 \pm 0.004$ & & $3.97 \pm 0.23$ & 25.3 & & G0V & $0.595$ & $0.13 \pm 0.05$ & $-3.67 \pm 0.34$ & $\lesssim 1.37 \pm 1.12$ \\ 
 BG2 & & 08:13:14.56 & -15:25:48.8 & $0.46 \pm 0.02$ & $2052 \pm 68$ & & $11.170 \pm 0.001$ & $1.462 \pm 0.001$ & & $-0.39 \pm 0.07$ & 24.3 & & K giant & $\dots$ & $\dots$ & $\dots$ & $\lesssim  0.05$ \\ 
 BG3 & & 08:13:14.66 & -15:25:44.3 & $0.08 \pm 0.14$ & $11602 \pm 1867$ & & $18.261 \pm 0.002$ & $0.445 \pm 0.022$ & & $2.94 \pm 0.35$ & 25.9 & & F1V & $0.330$ & $0.79 \pm 0.08$ & $-4.28 \pm 0.50$ & $\lesssim 0.08 \pm 0.09$ \\ 
 BG4 & & 08:13:14.23 & -15:25:38.9 & $0.37 \pm 0.07$ & $2773 \pm 420$ & & $17.037 \pm 0.001$ & $0.852 \pm 0.008$ & & $4.82 \pm 0.33$ & 30.0 & & G5V & $0.680$ & $-0.05 \pm 0.05$ & $-3.71 \pm 0.47$ & $\lesssim 0.72 \pm 0.82$ \\ 
 BG5 & & 08:13:16.45 & -15:25:48.4 & $0.27 \pm 0.50$ & $2964 \pm 317$ & & $19.882 \pm 0.005$ & $1.455 \pm 0.084$ & & $7.52 \pm 0.23$ & 14.2 & & K5V & $1.150$ & $-0.76 \pm 0.10$ & $-3.35 \pm 0.37$ & $\lesssim 0.28 \pm 0.26$ \\ 
 BG6 & & 08:13:17.25 & -15:25:47.2 & $0.03 \pm 0.50$ & $4872 \pm 612$ & & $19.771 \pm 0.005$ & $1.214 \pm 0.105$ & & $6.33 \pm 0.27$ & 24.0 & & K3V & $0.990$ & $-0.55 \pm 0.14$ & $-3.35 \pm 0.37$ & $\lesssim 0.17 \pm 0.16$ \\ 
 BG7 & & 08:13:17.30 & -15:25:40.8 & $0.39 \pm 0.07$ & $2851 \pm 473$ & & $16.977 \pm 0.001$ & $0.785 \pm 0.009$ & & $4.70 \pm 0.36$ & 28.7 & & G1V & $0.622$ & $0.08 \pm 0.07$ & $-3.67 \pm 0.34$ & $\lesssim 1.01 \pm 0.87$ \\ 
\enddata 
\tablenotetext{a}{We provide Gaia EDR3 coordinates at epoch J2000 with equinox J2000 and the \cite{2021AJ....161..147B} distances. } 
\tablenotetext{b}{We compute $G$-band absolute magnitudes by combining the Gaia DR2 photometry and Gaia EDR3 distances.} 
\tablenotetext{c}{Angular separations between J2000 coordinates of background sources and the X-ray detection 2RXS~J081318.4$-$152252.} 
\end{deluxetable} 
\end{longrotatetable}

\vfill
\eject
\end{document}